\documentclass[%
reprint,
superscriptaddress,
 amsmath,amssymb,
 prx,
]{revtex4-2}
\usepackage{graphicx}
\usepackage{float}
\usepackage{braket}
\usepackage[T1]{fontenc}
\usepackage{xcolor}
\usepackage{hyperref}
\usepackage{siunitx}

\newcommand{\calciumforty}{$^{40}\mathrm{Ca}^+$}
\newcommand{\m}[1]{\mathrm{#1}}

\newcommand{\latticename}{QSA}
\usepackage{tikz}
\newcommand*\circled[1]{\tikz[baseline=(char.base)]{
    \node[shape=circle,draw,inner sep=.5pt] (char) {#1};}}
\usepackage[normalem]{ulem}

\begin{document}

\title{Demonstration of two-dimensional connectivity for a scalable error-corrected ion-trap quantum processor architecture}

\author{M. Valentini\textsuperscript*}
\author{M. W. van Mourik\textsuperscript*}
\affiliation{Institut f\"{u}r Experimentalphysik, Universit\"{a}t Innsbruck, Technikerstraße 25/4, 6020 Innsbruck, Austria}
\altaffiliation[]{These authors contributed equally to this work.}
\author{F. Butt}
\affiliation{Institute for Theoretical Nanoelectronics (PGI-2), Forschungszentrum Jülich, 52428 Jülich, Germany}
\affiliation{Institute for Quantum Information, RWTH Aachen University, D-52056 Aachen, Germany}
\author{J. Wahl}
\author{M. Dietl}
\author{M. Pfeifer}
\author{F. Anmasser}
\affiliation{Institut f\"{u}r Experimentalphysik, Universit\"{a}t Innsbruck, Technikerstraße 25/4, 6020 Innsbruck, Austria}
\affiliation{Infineon Technologies Austria AG, Siemensstraße 2, 9500, Villach, Austria}

\author{Y. Colombe}
\author{C. Rössler}
\affiliation{Infineon Technologies Austria AG, Siemensstraße 2, 9500, Villach, Austria}
\author{P. C. Holz}
\affiliation{Alpine Quantum Technologies GmbH, Technikerstraße 17/1, 6020 Innsbruck, Austria}
\author{R. Blatt}
\affiliation{Institut f\"{u}r Experimentalphysik, Universit\"{a}t Innsbruck, Technikerstraße 25/4, 6020 Innsbruck, Austria}
\affiliation{Alpine Quantum Technologies GmbH, Technikerstraße 17/1, 6020 Innsbruck, Austria}
\affiliation{Institute for Quantum Optics and Quantum Information, Austrian Academy of Sciences, Technikerstraße 21a, 6020 Innsbruck, Austria}
\author{A. Bermudez}
\affiliation{Instituto de Física Teórica UAM-CSIC, Universidad Autónoma de Madrid, Cantoblanco, 28049, Madrid, Spain}
\author{M. M\"{u}ller}
\affiliation{Institute for Theoretical Nanoelectronics (PGI-2), Forschungszentrum Jülich, 52428 Jülich, Germany}
\affiliation{Institute for Quantum Information, RWTH Aachen University, D-52056 Aachen, Germany}
\author{T. Monz}
\affiliation{Institut f\"{u}r Experimentalphysik, Universit\"{a}t Innsbruck, Technikerstraße 25/4, 6020 Innsbruck, Austria}
\affiliation{Alpine Quantum Technologies GmbH, Technikerstraße 17/1, 6020 Innsbruck, Austria}
\author{P. Schindler}
\affiliation{Institut f\"{u}r Experimentalphysik, Universit\"{a}t Innsbruck, Technikerstraße 25/4, 6020 Innsbruck, Austria}

\begin{abstract}
A major hurdle for building a large-scale quantum computer is increasing the number of qubits while maintaining connectivity between them.
In trapped-ion devices, this connectivity can be achieved by moving subregisters consisting of a few ions across the processor. 
Here, we focus on an architecture, which we refer to as the Quantum Spring Array (QSA), that is based on a rectangular two-dimensional lattice of linear strings of ions. 
Connectivity between adjacent ion strings can be controlled by adjusting their separation.
This requires control of trapping potentials along two directions, one along the axis of the ion string and one radial to it.
In this work, we investigate key elements of the QSA architecture along both directions: We show that the coupling rate between neighboring lattice sites increases with the number of ions per site and the motion of the coupled system can be resilient to electrical noise, both being key requisites for fast and high-fidelity quantum gate operations. 
The coherence of the coupling is assessed and an entangling gate between qubits stored in radially separated trapping regions is demonstrated. Moreover, we demonstrate control over radio-frequency signals to adjust the radial separation, and thus the coupling rate, between strings.
We further present constructions for the implementation of parallelized, transversal gate operations, and map the QSA architecture to code primitives for fault-tolerant quantum error correction, providing a step towards a quantum processor architecture that is optimized for large-scale operation.
\end{abstract}

\maketitle
\section{Introduction}
\label{sec:introduction}
Ion traps are among the most promising platforms to host quantum computers and simulators. 
A prominent approach towards scaling ion-based quantum computers to a large numbers of ions is that of the Quantum Charge Coupled Device (QCCD) architecture \cite{Kielpinski2002,Pino2021}, in which different zones in a microfabricated surface ion trap are used for dedicated tasks such as qubit storage and interaction. 
In the QCCD architecture, independent trapping regions contain linear strings of ions confined in an individual potential well. Each ion string is  considered a quantum subregister and connectivity between qubits in different subregisters requires physical transport of the ions corresponding to the respective subregisters. These reorganization operations require usually  splitting and merging of ion crystals~\cite{Palmero2015,Kaufmann2014}.
For a scalable system, these one-dimensional trapping regions need to be transferred into a higher-dimensional arrangement of interconnected qubit registers. In the QCCD approach, any higher-dimensional architecture requires additional operations such as ion-crystal rotations~\cite{VanMourik2020,Kaufmann2017}, and transport operations through junctions~\cite{Wright2013a,Shu2014,Burton2023}. 

Altering the configuration of the subregisters is challenging to implement: during splitting and merging operations, ion strings are subjected to a decreased confining potential, making them more susceptible to motional excitation due to electrical noise, which limits the performance of subsequent operations~\cite{Kaufmann2014}. Also, these operations are very sensitive to electric stray fields caused by imperfect trap surfaces. Furthermore, ions that are transported through junctions cannot avoid experiencing an increase in residual RF field and lowered RF confinement~\cite{Wesenberg2009}, and junctions must be carefully designed to limit this type of effects. These transport operations require a significant portion of the total execution time of a quantum circuit~\cite{Pino2021,Moses2023}. While operations such as splitting, merging, rotation and transport through junctions have been demonstrated for ion chains with up to four ions~\cite{Moses2023}, the complexity increases for larger numbers of ions per register.

\begin{figure*}
    \centering    \includegraphics[width=\linewidth]{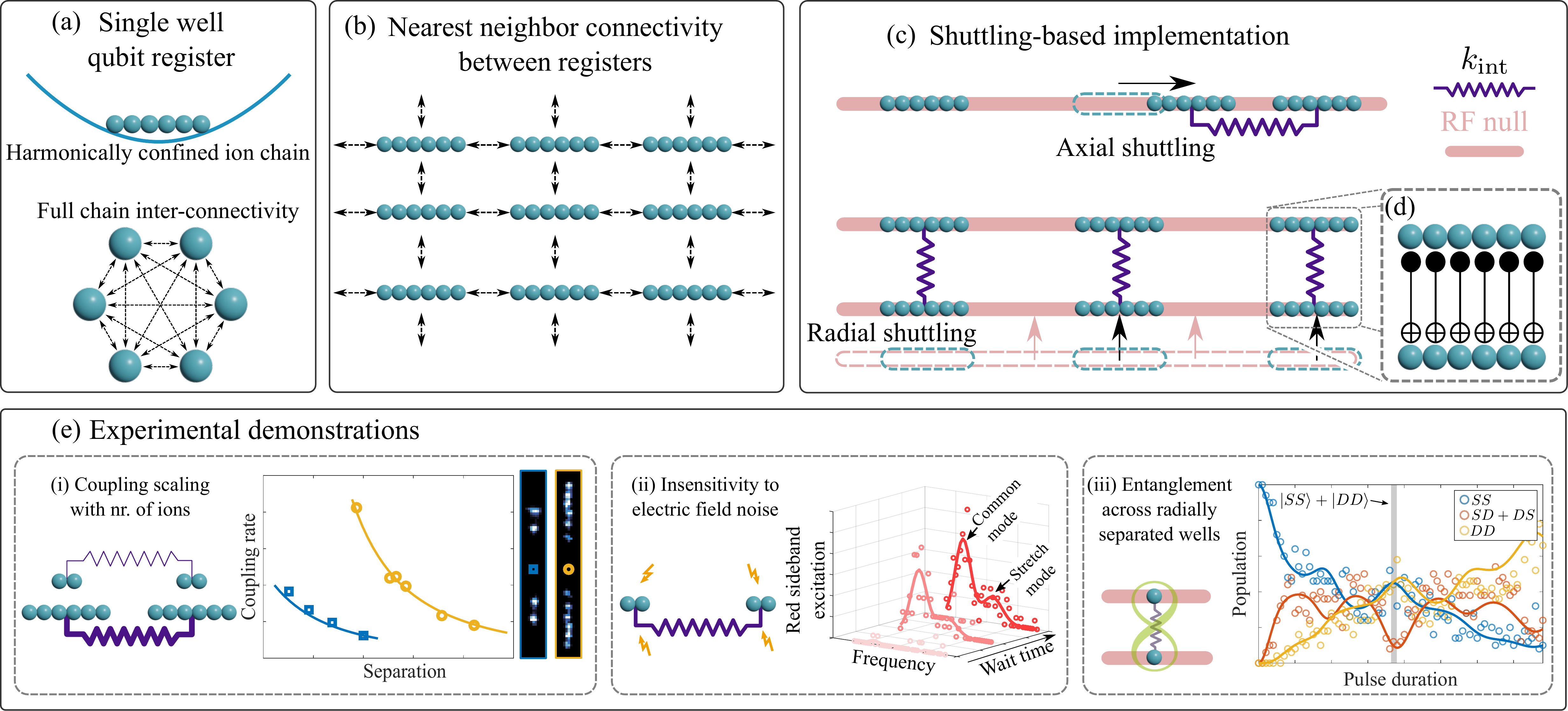}
    \caption{Overview of the main features of the proposed \latticename{} architecture. (a) A chain of ionic qubits are confined in a single well of an ion trap, with full connectivity between all qubits. (b) Multiple separate trapping regions are distributed  over a 2D lattice. 
    Trapping sites have connectivity between nearest neighbors.
    (c) A shuttling-based approach is used to bring ion chains close enough to couple them, with an interaction constant $k_{\m{int}}$, symbolically depicted in purple as springs. Axial shuttling is achieved by displacing an ion string along a linear trap's RF null, and radial shuttling is achieved by displacing the RF null itself. When coupled, all-to-all connectivity between ions in separate strings is enabled;
    the use of such an architecture for quantum error correction codes, and the ability to apply transversal operations between chains of ions, as depicted in (d), are both theoretically investigated in this work. (e) Experimental highlights discussed in this work include: (i) the favorable scaling of coupling rate with number of ions, (ii) the insensitivity to electric field noise of the double well stretch mode of the coupled-ion system compared to the common mode, and (iii) generation of an entangled state, $\ket{SS}+\ket{DD}$, in radially separated ions, mediated by an oscillatory exchange of motion between two separated wells.}
    \label{fig:graphical_abstract}
\end{figure*}

Here we focus on an alternate method to scaling up trapped-ion quantum processors that does not rely on transport of ions between trapping regions: a two-dimensional \emph{lattice} of ion traps \cite{Holz2020,Jain2020,Mielenz2016-mc,PhysRevA.107.L050601}. The well-established core building block on each lattice site in this architecture remains a one-dimensional chain of ions. Full ion-to-ion connectivity within such a chain, schematically shown in Fig. \ref{fig:graphical_abstract}(a), can be achieved by individually addressed laser pulses. Entanglement within a lattice site is mediated by the chain's common mode of motion. The quantum processor is then built up from a physical 2D lattice configuration of these chains~\cite{Schmied2009,Hakelberg2019}. Connectivity between neighboring lattice sites, and thus subregisters, is achieved through Coulomb interaction between the individual ion chains, as shown in Fig. \ref{fig:graphical_abstract}(b). This level of connectivity permits universal scalable quantum computation and thereby the execution of arbitrary quantum algorithms, but avoids the complications involved in multi-ion splitting, merging, and junction transport. We refer to this coupled-lattice architecture as a Quantum Spring Array (QSA).

In the \latticename{} architecture, the default configuration has negligible coupling between ion chains by setting a large-enough distance between them. Interaction between chains is enabled by moving two chains of ions close to each other, though without merging them into a single potential well. Due to the geometrical configuration of linear ion traps, the two orthogonal directions of shuttling require two different types of control. As schematically depicted in Fig.~\ref{fig:graphical_abstract}(c), these directions are denoted as axial and radial, referring to their orientation with respect to the 1D ion string. 

Coherent coupling between subregisters is a crucial element of scalable quantum information processing. The \latticename{} architecture avoids the necessity of splitting and merging of ion chains, and shuttling through junctions. 
This also relaxes the constraint that the subregisters are limited to a small numbers of ions. 
In fact, in the \latticename{} approach, larger register sizes are advantageous, as it enhances coupling between neighbors~\cite{Harlander2011}. 

In this work, we demonstrate these advantages by investigating interaction between ion chains along both the axial and radial directions. A theoretical foundation for these interactions is laid out in Section \ref{sec:double_well_physics}. For our experimental demonstrations, we use trapped \calciumforty ions on two separate types of surface traps, each with an electrode layout specifically suited for one of the two schemes.
While axial coupling and entanglement between ions in different potential wells has previously been observed \cite{Harlander2011,Brown2011,Wilson2014}, here we highlight the increased coupling rate that emerges from large ion crystals (Fig.~\ref{fig:graphical_abstract}(e)(i)) and show that the coupled ion strings contain usable motional modes that are resilient to electric field noise (Fig.~\ref{fig:graphical_abstract}(e)(ii)). These features are discussed for the axial direction in Section \ref{sec:axial}, and for the radial direction in Section \ref{sec:radial_entanglement}. Our work furthermore presents the first realization of coherent coupling and entangling gate operations (Fig.~\ref{fig:graphical_abstract}(e)(iii)) in radially separated linear traps, signifying an important step towards control of qubits in two dimensions (also Section \ref{sec:radial_entanglement}). On the route towards scalable control of 2D trap arrays, we demonstrate that through control of RF voltages, we are able to adjust the separation, and thus coupling rate, between neighboring ion strings (Section \ref{sec:radial_transport}). Finally, we introduce a new method for implementing parallel transversal entangling gates that is specifically designed for the presented 2D architecture and explore avenues to harness the potential of the \latticename{} architecture for scalable quantum computation, specifically within the context of quantum error correction (QEC) (Section \ref{sec:theory}). Here, the overhead of shuttling operations can be minimized by matching the QEC code to the given architecture. As an example, we investigate mapping of concatenated quantum codes onto the \latticename{} architecture, providing a pathway to error-corrected and universal logical quantum computation. 

\section{Theory: Interaction between separated ion chains}
\label{sec:double_well_physics}
This section provides an overview of the principles of coupled charges in separated harmonic potentials. We introduce the two orientations of coupling used in this work (axial and radial) and show how the coupling rate depends on the number of trapped charges for each orientation. We also numerically investigate the motional mode structure of coupled charges, and introduce a model that determines sensitivity of these modes to voltage noise on the trap electrodes.

Trapped ion qubits are usually stored in subregisters comprising of one-dimensional ion chains, whose axis is aligned with the so-called RF-null~\cite{Blatt2008,Monroe2013}. This region coincides with a minimum in the effective trapping potential experienced by ions in the radio frequency field created by the trap electrodes. Ion transport along the one-dimensional RF-null is referred to as axial shuttling. This can be achieved through control of the static trapping fields which are known as DC potentials. While some deviation of the ions' positions from the RF-null is permissible~\cite{VanMourik2020}, excursions of several micrometers are detrimental to the performance of ions as a string of qubits, since the larger amplitude of the RF-field destabilizes ion chains \cite{Berkeland1998,DeVoe1989}. Thus, two-dimensional ion transport requires a second distinct strategy of manipulation of trapping potentials: We implement radial shuttling by altering the position of the RF-null itself. The transport strategies are schematically depicted in Fig. \ref{fig:schematic_axial_and_radial_coupling}(a).

In both axial and radial transport strategies, the aim is to bring two (pseudo)potential minima, each containing a group of ions, towards each other, though without merging them into a single well. This allows separated ion chains to experience an enhanced coupling through Coulomb interaction, facilitating entangling gate operations. In this section, we describe the formalism for coupling of ion chains in separate potential wells as a function of the spatial distribution and charge of the ion chains, and orientation of their modes of oscillation.

\subsection{Coupling rate of separated charges}
\label{subsec:point_charge_model}
An approximate way to model the coupling of separated ion chains is to assume the interaction potential is that of a dipole-dipole interaction of point charges~\cite{Harlander2011,Brown2011}. The interaction potential is given by
\begin{equation}
    \label{eq:interaction_potential}
    V_{\mathrm{int}} = \frac{Q_1 Q_2}{4\pi\epsilon_0} \frac{\vec{r}_1\cdot\vec{r}_2 - 3(\vec{r}_1\cdot\hat{e}_d)(\vec{r}_2\cdot\hat{e}_d)}{d^3}
\end{equation}
with $Q_{\{1,2\}}$ the total charge in each of the two wells, $\epsilon_0$ the dielectric constant in vacuum, and $\vec{r}_{\{1,2\}}$ the displacement of the chains from their equilibrium positions. Here, $d$ is the distance between charges, and $\hat{e}_d$ the unit vector in the direction of separation.

\begin{figure}
    \centering
    \includegraphics[width=\linewidth]{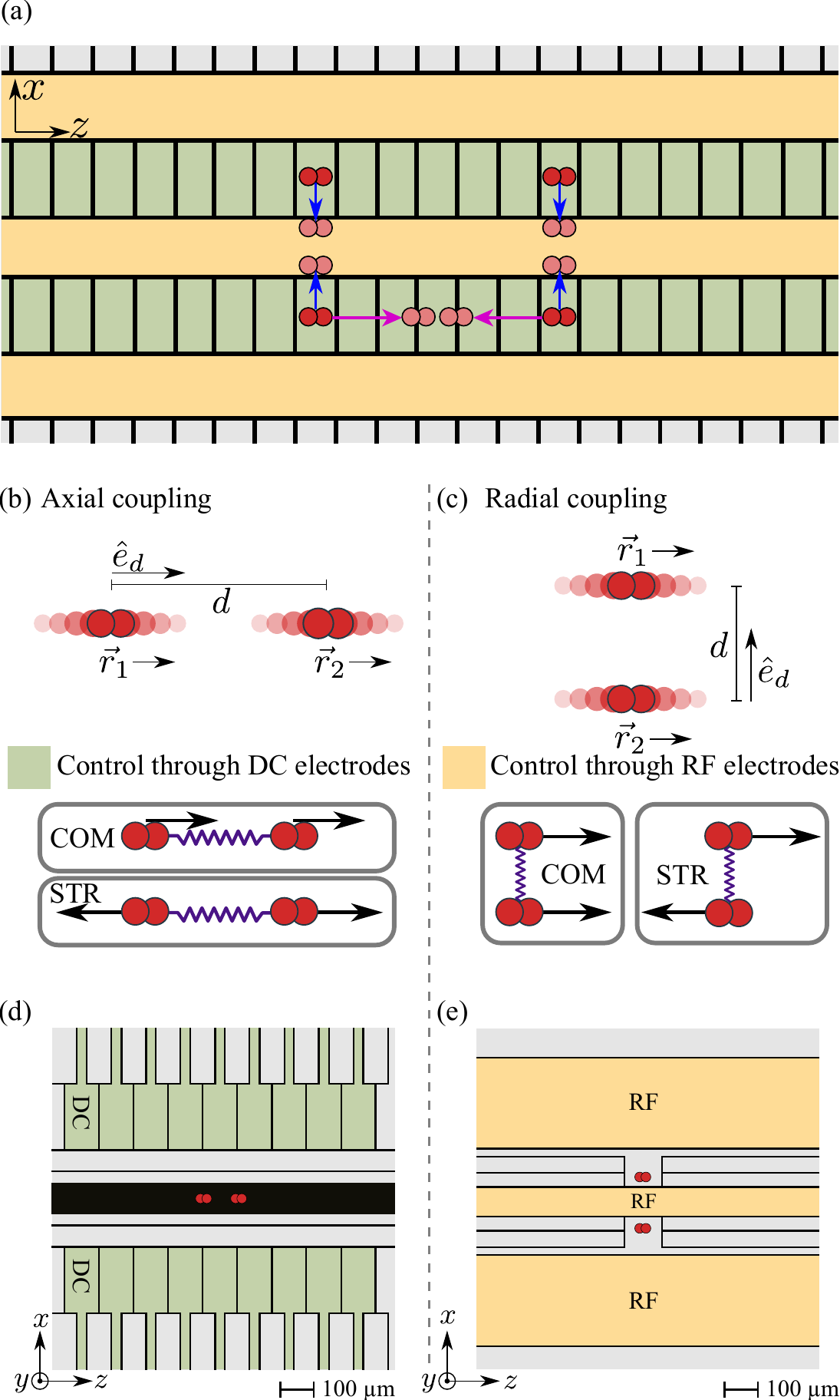}
    \caption{(a) Schematic depiction of an ion trap layout that enables axial (purple arrows) and radial (blue arrows) shuttling to adjust the separation between trapping sites in two dimensions. The direction of separation $\hat{e}_d$ lies along the (b) axial and (c) radial direction with respect to the ion chain's axial mode of oscillation leading, respectively, to axial and radial coupling of ion chains. In both cases, coupled ions exhibit shared axial modes of motion, and can oscillate in phase (COM - common), or out of phase (STR - stretch). In this work, the two means of coupling are investigated in separate ion traps, whose layouts are shown in (d) and (e).}
    \label{fig:schematic_axial_and_radial_coupling}
\end{figure}

In this work we consider the well-to-well interaction of ions through their \emph{axial} motion, the mode of motion that aligns with the axis of the RF-null(s). Here, we choose to define the axial direction to lie along the $z$-axis. We choose this mode instead of radial modes of motion, since it typically has longer coherence times as it is defined by static voltages, and has a stronger coupling to laser fields than the radial modes. The stronger coupling to the laser field is due to the higher Lamb-Dicke parameter associated with the lower trap frequency. For the trap geometries presented in this work, the axial modes of each well are parallel, such that $\vec{r}_1\cdot\vec{r}_2=|\vec{r}_1||\vec{r}_2|$. The axis of well-to-well separation, given by $\hat{e}_d$, is parallel to the axial modes in axial shuttling and perpendicular to the axial modes in radial shuttling. In other words, the interaction is represented by dipole-dipole coupling with dipoles oscillating either in the direction of separation or perpendicular to it. These coupling directions, shown schematically in Fig. \ref{fig:schematic_axial_and_radial_coupling}(b) and (c), are referred to as axial and radial coupling (of axial modes), respectively. The second fraction in Eq.~\ref{eq:interaction_potential} reduces to $\kappa|\vec{r}_1||\vec{r}_2|/d^3$, with $\kappa=-2$ for axial coupling and $\kappa=1$ for radial coupling. The interaction potential can thus be written as $V_{\m{int}}=k_{\m{int}}|r_1| | r_2|$, where $k_{\m{int}}$ is an interaction constant analogous to a spring constant, given by 
\begin{equation}
\label{eq:interaction_constant}
k_{\m{int}} =\kappa\frac{Q_1 Q_2}{4\pi\epsilon_0 d^3}
\end{equation}

The motion of a collection of $n$ trapped ions with mass $m$ and charge $q$ in a single potential well can be expressed as that of a harmonic oscillator. The curvature $\phi_z$ of the confining potential defines the uncoupled common oscillation frequency along the axial direction as $\omega_z=\sqrt{q\phi_z /m}$.
Thus, two chains of ions trapped in separate wells can be described as two coupled harmonic oscillators, with an interaction potential given by Eq.~(\ref{eq:interaction_potential}).

We consider the case that the collection of ions in both wells contain the same number $n$ of identical ions, and thus have the same overall charge $Q = n q$ and mass $n m$. Furthermore, we assume the axial curvature $\phi_z$ to be identical in both wells. 
Neglecting internal degrees of freedom within individual ion chains, the total potential energy relevant to the system's axial modes of motion is given by $V_z = \phi_{z} (r_{z,1}^2 + r_{z,2}^2) /2 + V_{\m{int}}$, where $r_{z,1}$ and $r_{z,2}$ represent the axial displacement of the first and second chain. Mode analysis in the basis $(r_{z,1},r_{z,2})$ results in two distinct eigenmodes, $(1,1)$ and $(1,-1)$ in which ions in the separate wells oscillate in and out of phase with each other, respectively.
These modes are schematically depicted in Fig. \ref{fig:schematic_axial_and_radial_coupling}(b) and (c). In both cases, both chains of ions contribute equally to the mode, making these modes ideal channels for the transfer of quantum information between wells. The frequencies of these modes are given by
\begin{equation}
    \label{eq:couplingrate_exact}
    \omega_{\m{com},\m{str}} = \sqrt{\omega_z^2\pm \frac{k_{\m{int}}}{n m} }
\end{equation}
with a `$+$' for the in-phase (common, COM) mode, and a `$-$' for the out-of-phase (stretch, STR) mode. The difference in frequency of these modes, $\Omega_c=|\omega_{\m{com}} - \omega_{\m{str}}|$, is commonly referred to as the coupling rate. For a small interaction potential, $|k_{\m{int}}| \ll m n \omega_z^2$, the coupling rate can be expressed as \cite{Brown2011,Harlander2011}
\begin{equation}
    \label{eq:couplingrate}
    \Omega_{c} = \frac{|\kappa| nq^2}{4\pi\epsilon_0 m \omega_z d^3}.
\end{equation}
The coupling rate presents an upper bound to the rate at which motional information can be exchanged between wells, and is in practice an upper bound for the rate at which entanglement between wells can be achieved. 

The model of Eq.~\ref{eq:couplingrate} predicts a linear increase in coupling with an increasing number of ions $n$. Entangling operations therefore benefit from being implemented on larger registers of qubits if the main error source is qubit dephasing, as the duration of the operation decreases, reducing the computation time and the susceptibility to decoherence.
However, the presented model for coupling strength is only exactly valid for two separated \emph{point} charges, as it neglects the spatial extent of the ion chain, which becomes relevant when the extent of the chains is comparable to their separation.
Numerical methods, outlined below, are required to obtain coupling rates for arbitrary numbers of ions per well. 

\subsection{Coupling rate of ion strings}
\label{subsec:coupling_ion_strings}
A 2D array of trapped ion chains requires a trapping potential that is periodic in two dimensions. We  analyze a simpler model which more closely resembles the experimental setting discussed in further sections: we consider a double-well potential along either the axial or radial axis.
We use the coordinate system $\{x,y,z\}$, as depicted in Fig. \ref{fig:schematic_axial_and_radial_coupling}(d) and (e), and denote positions as $\vec{r}=(r_x,r_y,r_z)$. The lowest-order polynomial representation of a symmetric double-well potential that confines ion chains in two separated trapping regions in three dimensions is given by 
\begin{equation}
    \label{eq:V_ax}
    V_{\m{ax}}=\frac{1}{2}\alpha r_z^2 + \frac{1}{4!}\beta r_z^4 + \frac{1}{2}\phi_x r_x^2 + \frac{1}{2}\phi_y r_y^2
\end{equation}
for axial coupling as in Fig. \ref{fig:schematic_axial_and_radial_coupling}(b), and 
\begin{equation}
    \label{eq:V_rad}
    V_{\m{rad}}=\frac{1}{2}\alpha r_x^2 + \frac{1}{4!}\beta r_x^4 + \frac{1}{2}\phi_y r_y^2 + \frac{1}{2}\phi_z r_z^2
\end{equation}
for radial coupling as in Fig. \ref{fig:schematic_axial_and_radial_coupling}(c).
In both cases, $\alpha$ and $\beta$ are the second and fourth order curvatures along the double-well axis, and $\phi_{\{x,y,z\}}$ are the second order curvatures along the remaining axes. The parameters $\alpha$ and $\beta$ define a double well that confines positively charged particles when $\beta>0$ and $\alpha<0$. The separation of the wells is uniquely defined by these parameters, as $d=\sqrt{-24\alpha/\beta}$, and each well has a local second order coefficient $\alpha_w=-2\alpha$. 
The local curvature sets the secular frequency $\omega$ of an ion trapped along the axis of the double well, $\omega = \sqrt{q\alpha_w/m}$, though it should be noted that if the well contains multiple ions, the motional frequency can deviate from this value due to the spatial extent of the ion crystal in an anharmonic potential.
In our surface traps, we can independently control the curvature parameters $\alpha$, $\beta$, and $\phi_{\{x,y,z\}}$ through a predetermined and calibrated set of DC and RF electrode voltages.

To determine the coupling rate between ions in separate wells, we first determine the equilibrium positions of all ions in both strings, $\vec{r}^{(i)}_{\m{min}}$. These positions are found by numerically minimizing the system's total potential energy $V_T$, given by
\begin{equation}
\label{eq:vt}
    V_T = \sum_i \left( V_{\m{ax}/\m{rad}}(\vec{r}_{\m{min}}^{(i)}) + \frac{1}{2}\sum_{j\neq i} V_{\m{Coul}}(\vec{r}_{\m{min}}^{(i)},\vec{r}_{\m{min}}^{(j)})\right)
\end{equation}
with $V_{\m{Coul}}(\vec{r}_{\m{min}}^{(i)},\vec{r}_{\m{min}}^{(j)})$ the Coulomb interaction between ions $i$ and $j$. Mode vectors $\vec{\nu}^{(l)}$ of modes indexed $l$ are found by calculating the eigenvectors of the Hessian $H$ of $V_T$, with terms 
\begin{equation}
    H_{i,j} = \left.\frac{\delta^2 V_T}{\delta r_k^{(i)} \delta r_k^{(j)}}\right|_{\vec{r}_{\m{min}}},\label{eq:hessian}
\end{equation}
for all ions $i,j$ in all directions $k\in\{x,y,z\}$.
The square-roots of the eigenvalues of $H/m$ are the mode frequencies, $\omega^{(l)}$. For the potentials given by Eq. \ref{eq:V_ax} and \ref{eq:V_rad}, each eigenvector represents displacements of ions in only a single of three axes and can thus be classified as being axial modes (along the axis of the ion string, the $z$ axis) or radial modes (perpendicular to the ion string, the $x$ and $y$ axes). 

The mode structure that arises from two strings of ions in a double well is similar to the set of modes that an ion string would have if only one of the wells were occupied. However, for each isolated single-well (sw) mode $l$ with mode vector $\vec{\nu}_{\m{sw}}^{(l)}$, there exist two double-well (dw) modes, which can be expressed as  superpositions of the modes in the two wells.
In one of these modes, $\vec{\nu}_{\m{dw},\m{ip}}^{(l)}$, ion pairs in separate wells oscillate in-phase (ip) with each other, while in the other mode they oscillate out-of-phase (oop), $\vec{\nu}_{\m{dw},\m{oop}}^{(l)}$. The frequencies of these modes, $\omega_{\vec{\nu},\m{ip}}^{(l)}$ and $\omega_{\vec{\nu},\m{oop}}^{(l)}$, are similar to those of the isolated single-well frequency but, crucially, have a non-zero difference, $|\omega_{\vec{\nu},\m{ip}}^{(l)} - \omega_{\vec{\nu},\m{oop}}^{(l)}| = \Omega_{\vec{\nu}}^{(l)}$. As in the point charge model of the previous section, $\Omega_{\vec{\nu}}^{(l)}$ is a coupling rate that sets an effective lower bound to information transfer time between wells, for a chosen mode $\vec{\nu}^{(l)}$. As an example, in Appendix \ref{sec:app_modestructure} we discuss in detail the axial mode structure of two axially and radially coupled ion strings.

In this work we look, in particular, at the eigenvectors in which the two ion chains \emph{collectively} oscillate in phase and out of phase with each other, as depicted in Fig. \ref{fig:schematic_axial_and_radial_coupling}(b) and (c). These modes, $\vec{\nu}_{\m{dw},\m{ip}}^{(0)}$ and $\vec{\nu}_{\m{dw},\m{oop}}^{(0)}$, can be viewed as being derived from the lowest order, $l=0$, single-well mode $\vec{\nu}_{\m{sw}}^{(0)}$, typically referred to as the common or center-of-mass mode. A feature of using these modes for ion coupling is the simplification of the overall mode structure of two separated strings, which can  be effectively treated as two coupled harmonic oscillators, without the need to consider internal degrees of freedom within the chain.
Since the coupling rate scales inversely with the frequency, and axial modes have lower frequencies than radial ones, we use the axial modes of motion for the well-to-well coupling experiments presented in this work.

For these modes, the coupling rate $\Omega_c=\Omega_{\vec{\nu}}^{(0)}$ is given by the difference between the in-phase and out-of-phase frequencies, $|\omega_{\vec{\nu},\m{ip}}^{(0)} - \omega_{\vec{\nu},\m{oop}}^{(0)}|=|\omega_{\m{str}} - \omega_{\m{com}}|$. An effective interaction constant $k_{\m{int}}$ between the chains can be obtained from the inverse of Eq. \ref{eq:couplingrate_exact},
\begin{equation}
    k_{\m{int}} = \frac{1}{2}n m \Omega_c \sqrt{4\omega_z^2-\Omega_c^2}
\end{equation}
where $\omega_z=\sqrt{-2q\alpha/m}$ for axial coupling and $\omega_z=\sqrt{q\phi_z/m}$ for radial coupling.

\begin{figure}
    \centering
    \includegraphics[width=\linewidth]{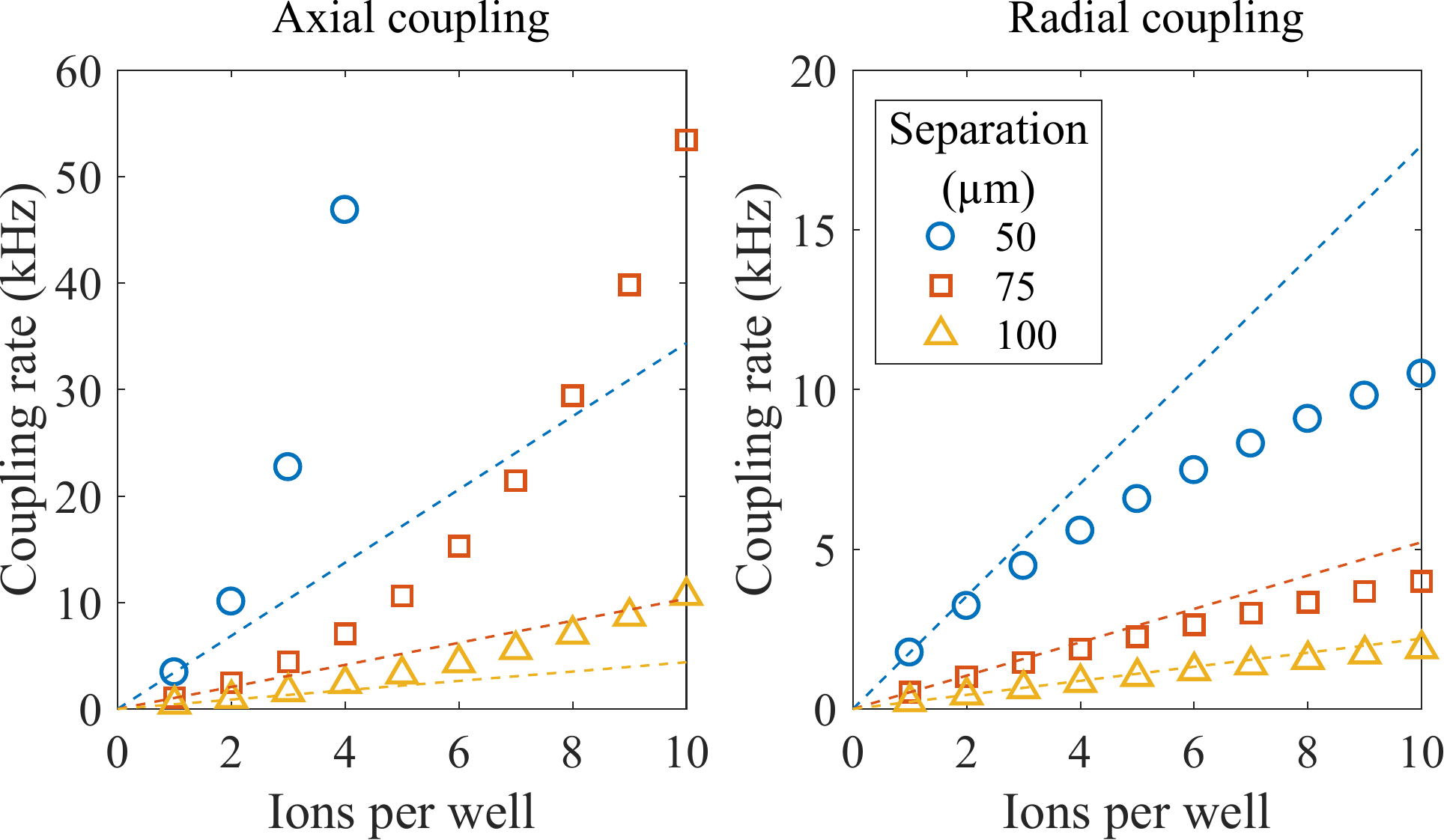}
    \caption{Simulated coupling rate as a function of number of ions per well, for various separations of the individual wells. The common mode frequency of the individual wells is set to \qty{400}{\kilo\hertz} for each point.
    If the axis of separation is along the ion chains' axial direction (axial coupling, left), the coupling of axial modes scales almost quadratically with number of ions, while for a separation along the radial direction (radial coupling, right) the scaling is sub-linear. Dashed lines are the coupling rates predicted from the point-charge model (Eq. \ref{eq:couplingrate}), which has a linear scaling with number of ions.}
    \label{fig:coupling_vs_ionnumber}
\end{figure}

Fig. \ref{fig:coupling_vs_ionnumber} displays coupling rates using the numerical approach, calculated as a function of the number of ions per well for various well-to-well separations, for coupling of axial modes along the axial and radial directions. The curvatures $\alpha$, $\beta$, and $\phi_{x,y,z}$ are set to ensure that the chains of ions align along the $z$-axis, as depicted in Figs. \ref{fig:schematic_axial_and_radial_coupling}(b) and (c). Furthermore, in each configuration, the potential curvatures are set to produce an uncoupled axial common mode frequency of each ion chain of \qty{400}{\kilo\hertz}. 

The numerical calculations presented in Fig. \ref{fig:coupling_vs_ionnumber} show that axial coupling benefits from a super-linear scaling with the number of ions, a notable increase with respect to the predicted linear scaling of the point-charge model~\eqref{eq:couplingrate}. This deviation is a natural consequence of the non-linear Coulomb interaction acting stronger on ions that are closer to each other than the well-to-well separation. This favorable scaling allows for coupling of larger registers of qubits in separate wells at time scales similar to that which can be achieved for fully merged and split crystals, while avoiding the complications involved in such operations. We further note that, using the numerical methods described above, we obtain that the influence of stray electric fields on coupling between ion chains in an anharmonic potential is negligible compared to the increase in coupling rate, with increasing ion number. For radial coupling, on the other hand, the scaling is sub-linear, since the distance between ions in different wells is on average larger than the point-model separation.

The potentials given by Eqs. \ref{eq:V_ax} and \ref{eq:V_rad} are experimentally realized with a set of DC and RF voltages that are predetermined using electrostatic simulations of our trap models. In practice, trap potentials will deviate from desired potentials due to limits in numeral precision in the electrostatic simulation, tolerances in the fabrication of the traps, and unaccounted stray fields. The most notable effect of this experimental imperfection is that the axial field curvatures at the position of the two ion chains may not be equal. As a result, the motional frequencies of uncoupled chains, $\omega_{z,1}$ and $\omega_{z,2}$, are not necessarily on resonance. The resulting mode frequencies deviate from those given by Eq. \ref{eq:couplingrate_exact}, and are instead given by 
\begin{equation}
    \label{eq:avoided_crossing_freqs}
    \omega_\pm = \sqrt{\frac{\Delta\omega^2}{4} + \omega_m^2 \pm \sqrt{\frac{k_{\m{int}}^2 }{n^2m^2} + \Delta\omega^2\omega_m^2}},
\end{equation}
in which we have used the difference frequency $\Delta \omega = \omega_{z,2} - \omega_{z,1}$ and the mean frequency $\omega_m = (\omega_{z,1}+\omega_{z,2})/2$. The resulting mode functions  are given by
\begin{equation}
    \label{eq:eigenmodes}
    \vec{\nu}_{\omega_{\pm}}^{(0)}=
    \frac{1}{\sqrt{1+(\sqrt{1+\chi^2}\mp\chi)^2}}
    \left(-\chi \pm \sqrt{1+\chi^2},1 \right),
\end{equation}
with $\chi=nm\omega_m \Delta\omega/k_{\m{int}}$.

For a large difference in frequencies of the two harmonic oscillators, $\Delta\omega\gg k_{\m{int}}/(n m \omega_m)$, (i.e., $\chi\to\infty$) eigenmodes are $(-1,0)$ and $(0,1)$ and thus the wells are not coupled. When the harmonic oscillators are tuned into resonance, $\Delta\omega=0$ (i.e., $\chi\rightarrow0$), the two coupled modes, $(1,1)/\sqrt{2}$ and $(-1,1)/\sqrt{2}$ at frequencies $\omega_{\m{com}}$ and $\omega_{\m{str}}$ emerge.

The frequency response of $\omega_+$ and $\omega_-$ as a function of $\Delta\omega$ exhibits an avoided crossing around $\Delta\omega=0$. The avoided crossing is a useful experimental feature, as it allows us to accurately determine the coupling rate, by measuring the motional frequencies $\omega_+$ and $\omega_-$ of the double-well system while varying $\Delta\omega$. The coupling rate at resonance is given by the minimum value of $|\omega_+ - \omega_-|$. 

\subsection{Mode heating of coupled ion strings}
\label{subsec:heating_theory}
Anomalous heating is a common affliction in ion traps, and is exacerbated in \emph{surface} traps, as ions are in relatively close proximity to electrodes which may carry voltage noise \cite{Brownnutt2015,Allcock2011,Dubessy2009,Turchette2000,Teller2021}. The rate of excitation of an individual mode is dependent on the projection of the  electric field noise at the position of the ions onto the mode's eigenvector. 

As described in Section \ref{subsec:point_charge_model}, the coupled ion strings exhibit an in-phase (common) and out-of phase (stretch) mode of oscillation. 
The common mode, with mode vector $(1,1)$ can be excited if field-noise is homogeneous over the two ion positions. The stretch mode with mode vector $(-1,1)$, on the other hand, cannot be excited with a homogeneous field and only responds to unequal fields at the ion positions.
Since the ion-electrode distance is typically larger than the separation between the coupled ion-chains, the electric field noise is predominantly homogeneous.
The stretch mode is therefore less susceptible to mode heating, making it a good choice for use in entangling operations. The fact that the stretch mode (and other higher-order modes) have lower heating rates is well-known for chains of ions in a single well~\cite{Kalincev2021}. However, for single-well chains of more than 2 ions, ions participate unequally in such modes, making them less suitable for entangling operations. In the double-well approach, the two chains of ions equally participate in the stretch mode, and furthermore benefit from the same low motional frequency as their bare common modes.

In the following, we present a model for estimating mode heating due to voltage noise. We first determine the ions' equilibrium positions $\vec{r}_{\m{min}}^{(i)}$ for a known single- or double-well potential, following the procedure outlined in Section \ref{subsec:coupling_ion_strings}. We then determine the axial electric field component $E_k(\vec{r}_{\m{min}}^{(i)})$ per volt for each independent DC electrode $k$, analyzed at each ion position. This field is obtained using finite-element electrostatic field simulations. We cast this field at multiple ion positions into the vector $\vec{E}_k$. The contribution of a noisy electric field to mode heating is determined by its projection onto that mode, given by $\vec{E}_{k,\parallel n} = (\vec{E}_k \cdot \vec{\nu}_n)\vec{\nu_n}$. The heating rate $\Gamma_{n,k}$ of a mode $n$ due to electrode $k$ is proportional to the square of the total force applied to the chain by $\vec{E}_{k,\parallel n}$, given by $(q\sum_i |E^{(i)}_{k,\parallel n}|)^2$, and inversely proportional with the mass of the chain, \cite{Brownnutt2015}. We thus have a prediction for heating rate per voltage noise, for each electrode, for a given trap geometry. The total heating rate of mode $n$ is the sum of that of each electrode, $\Gamma_n=\sum_k \Gamma_{n,k}$. We experimentally investigate heating of coupled ion strings due to electric field noise in Sections \ref{sec:axial_heating_rates} and \ref{sec:radial_heating}.

\section{Well-to-well coupling in the axial direction}
\label{sec:axial}
\begin{figure}
    \centering
    \includegraphics[width=\linewidth]{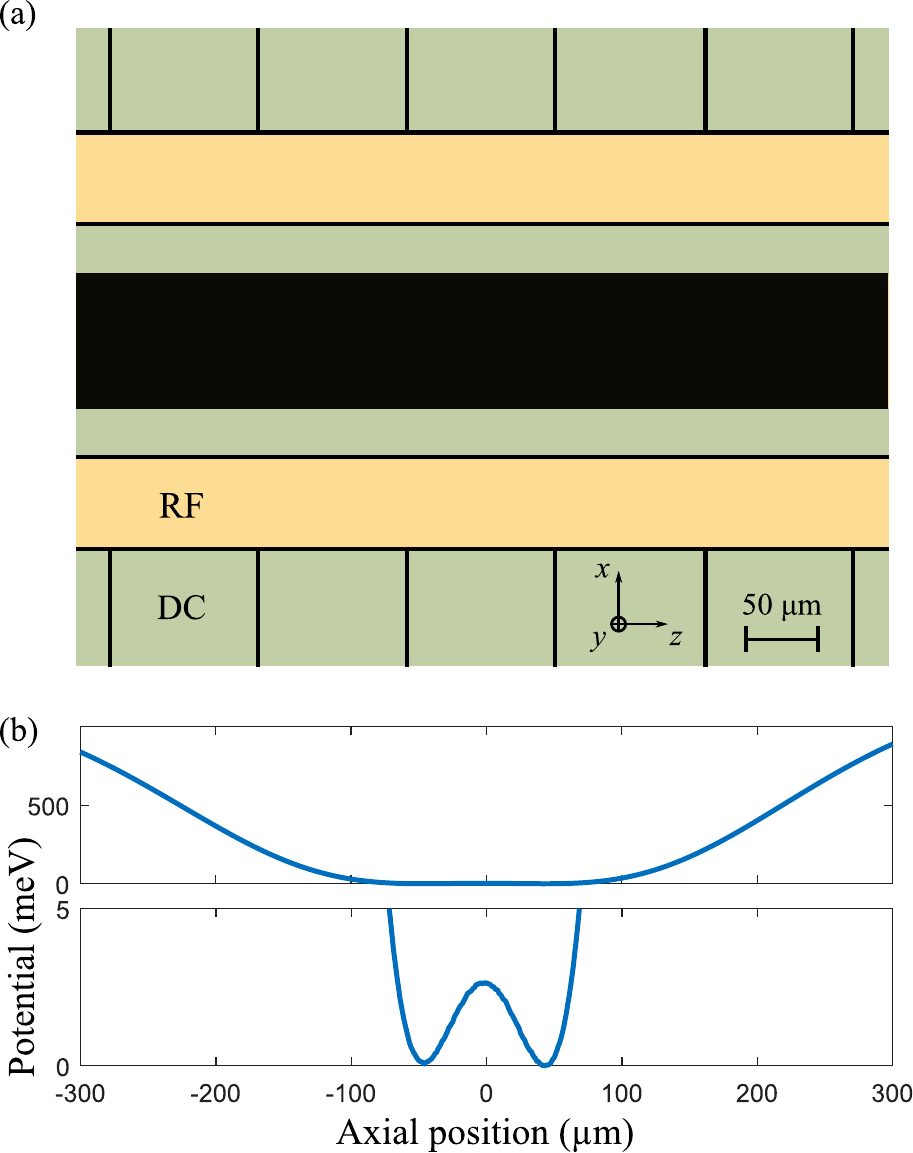}
    \caption{(a) Electrode layout of the trap used for the axial coupling experiments. (b) Double-well potential along the trap axis. The plots show the same data, at different scales. In this particular example, local trap frequencies are \qty{830}{\kilo\hertz}, with a double well separation of \qty{89}{\micro\meter}.}
    \label{fig:axial_coupling_trap_image}
\end{figure}

In this section, we experimentally investigate well-to-well coupling of ion strings along the axial direction. We show how the coupling rate scales with number of ions per well, observe coherent exchange of motion between wells, and investigate sensitivity of the double-well system to electric field noise. For these experiments, the slotted surface trap schematically shown in Fig. \ref{fig:axial_coupling_trap_image}(a) is used. A set of static voltages on electrodes that flank the trapping region (marked `DC' in Fig. \ref{fig:axial_coupling_trap_image}(a)) are used to generate a symmetric double-well potential along the $z$-axis (Fig. \ref{fig:axial_coupling_trap_image}(b)), given by the quadratic and quartic potential terms in Eq. \ref{eq:V_ax}, $\alpha$ and $\beta$.

\subsection{Coupling rate in the axial direction}
\label{sec:axialcoupling}
The coupling rate between ion chains in separate wells is obtained by experimentally determining the mode frequencies of the sets of ions in individual wells. As described in Section \ref{subsec:coupling_ion_strings}, these frequencies, $\omega_+$ and $\omega_-$, exhibit an avoided crossing when scanned across resonance, and the minimum frequency difference determines the coupling rate $\Omega_c$.

The motional frequencies of ions in both wells can be tuned by introducing a homogeneous field $E_z$ in the axial direction, which alters the double well potential to $V_{\m{ax}}=-E_z r_z + \alpha r_z^2/2 + \beta r_z^4/4!$ (neglecting field components in the $x$ and $y$ directions). As with the double-well potential, this additional field is produced by adding a predetermined and calibrated set of voltages to the trap electrodes. For harmonically confined ions, such a field would ordinarily not change the motional frequency. However, since the local trapping potential around each well is anharmonic, this field generates a change in motional frequencies of approximately equal magnitude but opposite sign in the two wells. Applying a homogeneous field therefore allows us to bring the two motional frequencies in and out of resonance, and is by approximation linearly proportional to varying the single well frequency difference $\Delta\omega$ in Eqs.~\ref{eq:avoided_crossing_freqs} and \ref{eq:eigenmodes}, while keeping the mean frequency $\omega_m$ constant. We can then deduce the coupling rate by observing the magnitude of the avoided crossing at $\Delta\omega=0$.

\begin{figure}
    \centering
    \includegraphics[width=0.8\linewidth]{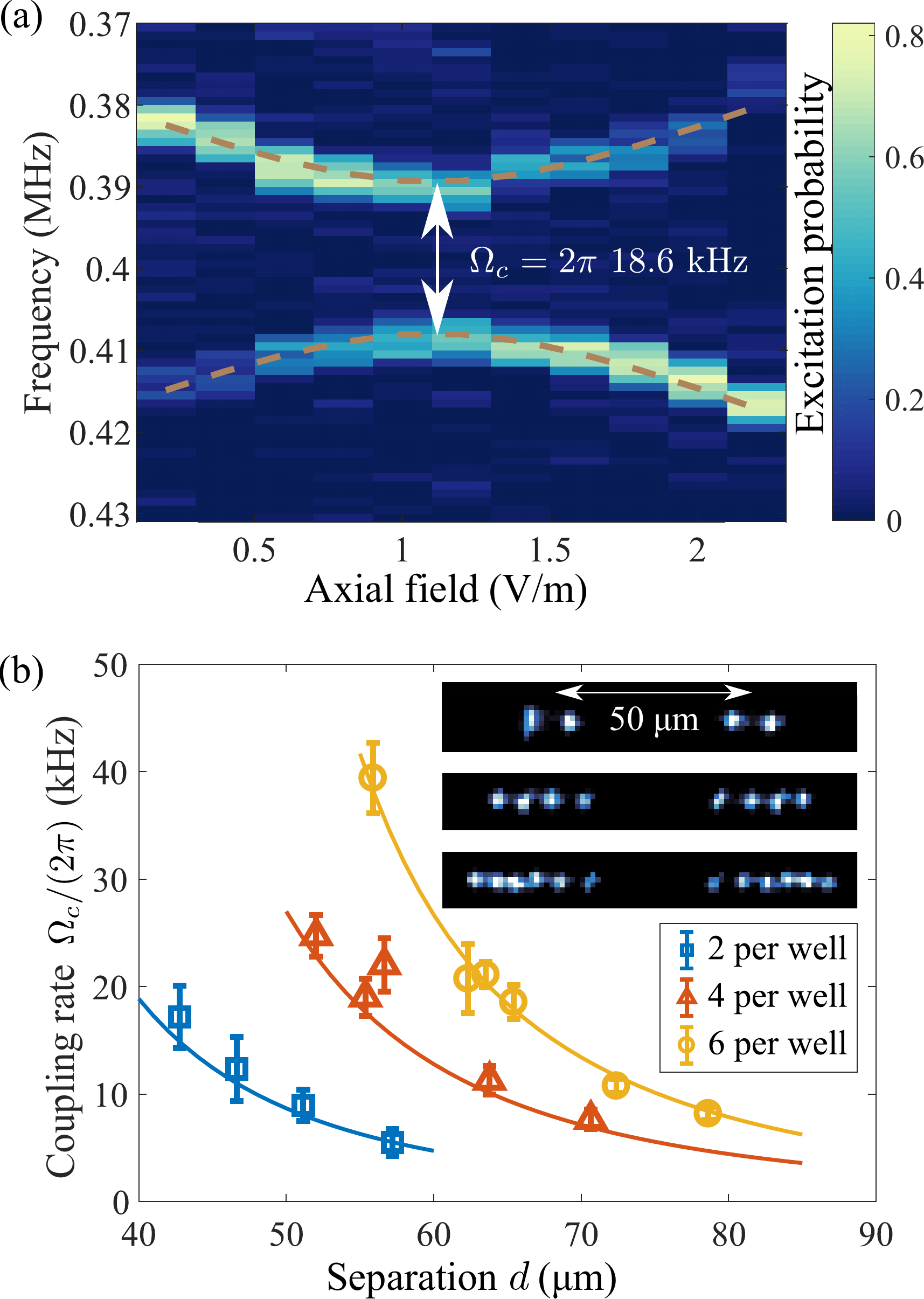}
    \caption{(a) Example of the avoided crossing in the coupling rate measurement for $n=6$ ions per well. (b) Measured coupling rate for various numbers $n$ of ions per well, and various separations. Lines are simulated values. Insets display chains of ions detected with an EMCCD camera. The error bars represent the 95$\%$ confidence interval coming from the the fits to the avoided crossing measurements using Eq.\ref{eq:avoidedcrossing_fit}.}
    \label{fig:coupling_measurements}
\end{figure}

Fig. \ref{fig:coupling_measurements}(a) displays an example of an avoided crossing measurement, obtained by sideband spectroscopy, in which we detect ions' excitation when optically probing near motional sidebands. The color mapping represents the mean excitation of all ions from only one of the two wells. The two wells each contain $n=6$ ions, and are separated by $d=\qty{65}{\micro\meter}$. The motional frequencies of the chains are about \qty{400}{\kilo\hertz}. We vary both the laser frequency (vertical axis, displayed as detuning from the carrier frequency) and the applied field (horizontal axis). The plot shows both how the detuning between wells, $\Delta \omega$, affects the motional frequencies as given by Eq. \ref{eq:avoided_crossing_freqs} and how it affects the magnitude of participation in each mode, as given by Eq. \ref{eq:eigenmodes}.
The magnitude of the applied stray field at which the avoided crossing occurs, $\approx\qty{1.11}{\volt\per\meter}$, is not zero.
This discrepancy is caused by stray fields from patch potentials on the trap surface, and limited numerical precision when calculating voltages to generate the double-well potential.

Motional frequencies for each axial field setting in Fig.~\ref{fig:coupling_measurements}(a) are obtained by fitting the data to a Gaussian function. The measured center frequencies are compared to a model function based on Eq. \ref{eq:avoided_crossing_freqs}, given by 
\begin{equation}
    \omega_{\pm}=a\sqrt{(\gamma-d)^2/4 + b \pm \sqrt{c^2 + b (\gamma-d)^2}},
    \label{eq:avoidedcrossing_fit}
\end{equation}
with $a\ldots d$ fit parameters, and $\gamma$ the applied field. The coupling frequency $\Omega_c$ is inferred from a least-squares regression between the measured data and the model, with 
\begin{equation}
    \Omega_c=\omega_+(\gamma=d) - \omega_-(\gamma=d)=a(\sqrt{b+c}-\sqrt{b-c}).
\end{equation}
In the example of Fig.~\ref{fig:coupling_measurements}(a), we find $\Omega_c=2\pi~\qty{19(2)}{\kilo\hertz}$.

Fig.~\ref{fig:coupling_measurements}(b) displays the measured coupling rate obtained from multiple avoided crossing measurements. Here, we vary the number of ions per well ($n=2$, 4 and 6), and the separation between wells. The mean motional frequency of the wells $\omega_m$ has been fixed for all configurations to be $2\pi~\qty{400}{\kilo\hertz}$. Note that since the common mode frequency in an anharmonic potential is dependent on the number of ions, the double-well potential parameters $\alpha$ and $\beta$ must be adjusted as a function of ion number to counteract this dependence.

The solid lines in Fig \ref{fig:coupling_measurements}(b) are numerically calculated values for the coupling rate. No free parameters are used in obtaining simulated values, solely the well-to-well separation and the mean common mode frequency. With an R-square value of 0.95, the simulated and measured values are in agreement.

At a separation of \qty{56}{\micro\meter}, we measure a coupling rate of $2\pi~\qty{39(3)}{\kilo\hertz}$ between two wells containing 6 ions each. In contrast to the dipole-dipole interaction model which predicts a linear scaling with number of ions, this is a sixteen-fold improvement over the predicted $2\pi~\qty{2.5}{\kilo\hertz}$ coupling that would be achieved for a configuration with 1 ion per well with the same separation and motional frequency.

\subsection{Exchange of motional excitation in the axial direction}
\label{sec:axial_phonon_exchange}
A prerequisite for coherent control of qubits in separate wells, such as entangling operations, is that motional excitation can be coherently exchanged between the two wells. This section details the experimental observation of coherent exchange of motional excitation of ion strings in separate wells.

As described in Section \ref{sec:double_well_physics}, we consider the motion characterized by the displacement of the string of ions in each well, $\vec{r}=(r_{z,1},r_{z,2})$. When on resonance, eigenmodes of oscillation are given by $\vec{\nu}_{\m{com}}=1/\sqrt{2}(1,1)$ and $\vec{\nu}_{\m{str}}=1/\sqrt{2}(1,-1)$, and have frequencies separated by the coupling rate, $\Omega_c$. A displacement of ions in the first well, $r_{z,1}=\lambda$ is written in terms of eigenmodes as $\lambda/\sqrt{2}(\vec{\nu}_{\m{com}} + \vec{\nu}_{\m{str}})$, which has a free evolution given by 
\begin{align}
   \vec{r}=\frac{\lambda}{\sqrt{2}} &\left[ \sin\left((\omega_m-\Omega_c/2)t\right)\vec{\nu}_{\m{com}}\right. \nonumber \\
   &\left. + \sin\left((\omega_m+\Omega_c/2)t\right)\vec{\nu}_{\m{str}}\right].
\end{align}
Applying trigonometric identities gives
\begin{align}
\label{eq:energy_exchange}
    \vec{r}=\lambda \left(\cos{\left(\frac{\Omega_c t}{2}\right)}\sin(\omega_m t),\right. \nonumber \\
    \left. -\sin{\left(\frac{\Omega_c t}{2}\right)}\cos(\omega_m t)\right)
\end{align}
from which we see both ion chains oscillating at the trap frequency $\omega_m$, though with an amplitude modulated at a rate $\Omega_c/2$. An excitation applied to ion chain 1 is therefore fully transferred to ion chain 2 after a time $t_{\mathrm{exc}}=\pi/\Omega_c$, and returned at $2t_{\mathrm{exc}}$. A similar treatment of exchange of motional excitation can be made with a quantum mechanical description of the coupled harmonic oscillator system \cite{Wilson2014}. In such a system, it can be shown that the mean local phonon number oscillates between wells.  

We directly observe this exchange of excitation in our experiment. We use a system of two ions per well, separated by \qty{58}{\micro\meter}, at a motional frequency of \qty{460}{\kilo\hertz}. We measure a coupling rate of $2\pi~\qty{5}{\kilo\hertz}$ with a measurement similar to that in Fig.~\ref{fig:coupling_measurements}(a).
The experimental protocol is as follows: Ions are first cooled to near the motional ground state using sideband cooling. After ground-state cooling, we determine the initial mean mode occupation to be approximately 0.7 phonons for both the common and stretch modes of the double-well system. 
After ground-state cooling, a \qty{397}{\nano\meter} pulse is applied to ions in only one of the wells (which we will call the first well), exciting them to the $4P_{1/2}$ level. Spontaneous decay to the $4S_{1/2}$ level induces random momentum kicks, exciting the motion of those ions. Spontaneous decay can also occur to the meta-stable $3D_{3/2}$ state with a branching ratio of about 1/16, where population is trapped, and the ion has left the heating cycle. A duration of \qty{5}{\micro\second} of the \qty{397}{\nano\meter} pulse ensures that $>99.9\%$ of population has been shelved to the $3D_{3/2}$ level, from where it is hidden from subsequent operations.

At this stage, the photon recoil has excited the motion of the ions in the first well, while those in the second remain near the motional ground state, corresponding to the initial condition of Eq. \ref{eq:energy_exchange}. We probe the motional excitation of the second well by analyzing optical excitation from red or blue sideband pulses, detuned from the $4S_{1/2}$ to $3D_{5/2}$  transition by the mean motional frequency. The mean phonon number in the second well is inferred by comparing the excitation of both sidebands to numerical models of excitation as a function of energy stored in the well. Since the ions in the left well are hidden from these analysis pulses, they do not need to be considered in the numerical models. The sideband excitation pulse is on for a duration of \qty{15}{\micro\second}. 
The total duration of motional excitation and analysis, \qty{20}{\micro\second}, is shorter than the expected energy exchange time, $t_{\mathrm{exc}}\approx\qty{100}{\micro\second}$, but is not negligible.
Since excitation transfer also occurs during these pulses, we expect a reduction in contrast in the amplitude of the phonon exchange measurement.

\begin{figure}
    \centering
    \includegraphics[width=0.8\linewidth]{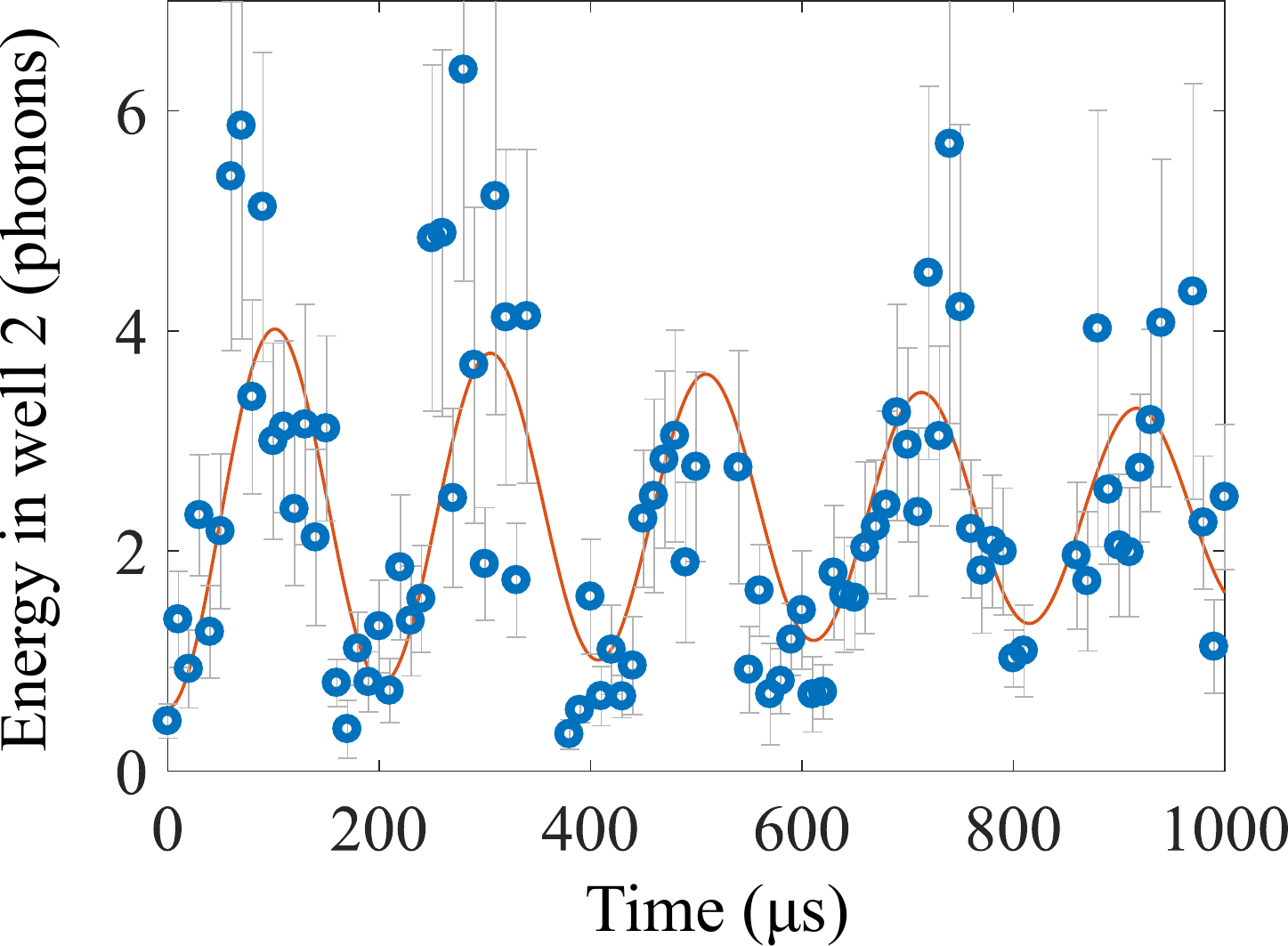}
    \caption{Demonstration of exchange of motional excitation between two axially coupled sets of two ions. After ions in one well are excited, the excitation of ions in the second well is measured as a function of delay time between excitation and measurement. The error bars represent a 95\% confidence interval in the least-squares regression between measured data and the numerical models.
    The solid line is a sinusoidal fit, with a decay parameter included. A coupling rate of $\Omega_c=2\pi~\qty{4.7(1)}{\kilo\hertz}$ is inferred from the rate of excitation transfer.}
    \label{fig:phonon_exchange}
\end{figure}

Fig. \ref{fig:phonon_exchange} displays the measured energy of the ions in the second well, $E_2(t)$, expressed in terms of phonons, $n_2(t) = E_2(t)/\hbar\omega_z$, as a function of wait time $t$ between excitation of the first well, and analysis of the second. The expected oscillatory exchange of energy is apparent. The red line is a least-squares fit between the measured data and the model 
\begin{align}
    \label{eq:phonon_exchange_model}
    n_2(t) = &\frac{1}{2}\left[ \left(n_1(0)-n_2(0)\right)\cos{(\Omega_c t + \phi )}\exp{(-t/\tau_d)}+\right.\nonumber \\
        &\left.n_1(0) +n_2(0)\right]
\end{align}
with $t$ the experimental wait time and all other variables free fitting parameters. $n_{1}(0)$ and $n_{2}(0)$ represent the first and second well's initial energy, expressed in phonon number, $\Omega_c$ is the coupling rate, $\phi$ is a phase offset caused by non-zero duration of the state preparation and analysis pulses, and $\tau_d$ represents a decay in contrast due to motional decoherence. From the fit we determine a coupling rate of $\Omega_c = 2\pi~ \qty{4.7(1)}{\kilo\hertz}$, in good agreement with the coupling rate measured through spectroscopy (as in the previous section) and with the numerically predicted value, $2\pi~\qty{4.9}{\kilo\hertz}$.

\subsection{Heating rates of axial modes}
\label{sec:axial_heating_rates}
Entangling gates used for quantum information require the motional states of the chain of ions to be coherent, since the motional modes act as a data bus. Motional heating due to electric field fluctuations diminishes motional coherence and is a common issue in surface traps~\cite{Brownnutt2015,Allcock2011,Dubessy2009,Turchette2000,Teller2021}. In this section, we investigate the properties of mode heating of ion crystals confined in a double-well potential.

\begin{figure}
    \centering    \includegraphics[width=.73\linewidth]{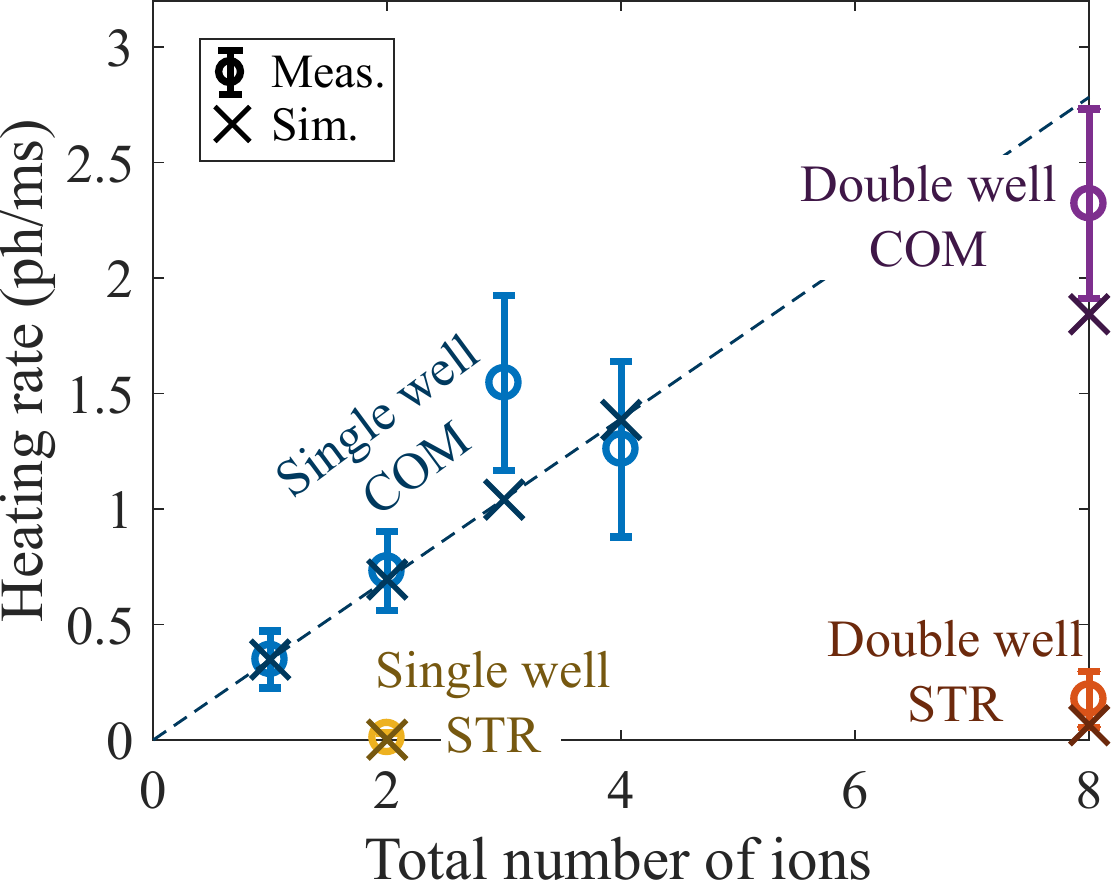}
    \caption{Heating rates for various configurations of ions, in a single well or double well. Circles are measured data, and error bars are propagated from 95\% confidence intervals in the least-squares regression between measured data and the model functions.
    Crosses are simulated values, using the single ion data as a calibration reference. The dashed line represents a linear scaling of heating rate with number of ions, following a single-well model, also using the single ion data as a reference.}
    \label{fig:heating_rates}
\end{figure}

Fig. \ref{fig:heating_rates} displays the results of heating rate measurements for various ion numbers measured via excitation of the motional sidebands~\cite{Brownnutt2015}. In all measurements the frequency of the measured mode is set to \qty{450}{\kilo\hertz}. For the double-well data, the wells are separated by \qty{50}{\micro\meter}. The blue data points are a set of control measurements in which ions are trapped in a single well. These measurements have been performed to confirm the expected linear scaling of the single-well heating rate with number of ions~\cite{Brownnutt2015}. The single particle data is furthermore employed to calibrate a technical noise model of electrode voltage noise, which is used to determine a simulated expected heating rate for the other measured configurations. The noise model is described in Section~\ref{subsec:heating_theory}. It assumes technical voltage noise that is uncorrelated between different electrodes but has the same noise amplitude on all electrodes. Results from the model are marked as crosses in Fig. \ref{fig:heating_rates}.

Double-well heating rates have been taken with four ions per well. The heating rate of the common mode of motion, 2.4(4) ph/ms (purple circle in Fig. \ref{fig:heating_rates}), is similar to that expected from an 8-ion single-well model, 2.7 ph/ms (dashed line). However, the measured heating rate lies lower than this value, which can be attributed to the spatial distribution of charges: The estimated heating rate when taking the spatial extent of the ion chain into account is 1.8 ph/ms (purple cross). The stretch mode, in contrast, has roughly an order of magnitude lower heating rate, 0.2(1) ph/ms (red circle), in agreement with the predicted value of 0.1 ph/ms (red cross). 

The fact that the stretch mode of motion exhibits a lower heating rate than the common mode is well-known for single-well ion crystals (exemplified by the yellow data point in Fig. \ref{fig:heating_rates}). We have demonstrated here that this concept continues to hold even when ion crystals are spatially separated, which highlights the benefit of using the stretch mode for well-to-well entangling operations, compared to merging and splitting protocols, which are inherently sensitive to mode heating.

The double-well stretch mode has the property that its motional frequency remains as low as the common mode frequency. In contrast, in single-well ion crystals the stretch mode has a factor $\sqrt{3}$ higher frequency. This, in combination with the fact that in a multi-ion stretch mode with more than two ions, ions unequally contribute to the mode (i.e., have unequal Lamb-Dicke parameters), makes the use of the stretch mode and higher order modes for entangling operations impractical. The double-well stretch mode is much less affected by these disadvantages even for large numbers of ions per well and is a good candidate for mediating entangling operations due to its low heating rate.

\section{Well-to-well coupling in the radial direction}
\label{sec:radial_entanglement}
In this section, we demonstrate coupling, phonon exchange, and entanglement of radially separated ions in two linear ion traps. Furthermore, we verify that the stretch mode of oscillation of coupled ions in a radial double-well potential is more resilient to electric field noise compared to the common mode.

\begin{figure}[t]
    \centering    \includegraphics[width=1\linewidth]{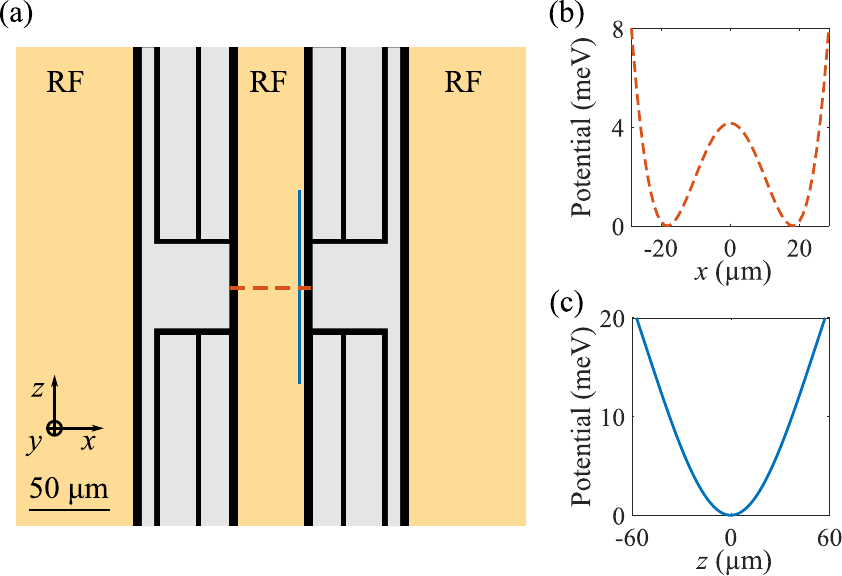}
    \caption{Overview of the trap used for radial coupling experiments. (a) Schematic trap design. The unlabelled electrodes in gray are the DC electrodes used for confinement and control of the two trapping sites.
    (b) Cross section of the total potential along the radial direction $x$, \qty{80}{\micro\meter} above the trap surface corresponding to the orange dashed line in (a), generated by applying an RF voltage of \qty{140}{\volt} peak-to-peak at \qty{19}{\mega\hertz} on the RF electrodes. The resulting radial frequency along $x$ is \qty{2.4}{\mega\hertz} in both trapping sites. (c) Cross section of the total potential along the axial direction $z$ (blue line in (a)), with an axial frequency of \qty{1}{\mega\hertz}.}
    \label{fig:2Strap}
\end{figure}

We realize the following experiments with the trap depicted in Fig. \ref{fig:2Strap}(a), which features 3 co-linear RF electrodes and a total of 12 DC electrodes distributed on a single aluminum metal layer on top of a fused silica substrate.
As discussed in Section \ref{sec:double_well_physics}, ion strings are stored at the minimum of the effective trapping potential generated by the RF electric field, commonly referred to as the null of the RF pseudopotential.
For the trap in Fig. \ref{fig:2Strap}(a), similar to those documented in Refs. \cite{Welzel2011,Tanaka2014,Tanaka2021,Holz2020}, the presence of three RF electrodes in a plane results in a pseudopotential landscape that is characterized by having two minima, as shown in Fig.\ref{fig:2Strap}(b). For our particular trap, we have determined the well-to-well separation to be $d\approx$ \qty{29}{\micro\meter} along the radial direction $x$, at a height above the surface of \qty{80}{\micro\meter}. 
A predetermined set of voltages on the DC electrodes allow us to generate individually controllable confining potentials along the axial direction of each of the linear traps, as shown in Fig. \ref{fig:2Strap}(c) for the right trapping site.

\subsection{Coupling of radially separated ions}
\label{sec:2S_coupling}
The coupling of axial modes of oscillation of ions located in a radially separated double-well potential is demonstrated by measuring the axial mode frequencies of a system of one ion in each well as a function of the well-to-well axial frequency detuning. As discussed in Section \ref{sec:axialcoupling}, upon scanning the motional frequencies of the two wells through resonance, the axial mode spectrum exhibits an avoided crossing, and the minimum frequency difference represents the coupling rate $\Omega_c$ when the two wells are resonant.
\begin{figure}
    \centering   
    \includegraphics[width=.9\linewidth]{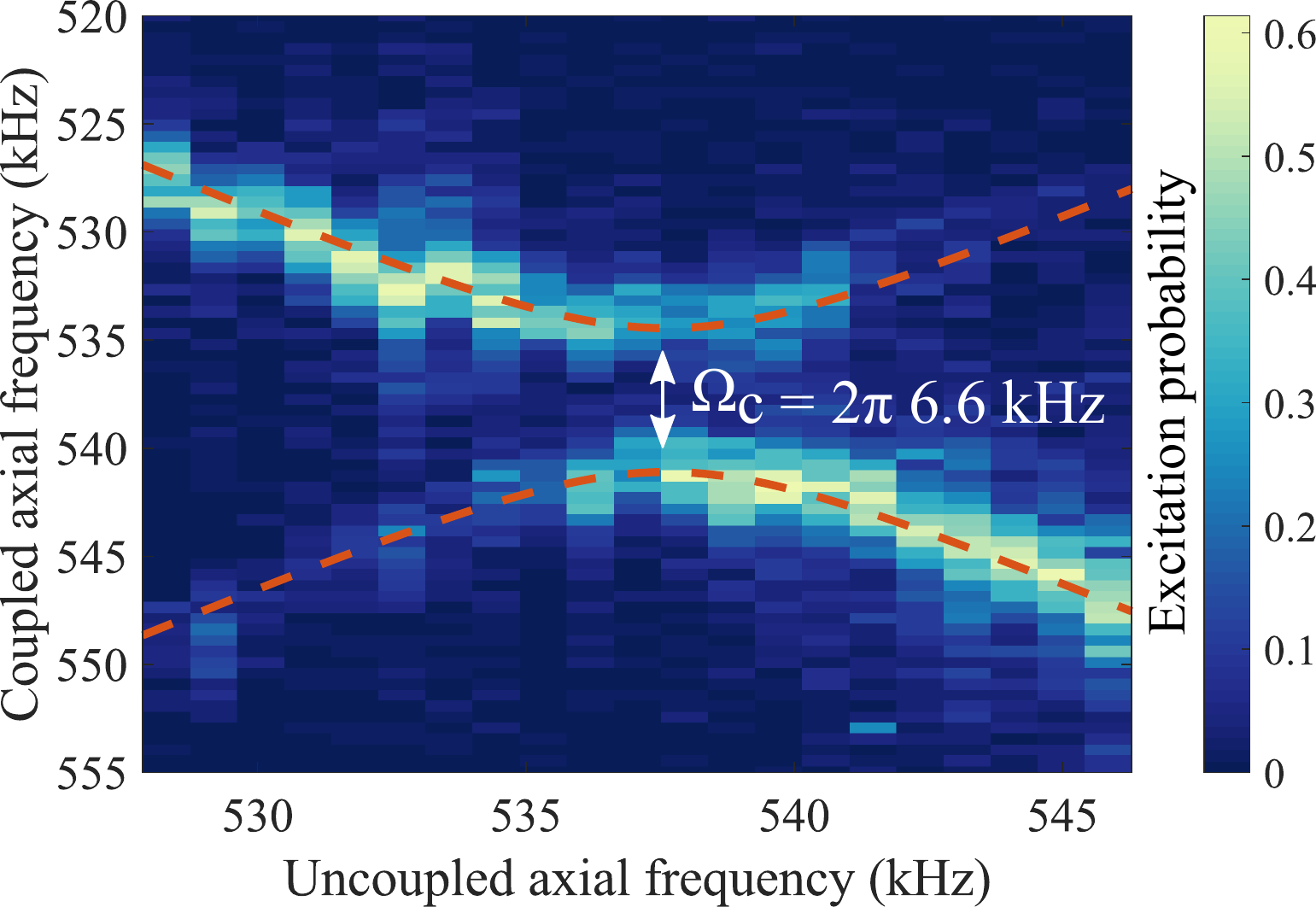}
    \caption{Avoided crossing measurement for $n=1$ ion per well at an axial frequency of \qty{540}{\kilo\hertz}. The measurement shows spectra of the ion on the left well as function of the expected left well axial frequency from the potential curvature, in the absence of well-to-well coupling.}   \label{fig:radial_avoided_crossing}
\end{figure}
We change the axial frequency in the two wells using two sets of DC voltages, each of which controls the potential curvature in one of the two potential wells, without affecting that of the other. We vary the axial curvature of both wells in opposite directions and obtain the frequency of the axial modes via sideband spectroscopy. 
We read out the electronic state of the ions by collecting their fluorescence light with a photomultiplier tube (PMT). By filtering out the light from the right well with an aperture, we observe the axial mode of only the ion in the left well, which splits into two as its mode frequency approaches resonance with that of the right well.

The measured spectra of the ion in the left well are shown in Fig.~\ref{fig:radial_avoided_crossing}, as a function of the expected 
left well axial frequency extracted from the potential curvature, in the absence of well-to-well coupling. We obtain the motional frequencies for each axial curvature by fitting experimental data with a double Gaussian function. As explained in Sec. \ref{sec:axialcoupling}, we compare the extracted mode frequencies with Eq.~\ref{eq:avoidedcrossing_fit} and obtain the coupling rate $\Omega_c$. For $n=1$ ion in each well trapped at an axial frequency of \qty{540}{\kilo\hertz}, we obtain a coupling rate $\Omega_c = 2\pi~\qty{6.6(3)}{\kilo\hertz}$. This value is in good agreement with that calculated using Eq.~\ref{eq:couplingrate}, $2\pi~\qty{7.0}{\kilo\hertz}$.

\subsection{Phonon exchange between radially separated ions}
\label{subsec:radial_phonon_exchange}
As introduced in Sec. \ref{sec:axial_phonon_exchange}, coherent exchange of motional excitation from one well to another is a prerequisite to perform well-to-well entangling operations. In this section we present phonon exchange between two ions located in two radially separated linear traps.
\begin{figure}
    \centering    \includegraphics[width=.88\linewidth]{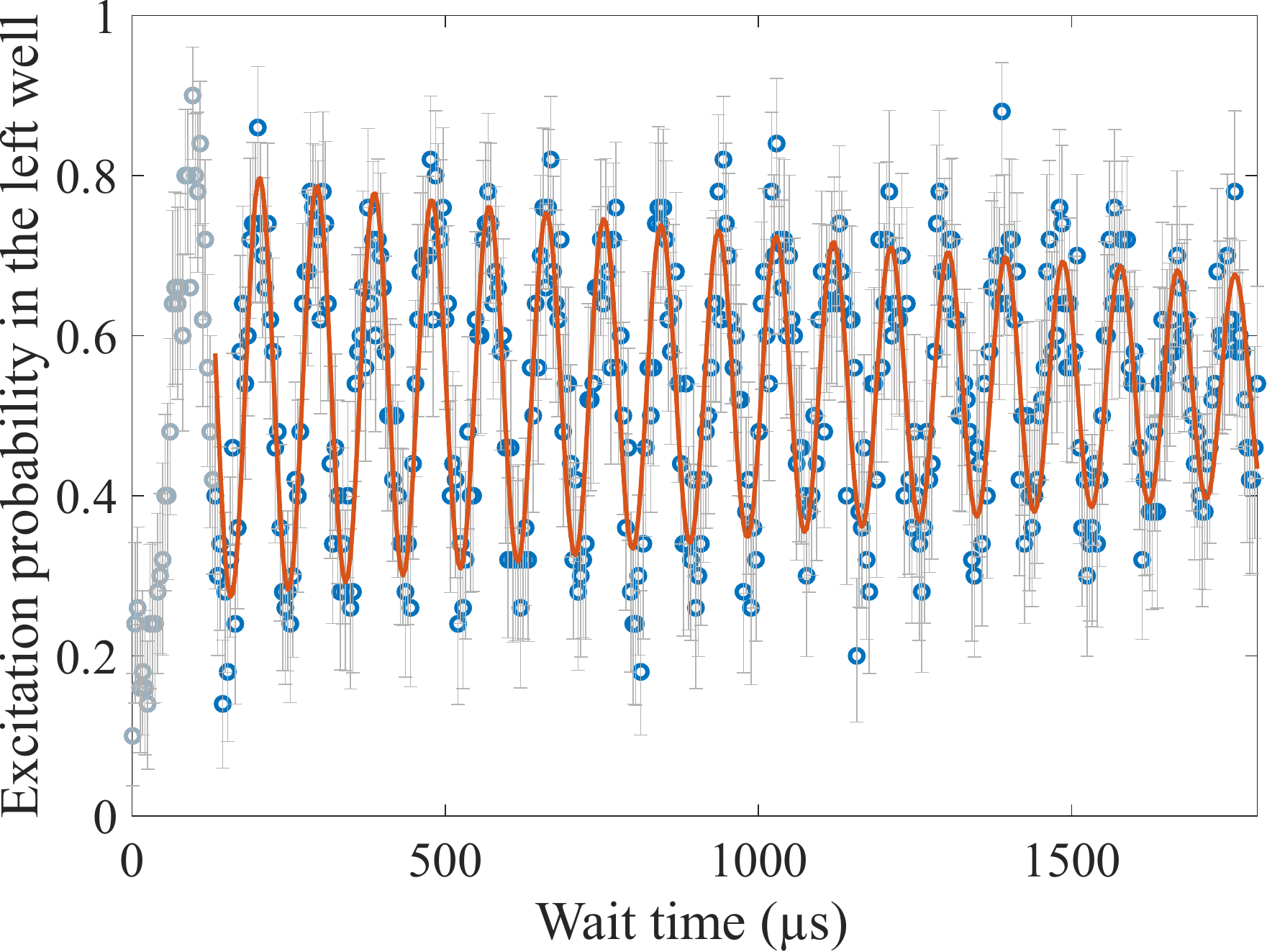}
    \caption{Demonstration of phonon exchange between axial modes of oscillation of two radially coupled wells. The wells have one ion in each. The solid line is a sinusoidal fit, with a decay parameter included. The greyed out data before \qty{130}{\micro\second} is not included in the fit as the motional frequencies of the two wells have not settled after switching DC voltages. From the rate of phonon transfer we infer a coupling rate $\Omega_c$ = 2$\pi$ \qty{10.92(2)}{\kilo\hertz}. The error bars represent the 95\% confidence interval of the mean excitation probability resulting from quantum projection noise \cite{PhysRevA.47.3554}.}  \label{fig:radial_phonon_exchange}
\end{figure}

The experiment consist of injecting one phonon in one of the two wells, and the observation of phonon transfer from one well to the other as a function of the wait time after the phonon injection. In contrast to the axial phonon exchange presented in Section \ref{sec:axial_phonon_exchange}, we do not have the means to individually address ions with laser beams in our setup. 

Instead, we realize selective phonon injection by implementing fast in-sequence voltage switching on the trap's DC electrodes, allowing us to tune the motional frequencies of the two wells in and out of resonance. The two wells become spectrally resolved when out of resonance, allowing us to perform ion-selective operations through sideband pulses. We have opted not to use this method of phonon injection in the axial motional excitation measurements of Section \ref{sec:axial_phonon_exchange} due to the limited bandwidth voltage switching induced by electrical filters in our DC lines.

We cool the ions to near the motional ground state via sideband cooling. Subsequently, we tune the two wells out of resonance, and inject one phonon in a target well with a blue sideband $\pi$-pulse.
We then tune the two wells on resonance, enabling phonon exchange between the two ions. After a given wait time, we detune the wells from resonance again, and use a blue sideband $\pi$-pulse on the target ion. If a phonon is present in the target well, this pulse reverses the excitation of the first pulse, de-exciting the ion. If, on the other hand, the phonon was transferred to the other well, no de-excitation occurs.

The result of the phonon exchange measurement for a system of one ion per well, at an axial frequency of \qty{540}{\kilo\hertz} is shown in Fig. \ref{fig:radial_phonon_exchange}.
Switching the wells into resonance does not happen instantaneously, as the adjusted voltages pass through electronic filters. Therefore, there is an intermediate time during which the two wells are near, but not on, resonance. We have independently determined that the motional frequencies settle to a steady state at approximately \qty{130}{\micro\second}, after which phonon exchange is expected to occur at a constant rate. Wait times prior to this are indicated by the faded region before \qty{130}{\micro\second} in Fig. \ref{fig:radial_phonon_exchange}.

The solid line in Fig.~\ref{fig:radial_phonon_exchange} represents a least-squares fit of a model equivalent to the one used in Sec.\ref{sec:axial_phonon_exchange}, Eq. \ref{eq:phonon_exchange_model}. 
Data from the first \qty{130}{\micro\second} are not included in the fit.
We infer a coupling rate $\Omega_c = 2\pi~\qty{10.92(2)}{\kilo\hertz}$. This value differs from the value obtained from the avoided crossing measurement in the previous section, $2\pi~\qty{6.6(3)}{\kilo\hertz}$. This discrepancy comes from the fact that the frequencies of the two wells were intentionally set to be marginally off-resonance. This detuning from resonance was introduced to counter a reduction in phonon exchange contrast caused by the non-zero switching time, in which partial phonon exchange occurs before resonance is reached. This effect is discussed in more detail in Appendix \ref{sec:app_DCswitching}.

The controlled exchange of quanta of motion between ions provides a means to map or transfer information about the ions' electronic state. This can, for example, be used in quantum logic spectroscopy \cite{Schmidt2005,Wolf2016,Guggemos_2019} or the generation of entanglement between qubits. For example, in Fig. \ref{fig:radial_phonon_exchange}, at wait times corresponding to 0.5 excitation probability, a phonon is in a superposition of being in the left and right well. The coherence of this motional superposition can be estimated by mapping the motion of the individual wells to the ions' electronic states, creating an entangled state. This protocol is demonstrated in Appendix \ref{sec:app_phononEx_entanglement}.

\subsection{Heating rates of radially-coupled axial modes}
\label{sec:radial_heating}
As has been introduced in theory in Section \ref{subsec:heating_theory}, the modes of a set of coupled ions in a double-well potential are unequally affected by electric field noise. In particular, for typical trapping geometries, the stretch mode of motion is considerably less sensitive to electric field noise compared to the common mode. This has been experimentally shown for ions coupled along the axial direction in Section \ref{sec:axial_heating_rates}. In this section, we discuss measurements of heating rates of modes emerging from ions coupled in the radial direction.

\begin{figure}
    \centering
    \includegraphics[width=\linewidth]{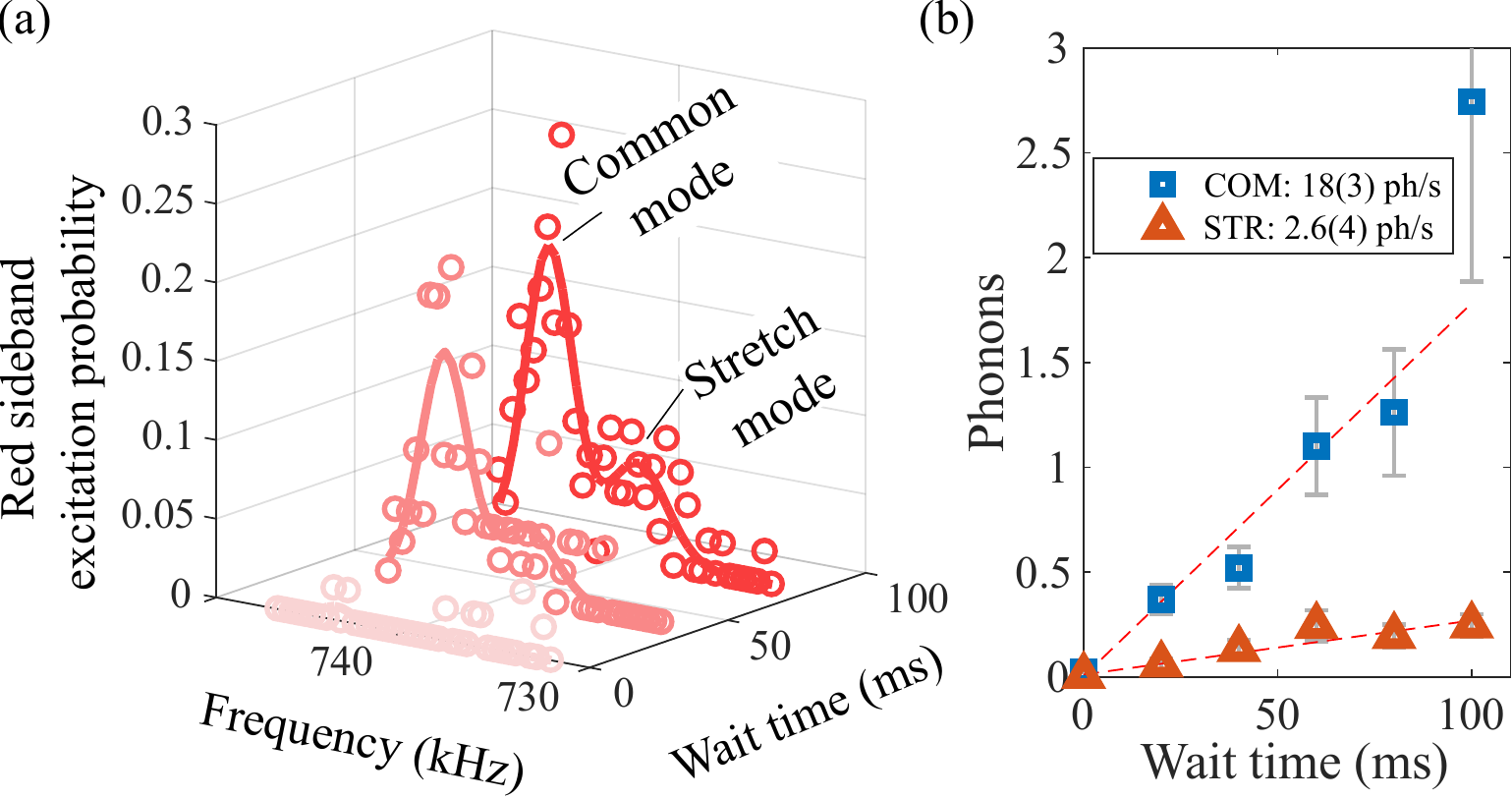}
    \caption{Results of heating rate measurements for radially separated ions. (a) Spectra around the red sideband transition depict the emergence of peaks as wait time between state preparation and sideband analysis increases. Notably, the stretch mode heats up less than the common mode. (b) Phonon numbers are extracted from spectra such as in (a) for various wait times. Error bars are 95\% confidence intervals from the least-squares regression between measured data and the model function.
    Heating rates of 18(3) and 2.6(4) ph/s are obtained from linear fits.}
    \label{fig:radial_heating_rate}
\end{figure}

The heating rate measurements are taken for one ion per well, with an axial motional frequency of \qty{740}{\kilo\hertz}. Fig. \ref{fig:radial_heating_rate}(a) shows examples of excitation spectra of the red sideband for various wait times. Spectra are obtained by detecting the state of only one of two ions, while the sideband pulse is applied to both. The emergence of peaks in the red sideband spectrum with an increase in wait time are indicative of an increase in phonon number. Notably, the common mode of motion increases at a higher rate than the stretch mode. 
By comparing data with numerical models, we obtain phonon numbers as a function of wait time, shown in Fig. \ref{fig:radial_heating_rate}(b). From linear fits of the data, we obtain heating rates of 18(3) ph/s and 2.6(4) ph/s for the common and stretch modes, respectively.

Simulations that assume uncorrelated voltage noise on the trap electrodes, as discussed in Section \ref{subsec:heating_theory}, predict that for ions separated by \qty{29}{\micro\meter}, the ratio of common and stretch mode heating rates is 16. Our measured ratio, 8.4, deviates from this prediction, which indicates that ion heating rates in our trap are not limited by the technical voltage noise described in our noise model. Regardless, the stretch mode heating rate is still lower than that of the common mode of motion of a single-well two-ion system, which we have measured to be 7 ph/s. This demonstrates the advantage of using well-to-well coupling for ion connectivity as opposed to merging ions into a single well.

\subsection{Entanglement of radially separated ions}
\label{subsec:radial_entanglement}
The generation of entanglement across quantum registers \cite{Wilson2014} is an essential tool for the proposed 2D trap array architecture. In this section, we present the realization of an entangling gate of two radially coupled ions, located in two distinct linear traps. We encode the entangled quantum information in the internal electronic states of the \calciumforty ions, $4S_{1/2}(m=-1/2)$ and $3D_{5/2}(m=-1/2)$, from here on using the shorthand notation $S$ and $D$.

To generate entanglement, we employ a M{\o}lmer-S{\o}rensen (MS) gate operation \cite{Sorensen1999}, one of the prevailing techniques for entanglement-generation for optical qubits. The driving mechanism of the gate is to employ a bichromatic laser pulse, where the frequency of each tone is detuned by an equal but opposite amount from a red and blue motional sideband transition. During the gate the collective vibrational motion of the ions is excited, undergoing a circular trajectory in phase space. To eliminate unwanted spin and motional entanglement at the end of the gate, the phase space trajectory has to form a closed loop. This condition is met at gate durations of $t=2\pi n_1/\delta_1$, with $n_1$ a non-zero integer with the same sign as $\delta_1$. 
In typical single-well two-qubit gates on the axial common mode, the detuning from the motional mode, and thus the gate time, can be freely chosen (within experimental limitations such as available laser power and qubit coherence time) as all other axial modes spectrally decoupled. 
In our double-well gate, we must account for the presence of two motional modes, separated in frequency by the coupling rate $\Omega_c/2\pi$. In Appendix \ref{sec:appendix_MS} we describe different ways of implementing an MS-gate in presence of two closely spaced motional modes, and present simulations that determine the most advantageous gate scheme, considering ion-heating as the most prominent source of decoherence in the system. 

\begin{figure}
    \centering
    \includegraphics[width=\linewidth]{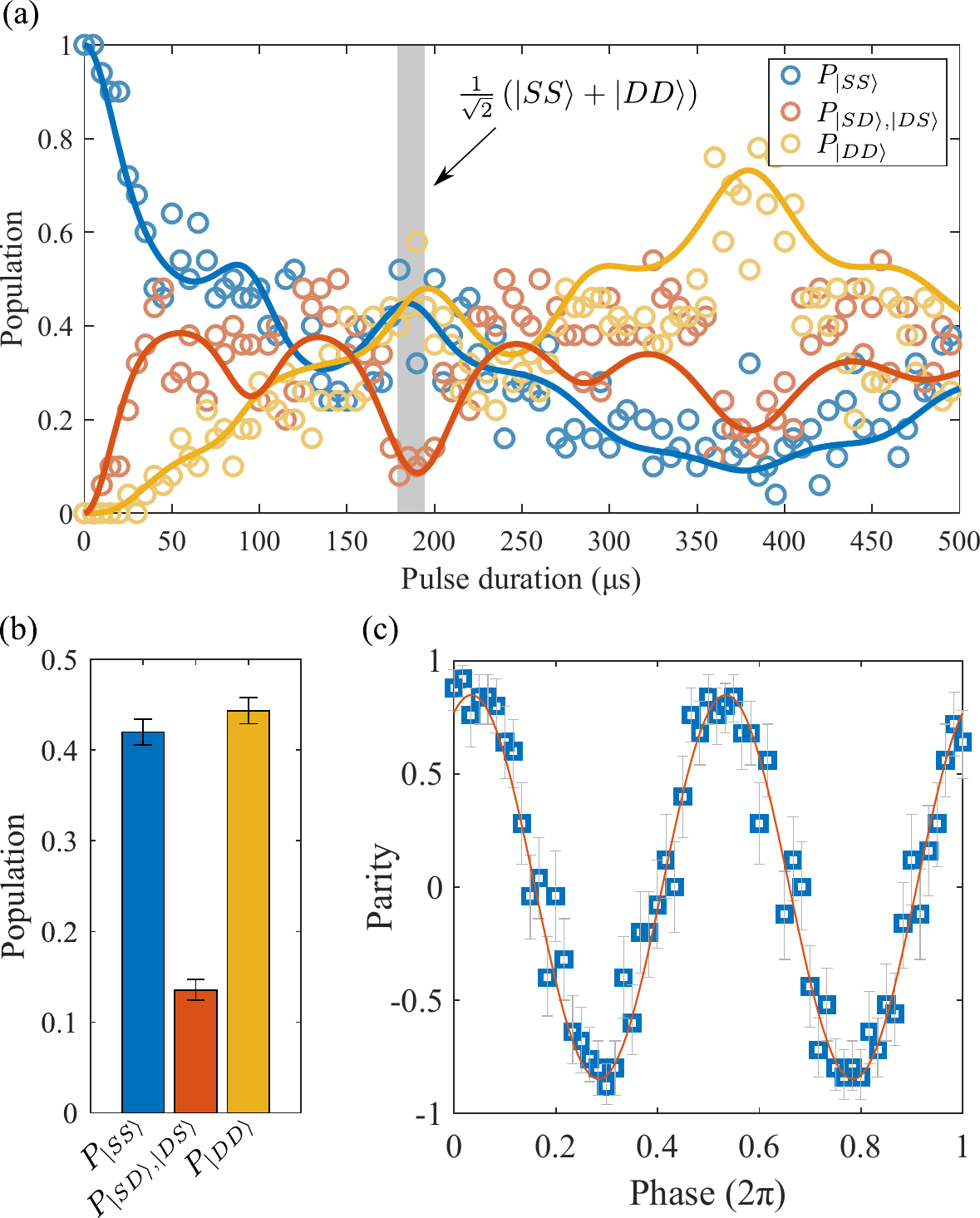}
    \caption{M{\o}lmer-S{\o}rensen (MS) gate on two radially separated ions. (a) Measured two-ion populations, as a function of gate duration. Solid lines are simulated data, taking into account estimated qubit decoherence measured in our system. (b) Mean populations, taken at a pulse duration of \qty{190}{\micro\second}, corresponding to the shaded area of (a). A mean population of $P=P_{\ket{SS}} + P_{\ket{DD}} = 0.86(1)$ is measured. Averages and their 95\% confidence intervals, shown by the error bars, are taken from 8700 repetitions of the experimental sequence. (c) The contrast in parity analysis indicates a visibility of $0.85(2)$. The error bars represent the 95\% confidence intervals for each measured phase and are extracted from quantum projection noise \cite{PhysRevA.47.3554}.}
    \label{fig:ms_gate_results}
\end{figure}

Among the methods of ensuring motional phase-space loops of multiple modes are closed at the end of a gate \cite{Milne2020,Choi2014,Green2015,Lu2019,Landsman2019}, a straightforward technique is to choose the detuning such that the gate duration obeys both $t=2\pi n_1/\delta_1$ and $t=2\pi n_2/\delta_2$, with $\delta_2=\Omega_c +\delta_1$ the detuning from the second mode, and $n_2$ another non-zero integer. As discussed in the Appendix~\ref{sec:appendix_MS}, this restriction consequently limits the gate time to a multiple of $2\pi/\Omega_c$.  
The gate presented in this section is applied to a double-well system with one ion per well with a coupling rate of $\Omega_c = 2\pi~\qty{5.3}{\kilo\hertz}$. Guided by the numerical simulations in Appendix~\ref{sec:appendix_MS}, we have chosen $\delta_1=-\Omega_c=\delta_{2}/2$ and set the gate duration to $t=-2\pi/\delta_1=-4\pi/\delta_2=\qty{190}{\micro\second}$. The results of the MS gate are shown in Figure \ref{fig:ms_gate_results}.

In Fig. \ref{fig:ms_gate_results}(a) we show the state evolution of both ions, as a function of the length of the bichromatic laser pulse. We observe state populations corresponding to the entangled state $1/\sqrt{2}(\ket{SS}+\ket{DD})$ at a pulse duration of \qty{190}{\micro\second}. In Fig. \ref{fig:ms_gate_results}(b) we show the mean populations of the two ions, calculated from $8700$ repetitions of the sequence at this wait time, from which we extract a mean population of $P=P_\m{SS}+P_\m{DD} = 0.86(1)$. 

In Fig. \ref{fig:ms_gate_results}(c) we show the parity of the final state as a function of the phase of the analysis pulse. The parity contrast corresponds to a visibility of $0.85(2)$, from which we calculate an entangled state fidelity of $F_\m{Tot} = 0.86(1)$~\cite{Sackett2000}, mainly limited by the \qty{729}{\nano\meter} laser linewidth. To confirm this, we characterized coherence times in our system via Ramsey spectroscopy, from which we conclude that optical decoherence, measured to be \qty{700(40)}{\micro\second}, is the limiting source of error. We have conducted numerical simulations of the gate, in which we consider optical dephasing to be the only source of decoherence. These simulations, displayed as solid lines in Fig.\ref{fig:ms_gate_results}(a), predict a gate fidelity of 0.85(1), in good agreement with our experimental result.

\section{Ion transport in the radial direction}
\label{sec:radial_transport}
In this section, we investigate a novel type of trapped-ion transport operation, transport along the radial direction. We show that we can control the distance between two radially separated ion chains by altering the RF voltages, and measure the coupling rate between chains of various numbers of ions depending on their separation.
\begin{figure}[t]
     \centering    \includegraphics[width=1\linewidth]{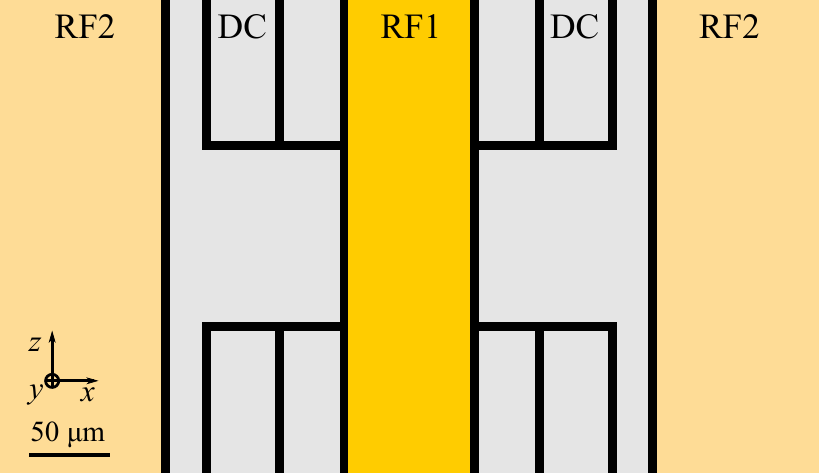}
    \caption{Schematic design of the trap used for radial transport. The inner and outer RF electrodes are labelled RF1 and RF2 respectively, as they are driven by two independent RF signals. The electrodes in gray are the DC electrodes used for confinement and control of the two trapping sites.} 
    \label{fig:4Strap_pot}
\end{figure}

\subsection{Concept and experimental implementation}
Radial transport of ion strings is a key element of the QSA architecture as altering the distance between linear ion traps enables radial connectivity between neighboring sites in the 2D ion trap array. 
In this section, we demonstrate control over the radial distance between two parallel linear traps by varying the RF confinement, and validate the theoretical model for radial coupling of axial modes of two ion chains, presented in Section \ref{sec:double_well_physics}, as a function of chain separation and number of ions per well.

We realize these experiments with the trap shown in Fig. \ref{fig:4Strap_pot}. The trap design is conceptually similar to the one introduced in Section \ref{sec:radial_entanglement}, but the inner and outer RF electrodes, RF1 and RF2, have widths of \qty{75}{\micro\meter} and \qty{255}{\micro\meter} and are separated by \qty{115}{\micro\meter}. This configuration produces a pseudopotential landscape with two RF minima separated by \qty{110}{\micro\meter}, when the electrodes RF1 and RF2 have the same RF amplitude. A more detailed description of the trap can be found in Appendix \ref{sec:appendix_RF}.

The aim of this experiment is to control the distance between these RF minima, which depends on the ratio between the voltage amplitudes of the signals applied to the inner and outer RF electrodes, $\zeta=V_{\m{RF}1}/V_{\m{RF}2}$. 
Moreover, the position of the RF pseudopotential minima depends on the relative phase between the RF signals, which should be in phase to avoid inducing excess micromotion \cite{MM_Itano_Wineland}.

Radial shuttling of ions is thus achieved by independently controlling both the amplitude and phase of the RF voltage applied to the inner and outer RF electrodes, using two RF sources and resonators. In practice, controlling the phase of both RF fields is necessary during RF transport, as the relative phase between the RF signals on the trap electrodes changes when varying the RF ratio $\zeta$. This change is a result of the coupling between the two resonators, and of an overall change in the RF circuitry parameters induced by temperature changes. The details about the double resonator circuitry are provided in Appendix \ref{sec:appendix_RF}. During RF transport, while the RF ratio is changed, the DC voltages used to confine the ions axially have to be adjusted to keep the DC and RF potential minima overlapped, and to compensate for changes in stray fields along the shuttling path.

\subsection{Characterization of radial shuttling}
We verify our control over the double-well pseudopotential by measuring how the well-to-well distance and radial trap frequencies depend on the RF ratio $\zeta$. We keep the axial motional frequency constant at $\omega_z = 2\pi~\qty{540}{\kilo\hertz}$ for all experimental settings.
\begin{figure}
    \centering
    \includegraphics[width=1\linewidth]{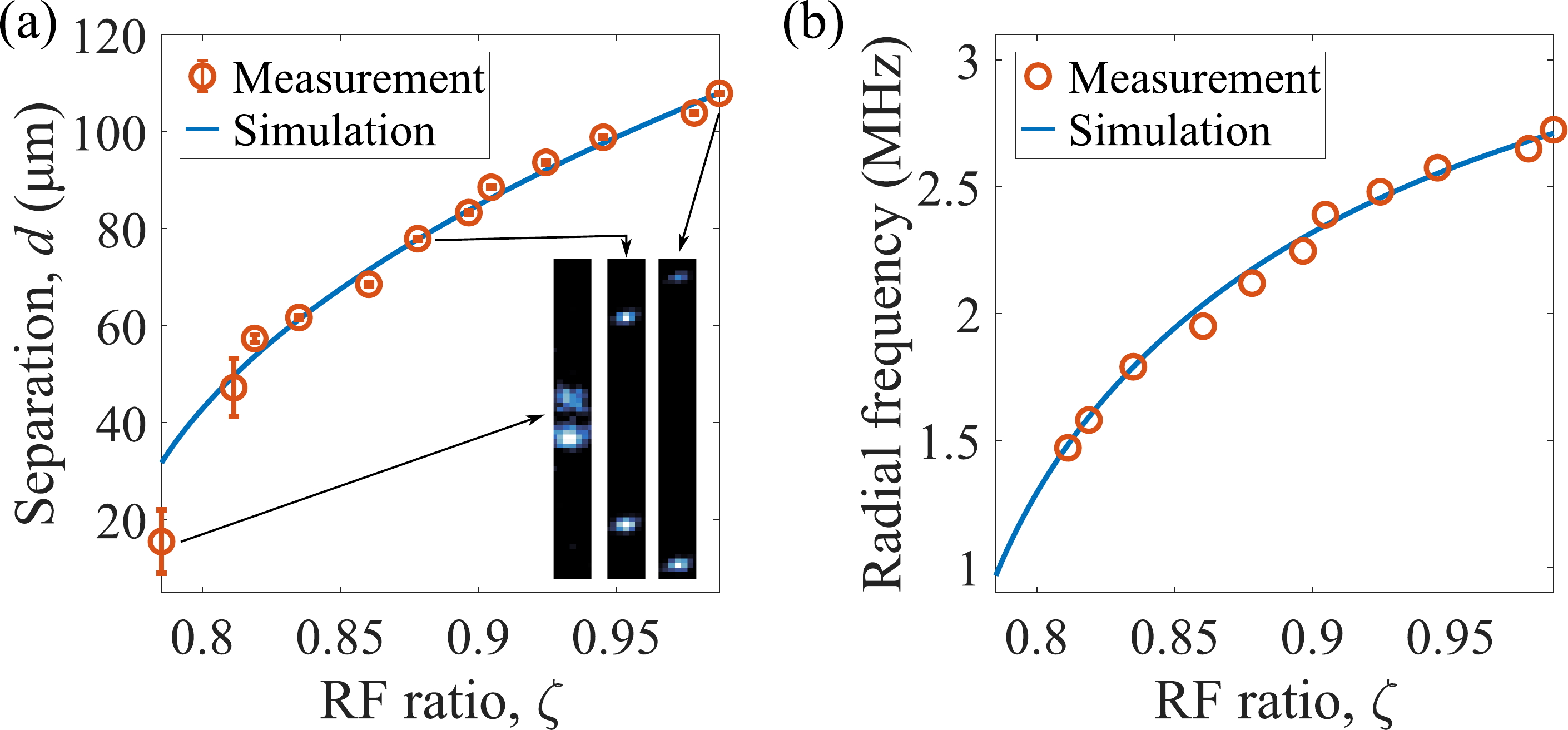}
    \caption{Characterization of radial transport. (a) Measurement of the well-to-well separation along the $x$ axis and comparison with simulations. The insets are camera images of ions at the corresponding values of RF ratio $\zeta$. In here, error bars are 95\% confidence intervals propagated from the least-squares regression between measured ion-images and the double-Gaussian model function used for fitting. 
    (b) Measurement and simulations of the radial frequency parallel to the trap surface, as a function of $\zeta$, for a constant axial frequency of $\omega_z$ = 2$\pi$ \qty{540}{\kilo\hertz}. The error bars are smaller than the markers, as the measurements' uncertainty is of the order of \qty{1}{\kilo\hertz}, and are thus not shown in the plot.}
    \label{fig:RF_shuttling}
\end{figure}

The distance between ions is determined by analyzing images of the ions. To obtain this distance, the magnification of the camera must be known. This is calibrated as follows: two ions are placed in a single well, and their axial trap frequency is measured using conventional sideband spectroscopy. The separation between ions, in pixels, is then related to the analytically expected separation for given trap frequencies \cite{James1998}.

In Fig.~\ref{fig:RF_shuttling}(a) we show how the well-to-well distance depends on $\zeta$ by adjusting the  separation between $d=\qty{15}{\micro\meter}$ and \qty{110}{\micro\meter}. 
The separation $d\approx\qty{15}{\micro\meter}$ is the lowest achievable separation below which the two wells merge into a single one. This limit is due to the discrete minimum step size of the RF sources of \qty{0.1}{\decibel}.
The results presented are in good agreement with the pseudopotential simulations up to a constant scaling factor of 1.039(5), which can be explained by systematic errors in the measurement of the RF voltage on the trap.

In Fig. \ref{fig:RF_shuttling}(b) we show the frequency dependence of the  radial mode of oscillation, aligned along the $x$ direction, as a function of $\zeta$. We determine this frequency by applying an oscillating signal that is resonant with the ions' motional frequency to one of the trap DC electrodes. This process, colloquially known as ``tickling'' \cite{Knoop_2014}, excites the motion of a trapped ion to an extent that it appears as a larger spot on the camera. This tickling method is used as opposed to the previously mentioned sideband spectroscopy, because it reduces ambiguity in determining radial modes.

In varying $\zeta$, only the inner RF voltage is changed, while the outer RF electrodes are kept at a constant voltage. The RF ratios of the data presented in Fig. \ref{fig:RF_shuttling}(b), include the scaling factor determined from the data in (a). 
We show that the measured radial frequencies decrease according to simulations. The presented measurements indicate that we understand and can control the RF double-well pseudopotential, allowing us to tune the radial distance between the two wells, and thus control the coupling rate of ion chains.

\subsection{Scaling of radial coupling rate}
\label{sec:radial_coupling_rate}
The coupling rate between ion chains in separate wells is determined experimentally by detecting the motional mode frequency of the ions in the two wells. As described in Section \ref{sec:double_well_physics}, the mode spectrum of two coupled oscillators splits into an in-phase (common) and out-of-phase (stretch) mode of oscillation. When the wells are tuned on resonance, the frequency difference of these modes is given by the coupling rate $\Omega_c$.

Ideally, we would determine the wells' motional mode frequencies via resolved \qty{729}{\nano\meter} laser spectroscopy, as described in Section \ref{sec:axialcoupling}. However, in this dual-resonator driven trap, the ion heating rates are on the order of tens of phonons per millisecond, which enhances the coupling to motional modes and complicates the overall mode spectra. This makes an accurate determination of coupling rates on the order of kHz using laser spectroscopy impractical.

Instead, we determine mode frequencies via the RF-tickling method described above. We select the DC electrodes such that the field produced by the applied RF signal has an overlap with both the in-phase and out-of-phase axial modes of oscillation of ions in the radial double-well. When the RF signal is on resonance with an axial mode, it increases its energy, and we can detect the driven motion by measuring the width of the ion crystals on a CCD camera. When the two wells are on resonance, a scan of the RF tickling frequency excites the coupled system's common and stretch modes at different frequencies, where the frequency difference indicates the coupling rate. If the wells are properly tuned on resonance, the ions in both wells are expected to participate equally in the excitation of both modes. 

\begin{figure}
    \centering    \includegraphics[width=1\linewidth]{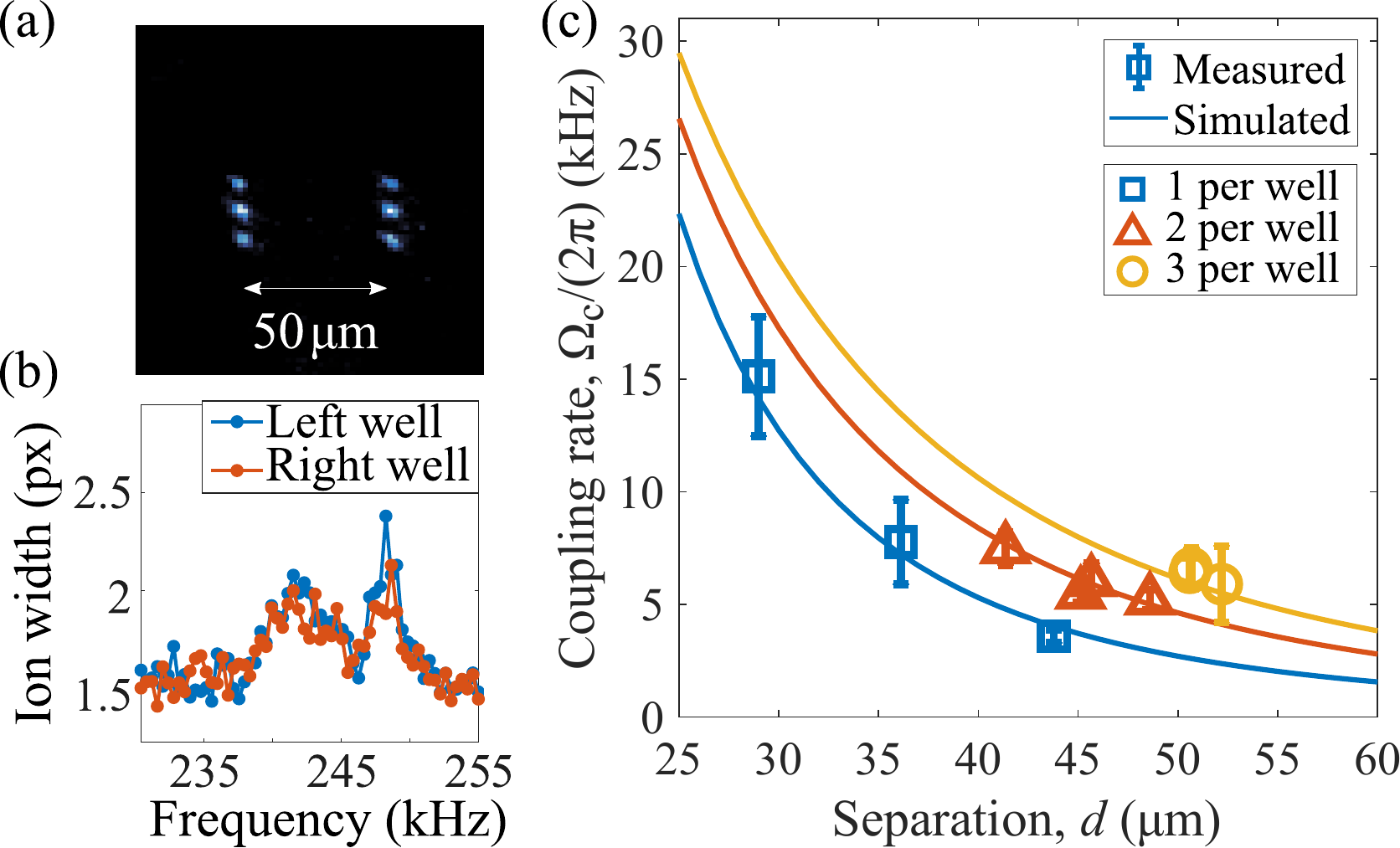}
    \caption{RF coupling measurement with RF tickling. (a) EMCCD image of ions trapped in a radial double well, with three ions per well.
    (b) Width of the ion's image on the EMCCD camera, as a function of the RF tickling frequency. The participation of the ion motion to the two coupled modes in both wells is detected. (c) Well-to-well coupling $\Omega_c$ as function of distance $d$ between wells, and number of ions per well. The measurements are taken for axial frequencies ranging from 238 to 312 kHz. The lines in the plot represent radial coupling simulations for different ion numbers at a fixed axial frequency of 262.2 kHz, i.e. the mean of the axial frequencies used in the measurements. The error bars represent the 95\% confidence interval propagated from the fit error to the measured spectra.}
    \label{fig:RF_coupling_explanation}
\end{figure}

We perform such measurement for the $n=3$ ions per well system shown in Fig. \ref{fig:RF_coupling_explanation}(a). The width of the ion's image, as detected on the EMCCD camera, is displayed as a function of the RF tickling frequency in Fig.~\ref{fig:RF_coupling_explanation}(b). Both modes of oscillation are present simultaneously in the two wells, indicating that they are coupled. The unequal excitation of the two modes, seen as differing widths of the peaks in Fig.~\ref{fig:RF_coupling_explanation}(b), is due to the unequal spatial overlap between the applied RF tickling field and the mode vector. The frequency separation between the modes gives the coupling rate, estimated to be $\Omega_c=2\pi~\qty{6.4(3)}{\kilo\hertz}$.

Following the above-mentioned approach, we collect the measured coupling rates for double-well configurations of up to $n=3$ ions per well, for different well-to-well distances and axial frequencies varying between \qty{238}{\kilo\hertz} and \qty{312}{\kilo\hertz}. The results are shown in Fig. \ref{fig:RF_coupling_explanation}(c), and are compared to simulated values, computed with the approach described in Section \ref{sec:double_well_physics}. We measure coupling rates as high as $\Omega_c = 2\pi~\qty{15}{\kilo\hertz}$ for one ion per well at a well-to-well distance of \qty{29}{\micro\meter}, and coupling rates between $2\pi~ \qty{5}{\kilo\hertz}$ and $2\pi~\qty{10}{\kilo\hertz}$ for configurations of two or more ions per well.
The measured coupling rates agree within $10\%$ with the simulated values.

Despite the good agreement with theory, the coupling measurement described in the previous paragraph does not unambiguously prove that the two wells are coupled, as we did not observe the full avoided crossing that indicates the emergence of the coupled modes. To confirm that the radially separated ions are coupled, we measure the relative phase of the ion motion in the two wells for both the inter-well common and stretch mode.
More specifically, we perform photon correlation measurements using a PMT and a time tagger, using the RF tickling signal as a trigger. The velocity of the ions in the two wells is modulated at the frequency of the RF tickling signal. Therefore, the ions' Doppler-shift dependent fluorescence rate is likewise modulated. Thus, when correlating PMT photon detection counts with the RF tickling signal, this modulation appears as a sinusoidal signal. These signals are obtained separately for the two wells by means of an adjustable aperture which is aligned to one of the two wells at a time. In our measurements, we determine both the amplitude and phase of the correlation signals. In the case of the common mode of motion, the detected phases are expected to be equal, whereas for excitation of the stretch mode, the signals are expected to be out of phase. 

\begin{figure}
    \centering
    \includegraphics[width=.9\linewidth]{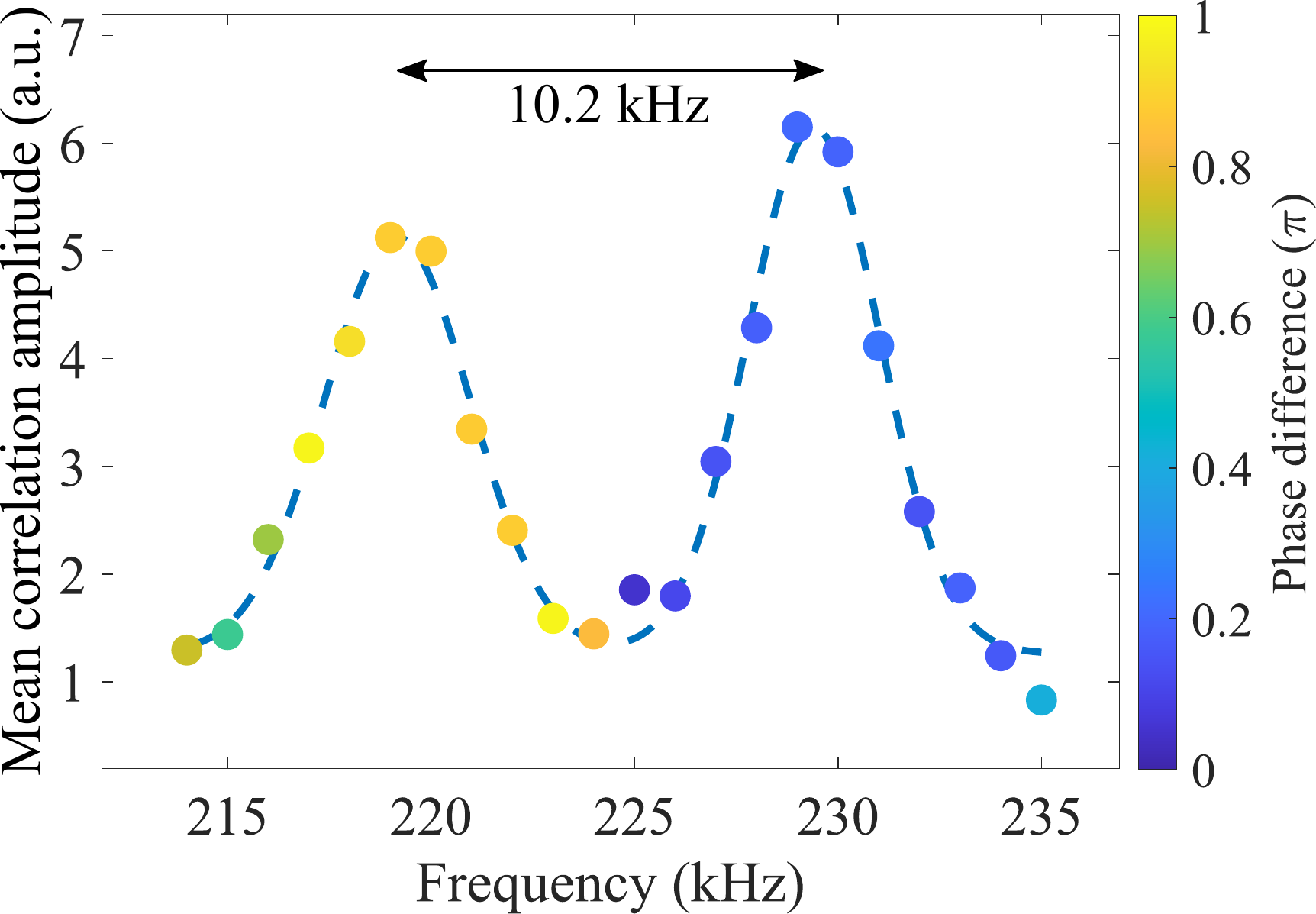}
    \caption{Phase resolved photon correlation signals of each well, as a function of the RF tickling frequency. The plot shows the mean of the correlation amplitude of the correlation signals collected in the two wells, and the dashed line is a guide-to-the-eye double-Gaussian fit. For the stretch mode, the ion in one well oscillates out of phase with respect to the one in the other well. For the common mode, the ions in the two wells oscillate in phase.}
    \label{fig:RF_coupling}
\end{figure}

In the photon correlation measurements, we trap one ion per well, keeping the wells separated by \qty{33}{\micro\meter}, and setting the axial frequency to $\omega_z\approx2\pi~\qty{225}{\kilo\hertz}$.
The results of these measurements are shown in Fig.~\ref{fig:RF_coupling}, which displays the mean amplitude of the correlation signals of the two wells, as a function of the frequency of the RF tickle frequency. The colors of the data points indicate the relative phase between the signals of the two wells. As expected, we observe one mode of oscillation in which the ions oscillate in phase with each other, and another in which they oscillate out of phase. The measured mode splitting agrees with the simulated value of $2\pi~\qty{10}{\kilo\hertz}$.

\section{Quantum error correction in the QSA architecture}
\label{sec:theory}
The previous sections focused on the experimental demonstration of the primitives of the \latticename{} architecture. Specifically, we demonstrated transfer of information between separated ion crystals in two dimensions, and entanglement of ions located in two different wells along the radial direction. In the following, we provide blueprints for future experimental implementations, showing how the two-dimensional connectivity of the \latticename{} architecture can be utilized for efficient quantum error correction (QEC) and fault-tolerant quantum computing (QC).

Scalable quantum computing relies on the ability to reliably detect and correct errors that may have corrupted the information stored in a qubit~\cite{knill1998resilient, aliferis2005quantum}.
To do so, quantum information is encoded across multiple physical qubits, which constitute a logical qubit. 
Errors on physical qubits can be corrected by performing QEC and, thereby, the information encoded in a logical state can be protected~\cite{gottesman1997stabilizer, terhal2015quantum}.
In the \latticename{} architecture, we propose to encode a logical state in a single chain of ions, stored in a single potential well. The capability to tune the interaction strength between separated ion chains, enabled by the combination of axial and radial shuttling, opens up the possibility to carry out operations between logical states in separate potential wells, and implement quantum error correction as well as logical operations on encoded states in a resource efficient manner.  
Operations between separate logical qubits are normally carried out in a sequential manner, as pairs of ions pertaining to distinct logical qubits are entangled consecutively. The \latticename{} architecture, on the other hand, opens up the possibility to perform parallel transversal entangling gates between two radially separated ion chains, enabling faster and modular implementations of error correction gadgets.

In the following, we first introduce a new method for implementing the parallel transversal gates by adopting a different entanglement scheme, and demonstrate the general feasibility of this implementation on the \latticename{} architecture with numerical simulations. We then explain how these gates can be utilized for QEC codes and fault-tolerant QC, specifying examples of possible code constructions and the layout of efficient QEC codes, tailored to the \latticename{} architecture.

\subsection{Transversal gate operations}
\label{subsec:tvgate} 
In this section, we present a scheme for the implementation of a transversal entangling gate that operates simultaneously between selected pairs of ions in radially separated potential wells. We start by exploiting an idea presented in the context of tunable frustration in ion spin ladders~\cite{bermudez2011frustrated, bermudez2012quantum}. In combination with refocusing sequences, this enables the implementation of parallel gate operations without requiring sequential addressing of specific ion pairs. 
The general idea is to exploit the anisotropy of the effective ion-ion interactions that underlie the aforementioned geometric-phase entangling gates. We show how these couplings can be designed in such a way that undesired intra-well couplings get geometrically canceled by either aligning the laser beams or tilting the crystal plane, while the effect of undesired inter-well terms that would break the transversality of the logical gate can be minimized by applying specific spin-echo sequences. 

We consider a double-well configuration, where ions are located in the $xz$-plane forming two strings along the $z$-direction, and each string resides in one of the two wells along $x$-direction, being thus radially separated.
To realize a transversal gate operation between two strings of ions, only the interactions between opposite ion pairs in separate wells should contribute, whereas all the other interactions should cancel out.
We consider a parameter regime that aligns with the capabilities of the architecture presented with $\phi_z < \alpha \approx \phi_y$ for the potential curvatures $\phi_y, \phi_z, \alpha$ as defined in Eq.~\ref{eq:V_rad}. In this setting, which is discussed in detail in Section~\ref{sec:double_well_physics} and Appendix~\ref{sec:appendix_MS}, we find two closely spaced axial modes of oscillation that are separated from higher-frequency modes. Achieving a transversal gate following the previous M\o lmer-S\o rensen scheme on optical qubits would require the usage of beam deflectors that couple the desired qubit pair to the motional modes that mediate the entangling gate, together with a sequential schedule that ensures that different qubit pairs do not talk to the motional modes simultaneously, as this would induce couplings that break transversality.  As discussed in Appendix~\ref{appendix:transversal_gates}, it is possible to find a promising transversal gate by switching to a light-shift  scheme~\cite{leibfried2003experimental,PhysRevLett.117.060504,PhysRevA.103.022427, PhysRevA.103.012603,PhysRevLett.127.130505}, which uses a different laser-beam arrangement. By applying a pair of quasi-resonant laser beams that are not co-propagating and induce a cross-beam ac-Stark shift, we can derive an effective Hamiltonian~\cite{bermudez2011frustrated, bermudez2012quantum} that describes the qubit-qubit interactions in the considered parameter regime with well-separated axial modes
\begin{align}
    H_{\mathrm{eff}} = \sum_k \sum_{j \neq k} J_{jk}^{\mathrm{eff}} Z_j Z_k.
    \label{eq:effective_Hamiltonian}
\end{align}

Here, $Z$ is the Pauli-Z operator, and  the effective coupling matrix $J_{ij}^{\mathrm{eff}}$  between ion $i$ and $j$ reads
\begin{align}
    J_{ij}^{\mathrm{eff}} \propto \cos{(\Delta\vec{k}_L \cdot \vec{r}_{ij}^{\,0})},\label{eq:effective_interaction_J_jk}
\end{align}
where the wavevector \hbox{$\Delta\vec{k}_L = \vec{k}_1 - \vec{k}_2$} characterizing the crossed-beam ac-Stark shift is given by the difference between the wavevectors of the individual laser beams $\vec{k}_1$ and $ \vec{k}_2$, and lies in the $yz$-plane. Finally, $\vec{r}_{ij}^0$ is the distance vector between the equilibrium positions of ion $i$ and $j$.

Notably, the interaction between pairs of ions depends on the geometric factor $J_{ij}^{\mathrm{eff}} \propto \cos{(\Delta\vec{k}_L \cdot \vec{r}_{ij}^{\,0})}$. This means that one can control the effective interactions between ions by choosing the effective wavevector $\Delta \vec{k}_L$ appropriately. 
More specifically, if the ions are localized along the $y$-direction, such that $r_{i,y}^0 \approx 0$ for all ions as a result of a large curvature $\phi_y$, and if the effective wavevector lies within the $yz$-plane, then the effective interaction $J_{ij}^{\mathrm{eff}}$  can be controlled by tuning the $z$-component of the wavevector since $\Delta\vec{k}_L \cdot \vec{r}_{ij}^{\,0} \approx \Delta k_{L, z} (z_{i}-z_j)$. 
The interaction between ions $i$ and $j$ can therefore be suppressed in a targeted way by choosing $\Delta k_{L, z} = (2p + 1) \cdot \frac{\pi}{2 |z_i - z_j|}$ for any integer $p$. One can now realize that, for adjacent ions $i$ and $j$, the interaction between the ion $i$ and its nearest neighbor $j$ along the $z$-direction, as well as its next-to-next-to nearest neighbor, etc, can be completely suppressed for equidistantly spaced ions within a well.

To demonstrate the concept of such a geometric suppression of intra-well couplings, we exemplarily consider two strings with two ions each. For the following theory analysis, we set the parameters $\omega_z/2\pi = $ \SI{0.613}{\mega\hertz}, $\omega_y/2\pi =$ \SI{2.1}{\mega\hertz}, $\omega_x/2\pi =$ \SI{2}{\mega\hertz}, where the frequencies are set by the local curvatures $\alpha = m \omega_z^2 / q$, $\phi_{x, y} = m \omega_{x, y}^2 / q$. We define $l_z/d = 6/31$ where $l_z$ is the typical length scale for the distance between ions along the $z$-direction $l_z = (e^2/m\phi_z)^{1/3}$. Analogously to case 3 in Appendix~\ref{sec:appendix_MS}, we choose a laser beatnote of $\delta = \Omega_c /2 $, where $\Omega_c = |\omega_{\mathrm{str}} - \omega_{\mathrm{com}}| $ is the frequency gap between the axial common mode and the axial stretch mode. Rather than exciting sidebands, this beam configuration induces a state-dependent dipole force that couples the qubits and the two nearby in-phase and out-of-phase common motional modes. As shown in Appendix \ref{sec:app_modestructure} for two radially coupled ion strings with 4 ions in each, we remark that, for sufficient distances between the two wells, the other motional modes are sufficiently separated in frequency space so that their contribution to the interaction and the residual spin-motion entanglement can be neglected. In this case, one only has to carefully time the gate so that the corresponding trajectories in phase space are closed, as already demonstrated for the M\o lmer-S\o rensen scheme.

\begin{figure}[!tb]
	\centering\includegraphics[width=\linewidth]{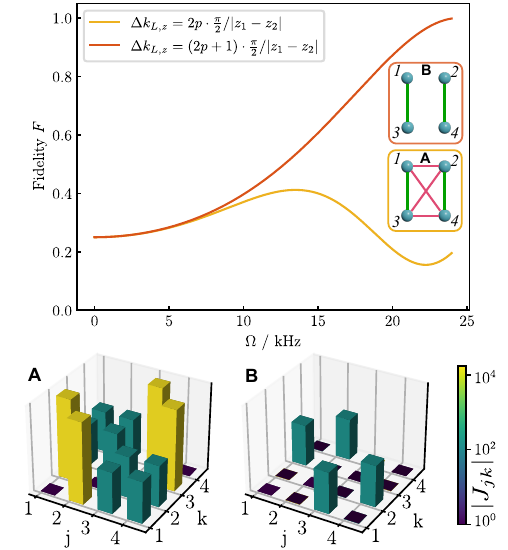}
 \caption{State fidelity for transversal entangling gate operation between 2x2 ions in radially separated wells as a function of the frequency $\Omega$ for a fixed gate time $ = 2\pi / \delta$. 
 The coupling matrix on the lower left (a) shows the effective coupling matrix for $\Delta k _{L, z} = 2p \cdot \frac{\pi}{2 |z_1 - z_2|}$, for which interactions between all pairs of ions participate. In this case, the state fidelity (orange) does not exceed $\approx 0.41$. 
 Undesired nearest-neighbor interactions along the z-direction are suppressed for $\Delta k _{L, z} = (2p + 1) \cdot \frac{\pi}{2 |z_1 - z_2|}$ (red), as shown in coupling matrix (b). Here, we exemplarily set $p = 25$. }
	\label{fig:four_ions_realistic_parameters}
\end{figure}

We consider an initial state $|\psi_{\mathrm{0}}\rangle = \otimes_{i=1}^{2n}|+_i\rangle$ with the number of ions per well $n = 2$ in this case, and the indexing of Fig.~\ref{fig:four_ions_realistic_parameters} (a). We define the state fidelity $F = |\langle \psi_{\mathrm{ideal}}| U_{\mathrm{eff}}|\psi_0\rangle |^2$ as the overlap with the ideally obtained state, where $U_{\mathrm{eff}} = e^{-{\rm i}H_{\mathrm{eff}}t}$ corresponds to the evolution due to the effective Hamiltonian $H_{\mathrm{eff}}$ (Eq.{\ref{eq:effective_Hamiltonian}}), with a final time $t$ in which both phase-space trajectories get closed. $|\psi_{\mathrm{ideal}}\rangle = U_{\mathrm{ideal}}| \psi_0\rangle = \prod_{j=1}^n \left(|+_{j}+_{j+n}\rangle -{\rm i} |-_{j}-_{j+n}\rangle \right)/\sqrt{2}$ is the state under the ideal transversal evolution \hbox{$U_{\mathrm{ideal}} = \prod_{j =1}^n {\rm e}^{-{\rm i}\frac{\pi}{4}Z_jZ_{j+n}}$}. 
We numerically determine the coupling matrices $J^{\mathrm{eff}}_{ij}$ for non-suppressed and geometrically-suppressed interactions between ions as given by Eq.~\ref{eq:effective_interaction}, which are shown in Fig.~\ref{fig:four_ions_realistic_parameters}~(a), (b). 
For a fixed gate time $t = 2\pi / \delta$, we vary the two-photon Rabi frequency $\Omega$, as defined by Eq.~\ref{eq:app_effective_Hamiltonian}, and numerically calculate the fidelity $F$, which is shown in the top panel of Fig.~\ref{fig:four_ions_realistic_parameters}. We find that, indeed, the next-nearest neighbor interactions are suppressed in the considered regime, thus implementing the desired transversal gate operation between the two pairs of qubits.

\begin{figure}[!tb]
    \centering\includegraphics[width=\linewidth]{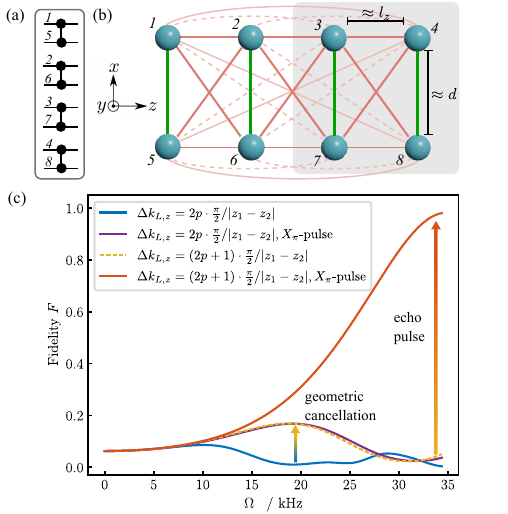}
    \caption{Transversal entangling gate operation between 2x4 ions in radially separated wells. (a) Parallel $(CZ)^{\otimes 4}$-gate operation, that can be realized with the presented sequence and additional single-qubit rotations. (b) Configuration of eight ions distributed in two radially separated wells. All ions interact with each other through Coulomb interaction. 
    The green lines correspond to the desired interactions that mediate the pairwise entangling gate along the x-direction. Nearest-neighbor and next-to-next-nearest neighbor interactions along the z-axis (red lines) can be geometrically canceled by choosing \hbox{$\Delta k_{L, z} =(2p + 1) \cdot \frac{\pi}{2 |z_1 - z_2|}$}. Residual next-nearest-neighbor interactions (dashed red lines) can be suppressed by employing a simple spin-echo sequence. 
    (c) State fidelity as a function of the frequency $\Omega$ for a fixed gate time $ = 2\pi / \delta$. For $\Delta k_{L, z} = 2p \cdot \frac{\pi}{2 |z_1 - z_2|}$, all qubits are coupled to each other and the fidelity drops to $F < 0.01$ (blue). Geometrically suppressing next-nearest neighbor interactions by setting \hbox{$\Delta k_{L, z} = (2p + 1) \cdot \frac{\pi}{2 |z_1 - z_2|}$} (orange), and then a single spin-echo pulse (red), increases the fidelity to a value of 0.99.}
	\label{fig:eight_ions_TV_gate}
\end{figure}

As a next step, we extend the presented concept to larger strings of ions, where not only next-nearest neighbor interactions contribute but also ions are not necessarily spaced equidistantly. 
First, a quartic axial contribution is added to the initial trapping potential $V_{\mathrm{rad}}^0$~(Eq.~\ref{eq:V_rad}) in order to achieve equidistant spacing between ions along this axis~\cite{lin2009large} $
V_{\mathrm{rad}} =  V_{\mathrm{rad}}^0 + \gamma r_z^4 / 4! $, where we define $d_z = \sqrt{4!\phi_z/\gamma}$ in alignment with Sec.~(2B). For fixed \,  parameters $\alpha , \phi_z$ and $d$, we minimize the inhomogeneity in the inter-well spacings by finding the optimal value of $l_z/d_z$. In the following, we exemplarily consider two strings of 4 ions each, and numerically find equidistant spacing along the $z$-axis for a quartic axial contribution with $l_z/d_z = 0.43$, for the above specified parameters. We simulate the effective Hamiltonian presented in Eq.~(\ref{eq:effective_Hamiltonian}) for these parameters and again determine the state fidelity, which is shown in Fig.~\ref{fig:eight_ions_TV_gate}. We find that, even if the nearest-neighbor and next-to-next-nearest-neighbor interactions along the $z$-axis are geometrically suppressed by setting  $\Delta k_{L, z} = (2p + 1) \cdot \frac{\pi}{2 |z_1 - z_2|}$, the fidelity does not exceed \hbox{$F\approx 0.17$}. This limitation is due to non-vanishing next-nearest-neighbor couplings depicted by dashed red lines in Fig.~\ref{fig:eight_ions_TV_gate}(a), which cannot be  canceled geometrically. However, these unwanted residual couplings now have a structure that allows to suppress them by adding a simple dynamical decoupling technique, which has been similarly used for Mølmer-Sørensen gates~\cite{sorensen1999quantum}. After an evolution time $t/2$, an $X_{\pi}$-pulse is applied only to the rightmost half of the qubits, here 3, 4, 7 and 8, which rotates these qubits by an angle $\pi$ about the $X$-axis. This inverts the sign of the next-nearest-neighbor interaction and cancels these unwanted contributions after another evolution time $t/2$. Overall, the described gate leaves only the desired transversal interactions along the $x$-direction between pairs of opposing ions belonging to different wells, resulting in a parallel gate operation as illustrated in Fig.~\ref{fig:eight_ions_TV_gate}(b). In doing so, the fidelity increases to a value of 0.99, as shown in red in Fig.~\ref{fig:eight_ions_TV_gate}(c). We note that this simple  refocusing boost  requires the first geometric-cancellation step as, otherwise, most of the couplings represented by the red solid lines in Fig.~\ref{fig:eight_ions_TV_gate}(a) would no longer vanish,  severely limiting the achievable fidelity.  We also note that this technique can be generalized to larger number of ions by increasing the complexity of spin-echo type composite sequences~\cite{khodjasteh2005fault, uhrig2009concatenated, august2017using, morazotti2024optimized}. 

The full entangling gate in this case takes 
\hbox{$ t\approx$ \SI{180}{\micro\second}}, which is a similar duration as the presented MS-gate with $ t= $\SI{190}{\micro\second}, and also on par with previously demonstrated entangling gates on trapped-ions~\cite{PRXQuantum.2.020343, postler2024demonstration}. We note, however, that our light-shift gate  acts simultaneously over two ion pairs, something that would require two-sequential MS gates addressing each pair of qubits one at a time which would, at the very least,  double the duration to $t=$ \qty{380}{\micro\second}. We expect this advantage to become even more pronounced as the qubit number per string $n$ increases. In fact, following our previous arguments about the gate speed being determined by the size of the avoided crossing between the in- and out-of-phase modes, we expect that its speed could be enhanced sub-linearly when increasing $n$ (see the right panel of Fig.~\ref{fig:coupling_vs_ionnumber}).
Moreover, previous works have shown that fidelities can be increased and gate times decreased by increasing the efficiency of the dynamical decoupling and by employing modulated laser fields to couple to multiple motional modes~\cite{Lu2019, Choi2014, Landsman2019} and to reduce the residual qubit-oscillator entanglement~\cite{Green2015, Milne2020}. A detailed analysis of the extension of this scheme to larger ions crystals, relevant noise sources and gate optimization is left for future work.

\subsection{Applications for QEC codes}

\begin{figure*}
    \centering
    \includegraphics[width=1\linewidth]{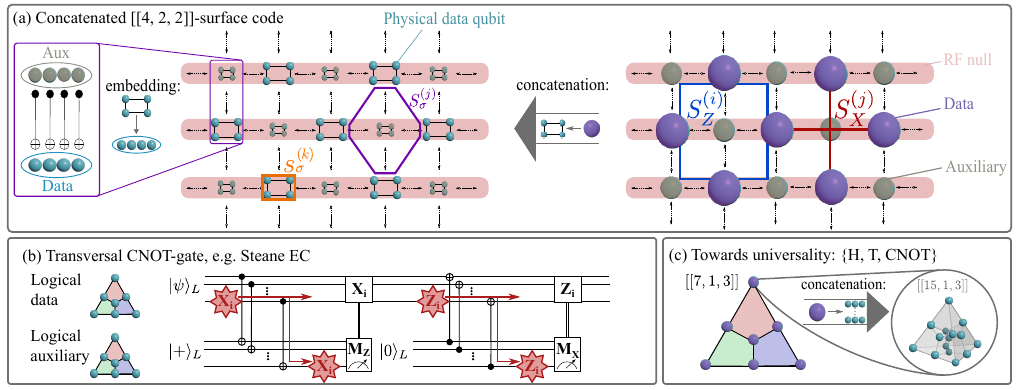}
    \caption{Applications of the 2D architecture for QEC codes. (a) Concatenated surface code: Every edge in the $[[n_c= (d_c^2 + (d_c-1)^2), 1, d_c]]$ surface code (right) is replaced by a square which contains four qubits~\cite{criger2016noise}. In doing so, one can construct a $[[n_c' = 4(d_c^2 + (d_c-1)^2), k_c' = 2, d_c' = 2d_c]]$ code on a square-octagon lattice, where each cluster of four qubits resides in one potential well (center). A single auxiliary qubit, depicted in grey (right), in a parity measurement may be replaced by four qubits encoding a logical state of the $[[4, 2, 2]]$ code ~\cite{criger2016noise} (embedding, left). (b) Transversal CNOT-gates enable efficient syndrome readout by means of Steane-type EC~\cite{steane1997active}. An auxiliary logical qubit is prepared in $|+\rangle_L$ using the same code as the data qubit and coupled to the data qubit with a transversal CNOT-gate. This copies $X$-errors from the data onto the auxiliary qubit. By measuring the auxiliary qubit projectively in the $Z$-basis, one can infer the respective syndrome. Analogously, $Z$-errors are corrected by preparing an auxiliary $|0\rangle_L$ and measuring in the $X$-basis. 
    These logical auxiliary qubits could be prepared in separate potential wells and readily be used as resource states in the 2D lattice architecture. (c) By concatenating the two-dimensional $[[7, 1, 3]]$ and the three-dimensional $[[15, 1, 3]]$ color code, one can implement a FT universal gate set \{H, T, CNOT\}~\cite{jochym2014using, chamberland2016thresholds, chamberland2017overhead}. }
    \label{fig:overview_QEC_applications}
\end{figure*}

To illustrate the potential power of the developed trapped-ion architecture for scalable FT QEC, we outline a few routes which can be naturally hosted and operated in the 2D lattice structure of coupled traps. Our simulations show the principle feasibility of implementing transversal gate operations in the proposed 2D lattice architecture.
These gates open up the possibility to carry out logical operations in a highly parallelized way: 
by coupling two ion-chains in separate wells, it is possible to, for example, apply CNOT-gates on pairs of qubits as depicted on the left side of Fig.~\ref{fig:overview_QEC_applications}(a). 
This implements a transversal two-qubit gate on specific codes, which can often only be performed sequentially~\cite{postler2022demonstration, postler2023demonstration, ryan2022implementing, Bluvstein2023}, but is a key feature for modular QEC, such as Steane-type error correction~\cite{steane1997active, aliferis2005quantum, chamberland2018new}. 
Here, logical auxiliary qubits are coupled to the logical data qubit with a transversal CNOT-gate, which effectively copies errors from the data to the auxiliary qubit, as depicted in Fig.~\ref{fig:overview_QEC_applications}(b). By measuring the logical auxiliary qubit, one can directly extract the error syndrome. 
Compared to other EC schemes, Steane-type error correction requires the lowest possible number of two-qubit gates and can outperform other state-of-the-art EC protocols such as flag-based EC, as has been demonstrated in recent experiments~\cite{postler2023demonstration, Bluvstein2023, huang2024comparing, cain2024correlated}. 
Transversal gates are not only being used for error correction and decoding~\cite{cain2024correlated} but also actively utilized for the parallelized implementation of logical operations~\cite{bluvstein2024logical} such as magic state injection~\cite{lacroix2024scaling}, distillation~\cite{rodriguez2024experimental} or lattice surgery~\cite{ryan2024high}, which are key components for universal quantum computing.\\

We now exemplarily consider the class of topologically-concatenated QEC codes, which are an ideally suited candidate for parallelly executed gate operations on fully controllable 2D lattices due to their inherent code structure. As illustrated in Fig.~\ref{fig:graphical_abstract}, subsets of ions are confined in separate wells on a 2D lattice, where each potential well can represent a logical qubit of the lower concatenation level. 
Within each of these wells, two-qubit gates can be carried out between arbitrary pairs of qubits. The different logical qubits are coupled by means of well-to-well entangling gates and form the second layer of concatenation. 
This architecture directly mimics the code structure of concatenated codes. One advantage of these quantum codes is the scaling of the total failure probability: below a certain threshold of the physical error rate, the failure probability is suppressed doubly-exponentially with each layer of concatenation, which enables practical FT quantum computing below this threshold~\cite{knill1998resilient, aliferis2005quantum}. One example is the concatenated surface code, as exemplarily illustrated in Fig.~\ref{fig:overview_QEC_applications}(a). 
When each physical qubit on a surface code lattice is substituted by an $[[n_c = 4, k_c = 2, d_c = 2]]$ code, defined by four-qubit stabilizers $s_{X}$ and $s_{Z}$, where $n_c = 4$ physical qubits encode $k_c = 2$ logical qubits with code distance $d_c = 2$, one obtains the $[[4, 2, 2]]$-concatenated code~\cite{criger2016noise}. 
By concatenating the surface code and the $[[4, 2, 2]]$ code, the stabilizer $S_Z^{j}$ ($S_X^{i}$) defined on vertex $i$ (plaquette $j$) of the surface code lattice (blue, red) is transformed into a weight-8 stabilizer $S_{\sigma}^{j} (S_{\sigma}^{i})$ with $\sigma = Z (X)$ (purple octagon) and the additional stabilizers $s_{\sigma}^{(k)}$ (orange square) of the $[[4, 2, 2]]$ code emerge on each replaced qubit. 

Furthermore, by concatenating codes with complementary sets of transversal and, therefore FT gates, it is also possible to implement a FT universal gate set~\cite{jochym2014using, chamberland2016thresholds, chamberland2017overhead}. Access to a FT universal set of gates allows the efficient approximation of arbitrary operations~\cite{kitaev1997quantum} on encoded states and poses a key requisite for error-corrected universal quantum computing. 
For example, the logical $H$- and CNOT-gate is transversal on the $[[7, 1, 3]]$ code, while the $[[15, 1, 3]]$ code has a transversal implementation of the non-Clifford $T$-gate, which is required for universal quantum computing. By replacing each qubit of the $[[7, 1, 3]]$ code with an instance of a $[[15, 1, 3]]$, as illustrated in Fig.~\ref{fig:overview_QEC_applications}(b), it is possible to access a full universal set of gates fault-tolerantly:
Single errors are prevented from propagating onto multiple qubits on the same encoded block by applying a non-transversal version of the T-gate on the $[[7, 1, 3]]$ code using only transversal gates of the concatenated $[[15, 1, 3]]$ code \cite{jochym2014using}. 
Moreover, concatenated codes have been found to efficiently protect against biased noise~\cite{aliferis2008fault}, which is often present in ion trap quantum processors~\cite{pal2022relaxation}.

Tab.~\ref{tab:physical_requirements} summarizes the minimal numbers of ions and registers required for the proposed QEC protocols. 
The embedding of these protocols on a fully connected two-dimensional lattice architecture, in principle, enables the implementation of scalable FT QEC and opens up a new route towards FT universal quantum computing on an ion trap quantum processor. 

\begin{table}[]
\centering
\caption{Required minimal number of ions per well and number of registers for different QEC protocols: Steane EC on the $[[7, 1, 3]]$ Steane code~\cite{steane1997active, postler2023demonstration}, which also corresponds to the smallest 2D topological color code~\cite{bombin2006topological, nigg2014quantum}; magic state injection (MSI) on the $[[7, 1, 3]]$~\cite{goto2016minimizing, chamberland2019fault, postler2022demonstration}; the implementation of a universal gate set by concatenating the $[[7, 1, 3]]$ code with the $[[15, 1, 3]]$ code~\cite{chamberland2017overhead}; and a general distance-$d$ $[[4, 2, 2]]$-surface code~\cite{criger2016noise}. The given numbers only correspond to the number of required data qubits and registers, and do not take auxiliary qubits into account.}
\label{tab:physical_requirements}
\begin{tabular}{r|c|c}
QEC protocol & \#Ions per well & \#Registers \\
\hline
Steane EC on $[[7, 1, 3]]$ & $7$& $2$\\
\hline
MSI on $[[7, 1, 3]]$ & $7$ & $2$\\
\hline
Universal gate set  &  $15$ & $7$ \\
\hline
$[[4, 2, 2]]$-surface code~& $4$& $ d_c^2 + (d_c -1)^2$\\
\end{tabular}
\end{table}

\section{Conclusions and Outlook}
\label{sec:conclusions}
\begin{table}[]
\centering
\caption{Summary of well-to-well coupling rates $\Omega_c$ and interaction constants $k_\m{int}$.}
\label{tab:coupling_values}
\begin{tabular}{c c c|c|c|c}
& & \begin{tabular}{@{}c@{}} Ions \\per\\ well\end{tabular} & \begin{tabular}{@{}c@{}} Separation \\ $d$ (\unit{\micro\meter})\end{tabular} & \begin{tabular}{@{}c@{}} Coupling \\  $\Omega_c$/(2$\pi$)\\ (\unit{\kilo\hertz})\end{tabular} & \begin{tabular}{@{}c@{}} Interaction \\ constant $k_{\m{int}}$ \\(\unit{\kilo\electronvolt\per\meter\squared})\end{tabular}\\ \hline
& Ref. \cite{Brown2011}& 1        & 40           &    3.1   & 26                             \\ \cline{2-6}
& & 1        & 54          &    1.9    & 16                             \\
& Ref.  \cite{Harlander2011} & 2        & 54           &    6   & 110                             \\
& & 3        & 54            &   14   & 370 \\ \cline{2-6}
Axial & Ref. \cite{Wilson2014} & 1 & 30 & 12 & 177 \\ \cline{2-6}
& & 2             & 42        &     17     & 230                              \\
 & This work & 4             & 52         &     25    & 650                              \\
& & 6             & 56          &    39    & 1600                             \\ \hline
& & 1 &  29 &  15& 61 \\
Radial & This work & 2 & 41 & 7.5 & 61 \\
& & 3 & 51& 6.4& 77\\

\end{tabular}
\end{table}

In summary, we have proposed and realized the building blocks of a so-called Quantum Spring Array (QSA) architecture, where independent ion-strings are arranged in a two dimensional array, and  connectivity is achieved via axial and radial ion transport. 
We have implemented and characterized coupling between chains of ions in separate wells in both axial and radial directions with respect to the axis along which ion crystals are confined. 
The coupling rates of our system are compared to the results from references~\cite{Harlander2011,Brown2011,Wilson2014} in Table \ref{tab:coupling_values}. 
The coupling rates measured in this work exceed those of previous works despite higher separation between wells, owing to the higher number of ions per well. We note that in references \cite{Brown2011,Wilson2014} the ion species under investigation is $^{9}$Be$^+$. Since the coupling rate depends on the mass of the ion species (see Eq. \ref{eq:couplingrate}),  a mass- and frequency independent comparison between the various results can be made by expressing the interaction between ions in separate wells in terms of the interaction strength (see $k_{\m{int}}$  in Eq. \ref{eq:interaction_constant}), which depends solely on the charge distribution.

Several features of well-to-well coupling of ions have been investigated in this work in two types of surface traps, each dedicated to generating a double-well potential along one of the two principle axes of the \latticename{}.
In a linear segmented surface trap, we have demonstrated axial well-to-well coupling rates of 39~kHz for 6 ions per well.
We have shown exchange of motional energy between the axially separated wells and a reduced heating rate in the double-well stretch mode of oscillation compared to that of the common mode. 
In a surface trap containing two co-linear trapping regions, the first implementation of radial coupling has been shown. We harnessed this radial coupling to show exchange of a single phonon between axial modes of oscillation of ions located in a radial double well potential. Furthermore, we have, for the first time, implemented an entangling gate of radially coupled ions located in two distinct linear traps. 

We demonstrated the capability to control the distance and thus the coupling rate between ions distributed in adjacent linear traps in the array.  The coupling rates for these interactions between radially separated wells are up to \qty{15}{\kilo\hertz} with of up to three ions per well. 
We observed a high heating rate in the radial transport configuration which is not fully understood. We speculate that it is caused by uncorrelated phase and/or amplitude fluctuations of the voltages applied to the RF electrodes. We expect that we can overcome this technical obstacle with low-noise electronics and resonators with higher quality factors that decrease the bandwidth at which the noise is affecting the motion. Nevertheless, we expect that ion heating in the double-RF drive configuration used for radial transport will be one of the main challenges for a large-scale implementation. Thus, investigating the origin and characterizing the effect of such noise is a sensible goal for further experimental research.

Larger 2D-trap arrays will feature many trapping sites distributed along several adjacent linear traps. Two RF signals are sufficient to control the relative spacing between all the linear traps in the array simultaneously. Still, more RF signals could be delivered to the trap if independent control of the spacing between specific linear traps in the array is needed. Moreover, since all the trapping sites pertaining to a linear trap will be moved simultaneously during the RF transport operation, to avoid excess micromotion, local compensation for stray fields will be needed. Further investigation is necessary to characterize the length over which locally addressable DC compensation is required, although it is expected that this length is comparable to the ion-surface distance.

In this work, two separate trap designs were used to demonstrate axial and radial coupling.
As a natural next step, we will focus on the design and fabrication of surface traps capable of controlling larger 2D arrays via both axial and radial shuttling operations. Their design will be similar to the one shown in Ref.~\cite{Holz2020}.
In addition, these traps will feature multiple trapping regions in both axes and will allow the implementation of integrated voltage control electronics and optical elements, as expected in all scalable ion-trap quantum computing architectures. 
More specifically to the QSA, the implementation of transversal entangling gates requires single ion addressing. We envision dedicated specific zones along the trap where single ion addressing is realized via a combination of waveguides and microlenses. These integrated optical components, together with the shuttling capabilities of the \latticename{}, would allow one to achieve the required connectivity for transversal gate operations between specified lattice sites. 

This type of connectivity represents one of the integral components of concatenated quantum error correction (QEC) schemes. Furthermore, the code family of quantum Low Density Parity Check (qLDPC) codes holds the promise of constant encoding rates, thus enabling low-overhead fault-tolerant quantum computing~\cite{gottesman2013fault, breuckmann2021quantum}. Although these codes generally rely on non-local connectivity, certain qLDPC codes can be embedded on a 2D architecture~\cite{strikis2023quantum, berthusen2025toward}.

With this work we have shown that the connectivity provided by
the \latticename{} architecture 
readily lends itself to natural embeddings of such QEC schemes.
In conjunction with developing these QEC techniques, we explored efficient implementations of parallel transversal gate operations between the physical qubits of each subregister. Such operations have the potential to significantly improve the efficiency of QEC primitives and thus to reduce the overhead that is associated with fault-tolerant operations. This will be the starting point for more detailed research on QEC codes and fault-tolerant operations that are optimized to the available physical operations.

\section{Acknowledgments}
\label{sec:acknowledgements}
We gratefully acknowledge support by the European Union’s Horizon Europe research and innovation program under Grant Agreement Number 101114305 (``MILLENION-SGA1'' EU Project), the US Army Research Office through Grant Number W911NF-21-1-0007, the European Union’s Horizon Europe research and innovation program under Grant Agreement Number 101046968 (BRISQ), the ECSEL JU (which is supported by
European Union’s Horizon 2020 and participating countries)
under Grant Agreement Number 876659 (iRel),
the Austrian Research Promotion Agency (FFG) under the project “OptoQuant” (37798980), the European support
“Important Projects of Common European Interest” (IPCEI) on Microelectronics, the Austrian Science Fund (FWF Grant-DOI 10.55776/F71) (SFB BeyondC), the Austrian Research Promotion Agency under Contracts Number 896213 (ITAQC) and 897481 (HPQC), the Austrian Science Fund (FWF) [10.55776/COE1] and the European Union – NextGenerationEU, the Office of the Director of National Intelligence (ODNI), Intelligence Advanced Research Projects Activity (IARPA) and the Army Research Office, under the Entangled Logical Qubits program through Cooperative Agreement Number W911NF-23-2-0216. We further receive support from the IQI GmbH. 
This research is also part of the Munich Quantum Valley (K-8), which is supported by the Bavarian state government with funds from the Hightech Agenda Bayern Plus. M.M. additionally acknowledges support by the Deutsche Forschungsge- meinschaft (DFG, German Research Foundation) under Germany’s Excellence Strategy ‘Cluster of Excellence Matter and Light for Quantum Computing (ML4Q) EXC 2004/1’ 390534769, and the ERC Starting Grant QNets through Grant No. 804247. A.B. additionally acknowledges support from PID2021- 127726NB- I00 (MCIU/AEI/FEDER, UE), from the Grant IFT Centro de Excelencia Severo Ochoa CEX2020-001007-S, funded by MCIN/AEI/10.13039/501100011033, from the the CAM/FEDER Project TEC-2024/COM 84 QUITEMAD-CM, and from the CSIC Research Platform on Quantum Technologies PTI- 001.
We thank the referees for their insightful comments.

\clearpage
\appendix

\section{Mode structure of two coupled ion strings}
\label{sec:app_modestructure}
In Sec. \ref{sec:double_well_physics} we developed the model describing the coupling of ion strings, but, throughout this work, we focus only on the lowest order coupled axial modes of motion. In this appendix, we discuss the full axial mode structure of two linear strings of ions with 4 ions in each, coupled along the axial and radial directions. The modes and mode frequencies are obtained by solving Eq. \ref{eq:vt} and Eq. \ref{eq:hessian}, as explained in the main text.

As an exemplary representative set of parameters, we set the two wells to have a separation of \qty{60}{\micro\meter} for the case of axial coupling, and \qty{50}{\micro\meter} for radial coupling. The wells have an axial curvature that corresponds to an axial frequency of \qty{400}{\kilo\hertz} if only a single \calciumforty ion was trapped.
\begin{figure}
    \centering
    \includegraphics[width=.8\linewidth]{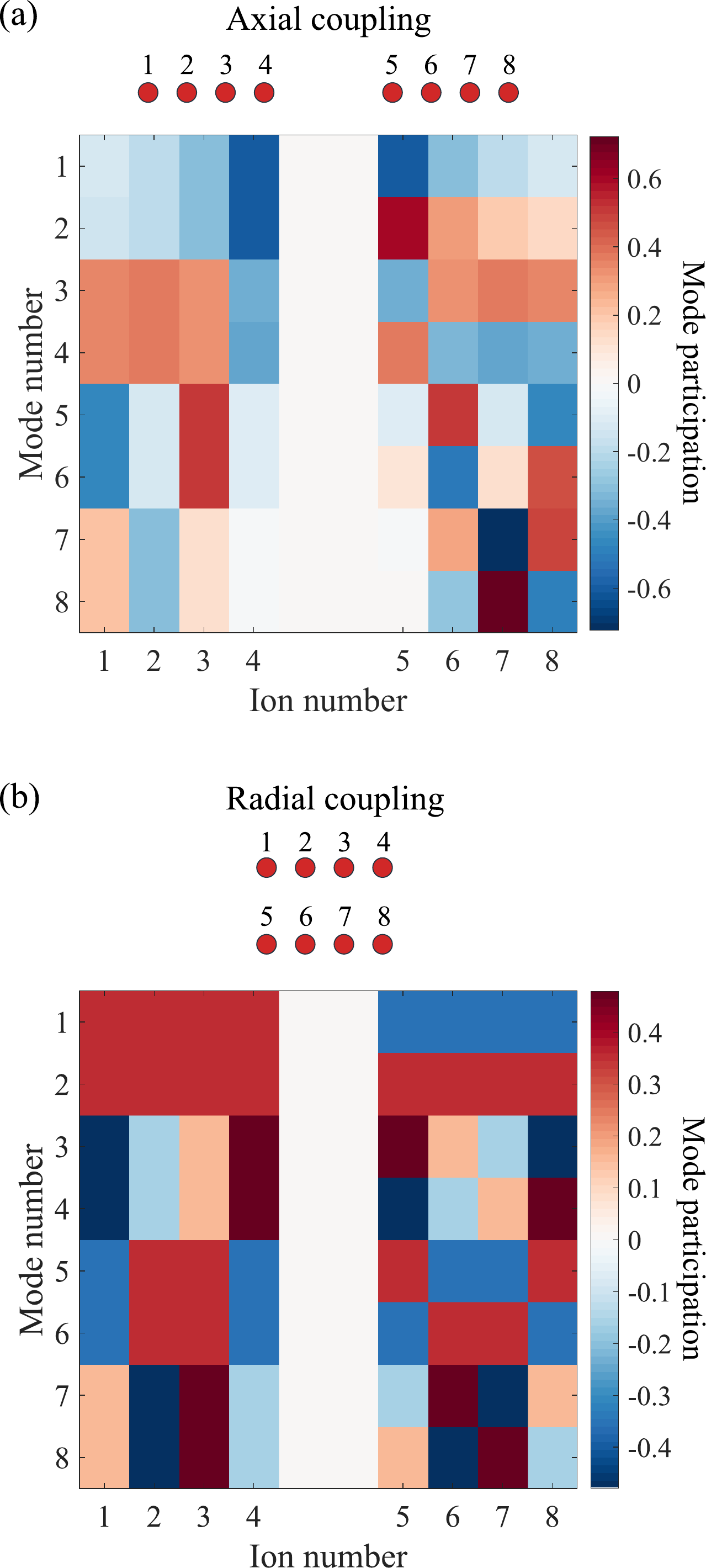}
    \caption{Normalized  interaction strengths of axial modes, for (a) axially coupled and (b) radially coupled ion strings of 4 ions each. Wells in (a) are separated by \qty{60}{\micro\meter}, while in (b) the well-to-well separation is of \qty{50}{\micro\meter}. The potential curvature is selected to result in an axial frequency \qty{400}{\kilo\hertz} for a single trapped ion in both cases.}
    \label{fig:modemat}
\end{figure}

The normalized ion interaction strength for axially coupled ion strings are depicted for each mode in Fig. \ref{fig:modemat}(a). Within a chain, each ion participates differently in each mode, also for lowest order axial common mode. This unequal participation stems from the anharmonicity of the axial double-well potential. Still, for a given mode, the participation is the same for each pair of ions that are located symmetrically with respect to the center of the double-well. In addition, the modes are grouped in pairs, where in each of the modes the participation of the ions is the same, except that opposing ions oscillate in-phase or out-of-phase.
Fig. \ref{fig:modemat}(b) shows the normalized interaction strengths of the ions in two radially coupled ion strings. For the lowest order modes (which are the radially coupled axial stretch and common modes of oscillation referred to as $\omega_{str, com}$ in the main text), all the ions participate equally in the oscillation and thus equally couple to a laser beam illuminating the entire chain, making the implementation of the transversal gate described in Sec. \ref{subsec:tvgate} feasible. Similar to the axial coupling case, modes manifest in groups of two.
\begin{figure}
    \centering
    \includegraphics[width=1\linewidth]{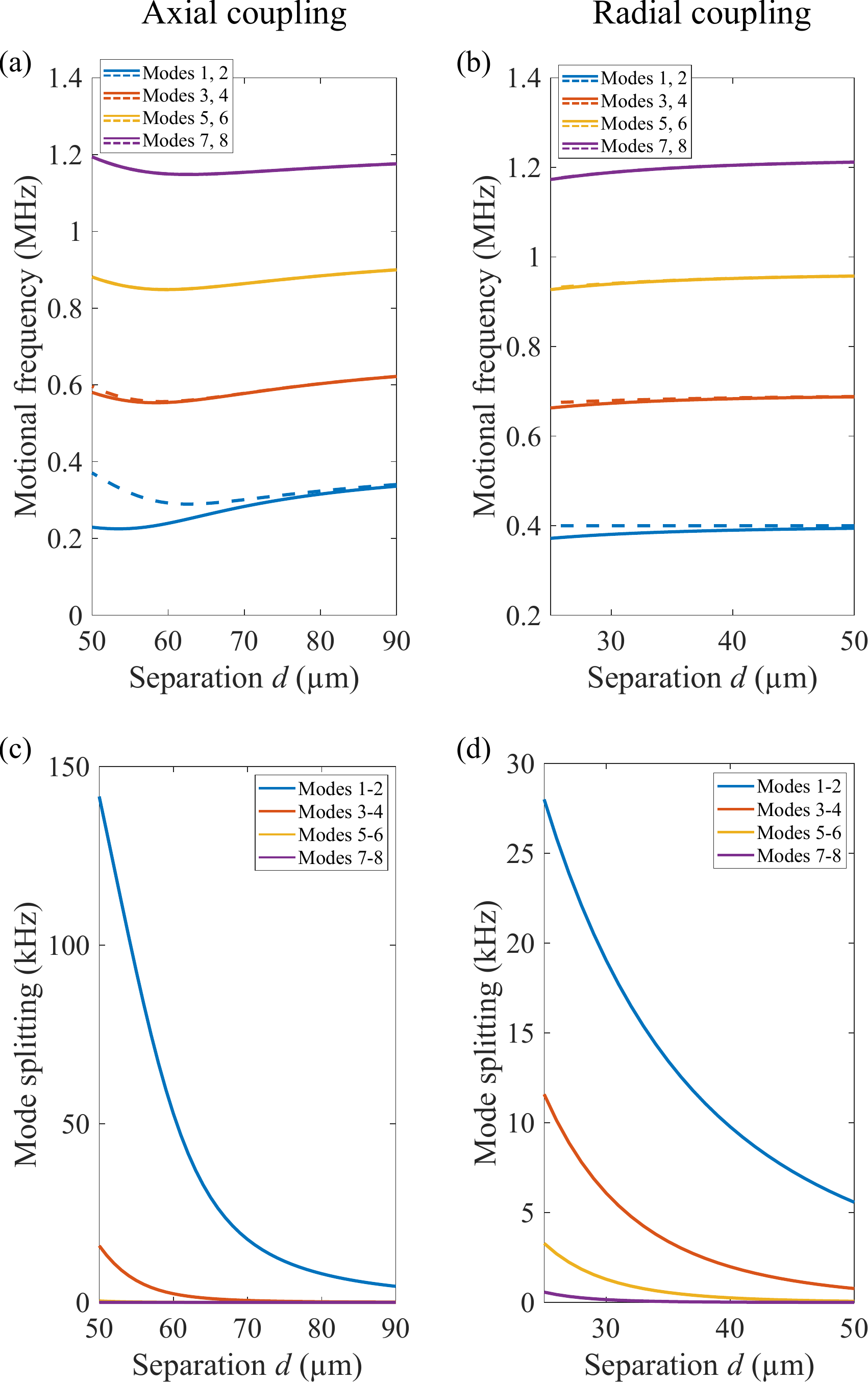}
    \caption{Motional spectrum for the axial modes as a function of well to well separation. (a) Motional frequency as a function of the well-to-well separation for the case of 2 ion strings with four ions in each distributed in an axial double-well potential. (b) Mode splitting for the case (a). (c) Motional frequency as a function of the well-to-well separation for the case of 2 ion strings with four ions in each distributed in a radial double-well potential. (d) Mode splitting for the case (c).}
    \label{fig:modesplit}
\end{figure}

Fig. \ref{fig:modesplit} displays the dependence of the mode spectrum as a function of the well-to-well separation for axial modes of two axially and radially coupled wells with 4 ions in each. Similarly to above, the axial curvature is set to correspond to an axial frequency of \qty{400}{\kilo\hertz} if only a single \calciumforty ion was trapped.

For the axial modes of the axially and radially coupled wells (Figs. \ref{fig:modesplit}(a) and (b)), the axial mode spectrum for large well-to-well separation is that of two independent wells, set to have the same axial curvature. The mode frequencies in the two wells are thus degenerate. As the separation decreases and the coupling between ion chains increases, each axial mode splits into two, where the splitting is higher for lower mode frequencies.
Figs. \ref{fig:modesplit}(c) and (d) show the dependence of the splitting between pairs of modes depicted, respectively, in Figs. \ref{fig:modesplit}(a) and (b). The mode splitting is larger for axially coupled wells in comparison to the radially coupled ones. This difference emerges from the spatial distributions of the ion chains in the axial and radial double-wells, described in Sec. \ref{sec:double_well_physics}, leading to a deviation from the distance scaling ($d^{-3}$) predicted from the coupled point-charge model.

\section{Phonon transfer with DC voltage switching}
\label{sec:app_DCswitching}
In Section \ref{subsec:radial_phonon_exchange} we have demonstrated an exchange of phonons between radially separated ions and Appendix \ref{sec:app_phononEx_entanglement} has shown that this exchange can be used to generate entanglement between those ions. In both cases, this exchange involves tuning the motional frequencies of the two wells out of resonance to allow addressed sideband pulses to inject a phonon in a chosen well, and into resonance to enable phonon exchange through well-to-well coupling. In both the phonon exchange and entangling measurements the rate of phonon exchange is higher than is predicted from basic theoretical models. This is attributed to having intentionally set the two wells to be near, but not in, resonance for the exchange. The reason for this choice is related to the fact that the switching in and out of resonance does not happen instantaneously, but is gradual due to our built-in electronic filters of DC electrodes to which the voltage changes are applied. The two wells therefore spend some non-negligible time near resonance before settling to a steady state on resonance. While the setting is at near-resonance, partial phonon exchange already takes place, and disrupts the full transfer that would occur at resonance. In this appendix, we use numerical models to detail this effect and further motivate why we have chosen to measure phonon transfer with the two wells not perfectly resonant.

\begin{figure}
    \centering
    \includegraphics[width=\linewidth]{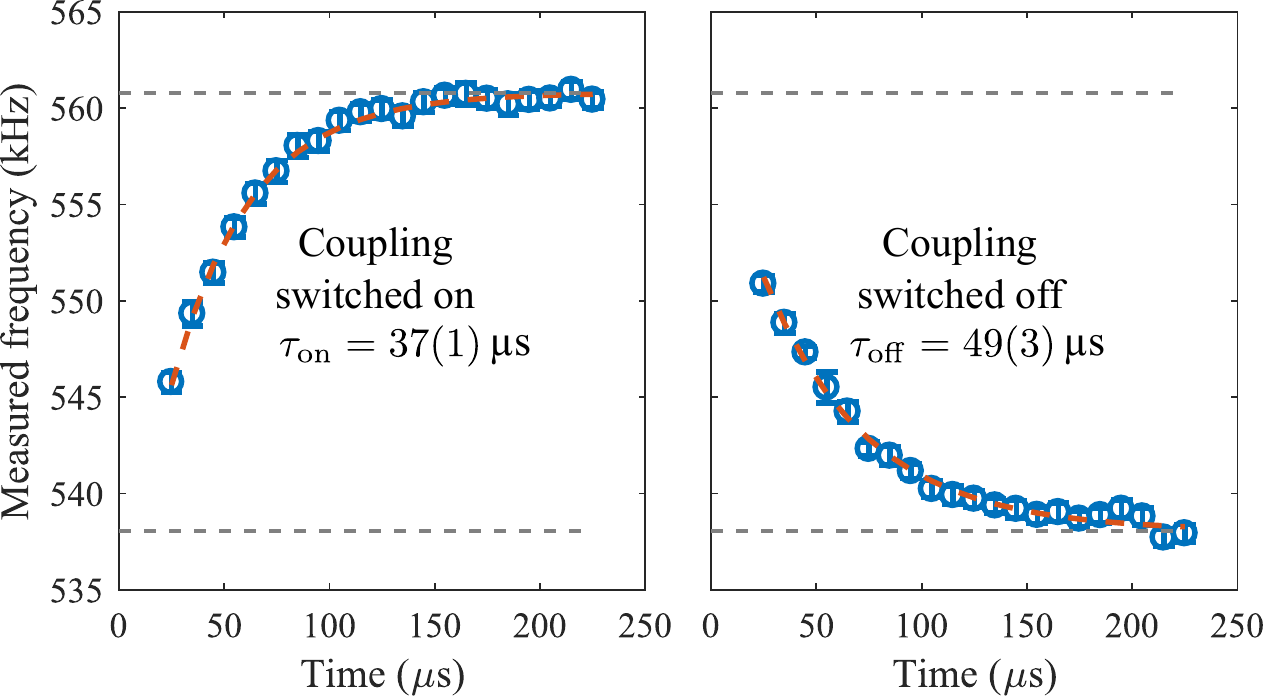}
    \caption{Characterization of switching time between voltage sets. The plot shows the measured motional frequency of one of the two wells, as a function of wait time after switching wells out of and into resonance, where the error bars represent the 95\% confidence interval of the fit to the motional spectra. Dashed lines are exponential decay fits, from which we extract characteristic switching times.}
    \label{fig:dc_switch_plot}
\end{figure}

One of the parameters of the calculation is the rate at which motional frequencies of the wells vary over time after DC voltages are switched. We experimentally determine this value by taking spectra around the motional sideband of a trapped ion as a function of wait time after switching DC voltages. The probe time (\qty{50}{\micro\second}) is not negligibly small compared to the rate at which motional frequencies change. Since the change in frequency during the probe time is approximately linear, it can be assumed that the measured frequency corresponds to the motional frequency at half the probe time, \qty{25}{\micro\second}. The results of this measurement are shown in Fig.~\ref{fig:dc_switch_plot}. The dashed lines are fits using an exponential decay model, $f=f_{\m{fin}}-(f_{\m{fin}}-f_{\m{init}})\exp(-t/\tau_{\m{on}/\m{off}})$, with $f_{\m{init}}$ and $f_{\m{fin}}$ the initial and final frequencies, $t$ the time after switching, and $\tau_{\m{on}/\m{off}}$ the characteristic decay time when switching on/off resonance. From the fits we determine $\tau_\m{on}=\qty{37(1)}{\micro\second}$ and $\tau_{\m{off}}=\qty{49(3)}{\micro\second}$.

To examine how the non-zero switching time affects the transfer of phonons between wells, we set up a model of coupled harmonic oscillators. The model solves the following differential equation
\begin{equation}
    \frac{d^2\vec{y}}{dt^2}= 
    \begin{pmatrix}
-(2\pi f_1(t))^2+\tilde{k}_{\m{int}} & \tilde{k}_{\m{int}} \\
\tilde{k}_{\m{int}} & -(2\pi f_2(t))^2+\tilde{k}_{\m{int}} 
\end{pmatrix}
\vec{y}
\end{equation}
where $\vec{y}=(y_1,y_2)$ are the oscillator amplitudes of oscillators 1 and 2, $f_1(t)$ and $f_2(t)$ their time dependent oscillation frequencies, and $\tilde{k}_{\m{int}} = k_{\m{int}} / m$ the (mass-independent) coupling between them. The vector $\vec{y}$ is initialized as $(1,0)$, representing a phonon occupying the first oscillator (which we will call ``well 1'') and no phonons in the second (well 2). Furthermore, we set $d\vec{y}/dt=(0,0)$. 

Using experimental parameters of the phonon exchange measurement of Section \ref{subsec:radial_phonon_exchange}, we initialize oscillation frequencies to $f_1(0)=\qty{520}{\kilo\hertz}$ and $f_2(0)=\qty{560}{\kilo\hertz}$, which represents the frequencies of the two wells in the detuned state. The on-resonance coupling rate is set to \qty{6.6}{\kilo\hertz}, which corresponds to $\tilde{k}_\m{int}=\qty{1.41d11}{\per\second\squared}$. At $t=0$, the DC voltages are switched, and the wells exponentially approach resonance with characteristic time $\tau_{\m{on}}=\qty{37}{\micro\second}$, though are set to settle to a detuning that is not necessarily on resonance, $\Delta f=f_2(t)-f_1(t)$ for $t\rightarrow\infty$.

\begin{figure}
    \centering
    \includegraphics[width=\linewidth]{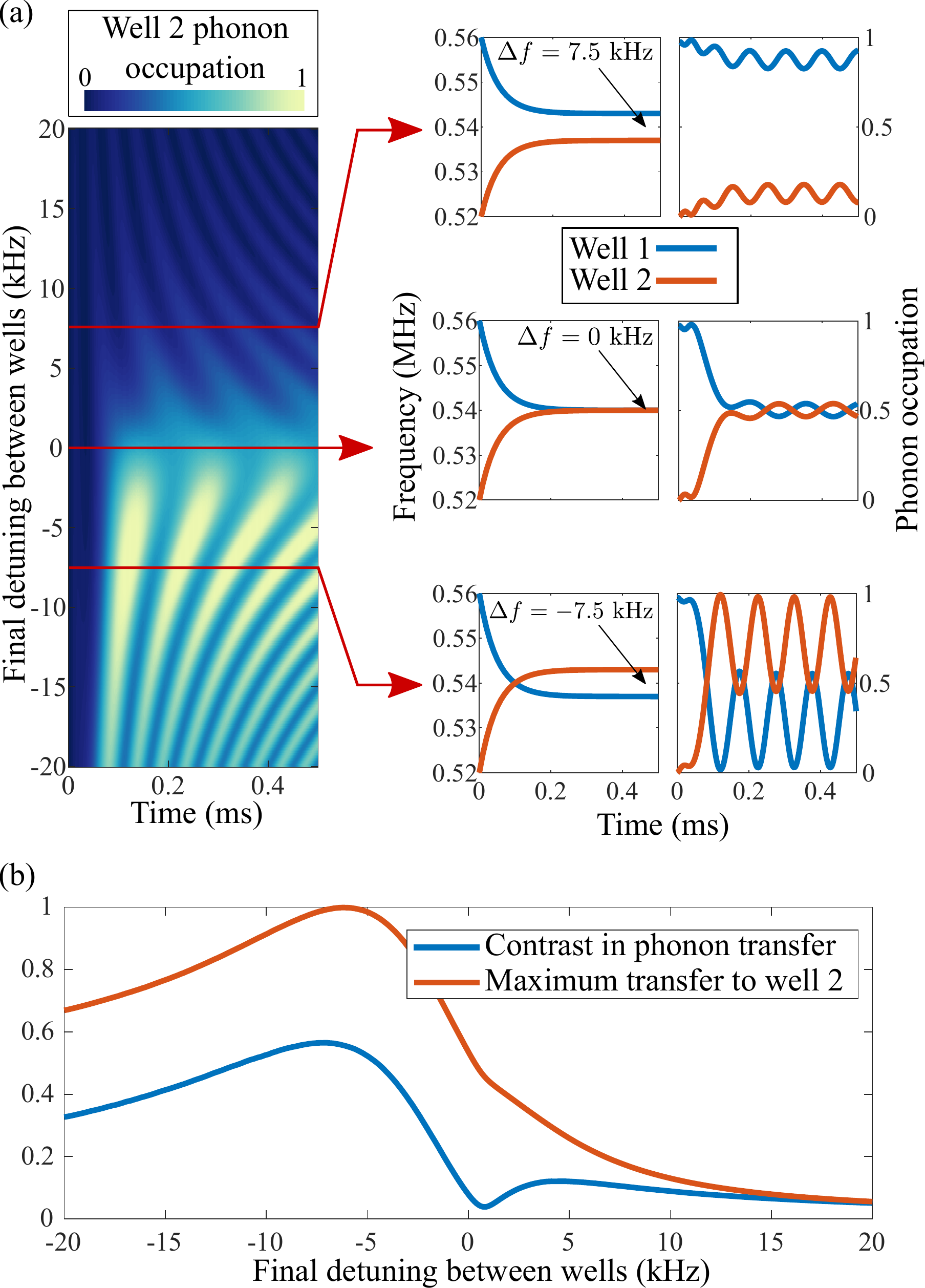}
    \caption{Results of coupled harmonic oscillator simulations, where two off-resonance oscillators are brought into resonance towards a steady state with a certain ``final'' detuning. (a) The image on the left displays the phonon occupation in well 2 for various final detunings, as a function of wait time. Plots on the right show characteristic examples at -7.5, 0, and 7.5 kHz. (b) Due to the finite switching speed, the optimal phonon transfer occurs at \qty{-6}{\kilo\hertz}.}
    \label{fig:app_dc_switch_results}
\end{figure}

Figure \ref{fig:app_dc_switch_results}(a) shows the results of such numerical calculations, shown as a function of time for several final detunings, $\Delta f$. The color plot on the left shows the phonon occupation of well 2 as a function of time, where phonon occupation is calculated as the square of the amplitude of motion. The plots on the right show three examples of phonon exchange, taken at final detunings of $\Delta f=$--7.5, 0, and 7.5 kHz. The example at \qty{-7.5}{\kilo\hertz} results in a transfer frequency of $\Omega_c=\qty{10}{\kilo\hertz}$, similar to what is obtained from the radial phonon exchange measurement, shown in Fig. \ref{fig:radial_phonon_exchange}.
The data illustrates that bringing the two wells perfectly on resonance does not result in maximal phonon exchange contrast. Instead, it is beneficial to overshoot resonance, and settle on a non-zero detuning of the two wells. 

Figure \ref{fig:app_dc_switch_results}(b) shows the contrast in phonon exchange as a function of the final detuning $\Delta f$. Also shown is the maximum phonon occupation that well 2 reaches. Results indicate that a maximum contrast in phonon exchange is reached at a detuning between the wells of about $\Delta f=\qty{-7}{\kilo\hertz}$. If an optimal transfer of motional information from well 1 to well 2 is desired, the final detuning should be set to about \qty{-6}{\kilo\hertz}. 

\clearpage
\section{Generation of entanglement through phonon exchange}
\label{sec:app_phononEx_entanglement}
In Section \ref{subsec:radial_phonon_exchange} we have shown that motional excitation can be coherently transported  between two radially separated ions. In this section, we show how this transfer can be used to generate entanglement, an essential tool for the proposed 2D trap array architecture.

We encode the entangled quantum information in the internal electronic states of the \calciumforty ions, $4S_{1/2}(m=-1/2)$ and $3D_{5/2}(m=-1/2)$, and use the shorthand notation $S$ and $D$. We use the basis $\ket{A_1 A_2,n_1 n_2}$ to describe the two-ion system, where $A_{k}\in\{S_k,D_k\}$ represents the electronic states, and $n_k$ the phonon occupation of ion $k=1,2$.
In the entanglement generation sequence, we consider the two wells on resonance in default operation, while we detune them out of resonance to enable individual ion addressing by applying either red or blue sideband pulses resonant with a target well.

\begin{figure*}
    \centering    \includegraphics[width=\linewidth]{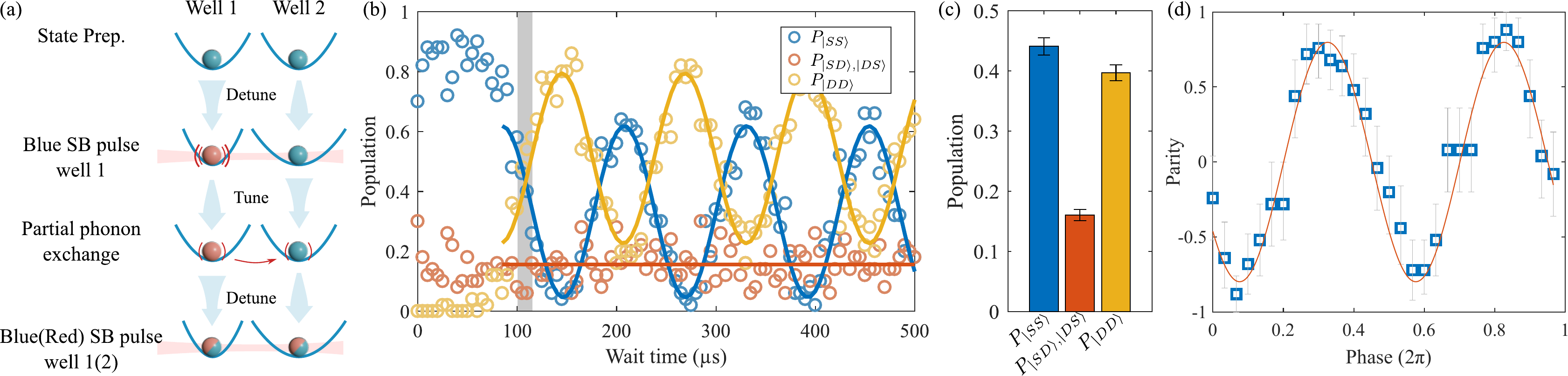}
    \caption{Entanglement of radially separated ions. (a) Measurement scheme, which relies on tuning the motional frequencies of the two wells out of resonance to allow ion-selective phonon injection and removal with red and blue sideband (SB) pulses.
    (b) Measured two-ion populations, as a function of duration that ions are kept in resonance, and phonon exchange can take place. (c) Mean populations, taken at a wait time of \qty{105}{\micro\second}, corresponding to the shaded area of (b). A mean population of $P=P_\textrm{SS}+P_\textrm{DD}=0.84(1)$ is measured. Averages and their 95\% confidence intervals, shown by the error bars, are taken from 5000 repetitions of the experimental sequence. (c) The contrast in parity analysis indicates a visibility of 0.79(3). Error bars represent the 95\% confidence interval given by quantum projection noise \cite{PhysRevA.47.3554}.}   \label{fig:radial_entanglement}
\end{figure*}

The experimental sequence, schematically shown in Fig.~\ref{fig:radial_entanglement}(a), is described in Tab. \ref{tab:entanglement_table}, where bsb and rsb denote respectively that a blue or a red sideband pulse is applied to ion $k=1,2$.
\begin{table}[H]
\centering
\caption{Sequence for entanglement generation through well-to-well coupling. Bsb and rsb are pulses applied to blue and red sidebands, respectively.}
\label{tab:entanglement_table}
\begin{tabular}{c|c|c}
\hline
Step & Operation & State \\
\hline \hline 
1 &  Initialize &  $\ket{S_1S_2,0_10_2}$\\ \hline
2 & \begin{tabular}{@{}c@{}}$\pi$-pulse\\ bsb, ion 1\end{tabular} & $\ket{D_1S_2,1_10_2}$                                                             \\ \hline
3 & Coupling on                                         &                                                                                   \\ \hline
4 & \begin{tabular}{@{}c@{}}wait \\ $t\approx\pi/(2\Omega_c)$\end{tabular} & $\frac{1}{\sqrt{2}}\biggl{(}\ket{D_1 S_2,1_1 0_2}+\ket{D_1 S_2,0_1 1_2}\biggr{)}$ \\ \hline
5 & Coupling off&\\ \hline
6 & \begin{tabular}{@{}c@{}}$\pi$-pulse\\ bsb, ion 1\\ rsb, ion 2 \end{tabular} & $\frac{1}{\sqrt{2}}\biggl{(}\ket{S_1 S_2,0_1 0_2}+\ket{D_1 D_2,0_1 0_2}\biggr{)}$ \\ \hline
\end{tabular}
\end{table}
After the entanglement sequence, we apply a carrier $\pi/2$-pulse, allowing us to alter the parity of the Bell state. The contrast in parity as a function of phase of this analysis pulse corresponds to the coherence of the entangled state.

In Fig. \ref{fig:radial_entanglement}(b) we show the state evolution of both ions under the entanglement operation, as a function of the time the ions are kept on resonance. Similarly to the phonon exchange measurement described in Section \ref{subsec:radial_phonon_exchange}, further explained in Appendix \ref{sec:app_DCswitching}, the two wells' motional frequencies require time to settle to a steady state after switching the trap's DC potentials. After settling, an oscillatory behavior between the $\ket{S_1 S_2}$ and $\ket{D_1 D_2}$ states is observed. Notably, we observe the entangled state $1/\sqrt{2}(\ket{D_1 D_2}+\ket{S_1 S_2})$ at a wait time of \qty{105}{\micro\second}. In Fig. \ref{fig:radial_entanglement}(c) we show the mean populations of the two ions, calculated from $5000$ repetitions of the sequence at this wait time, from which we extract a mean population of $P=P_\m{SS}+P_\m{DD} = 0.84(1)$.

In Fig. \ref{fig:radial_entanglement}(d) we show the parity of the final state as function of the phase of the analysis pulse. The parity contrast corresponds to a visibility of $0.79(3)$, from which we calculate an entangled state fidelity of $F_\m{Tot} = 0.82(2)$~\cite{Sackett2000}. 

\clearpage
\section{Dual RF circuitry}
\label{sec:appendix_RF}
\begin{figure}
    \centering    
    \includegraphics[width=\linewidth]{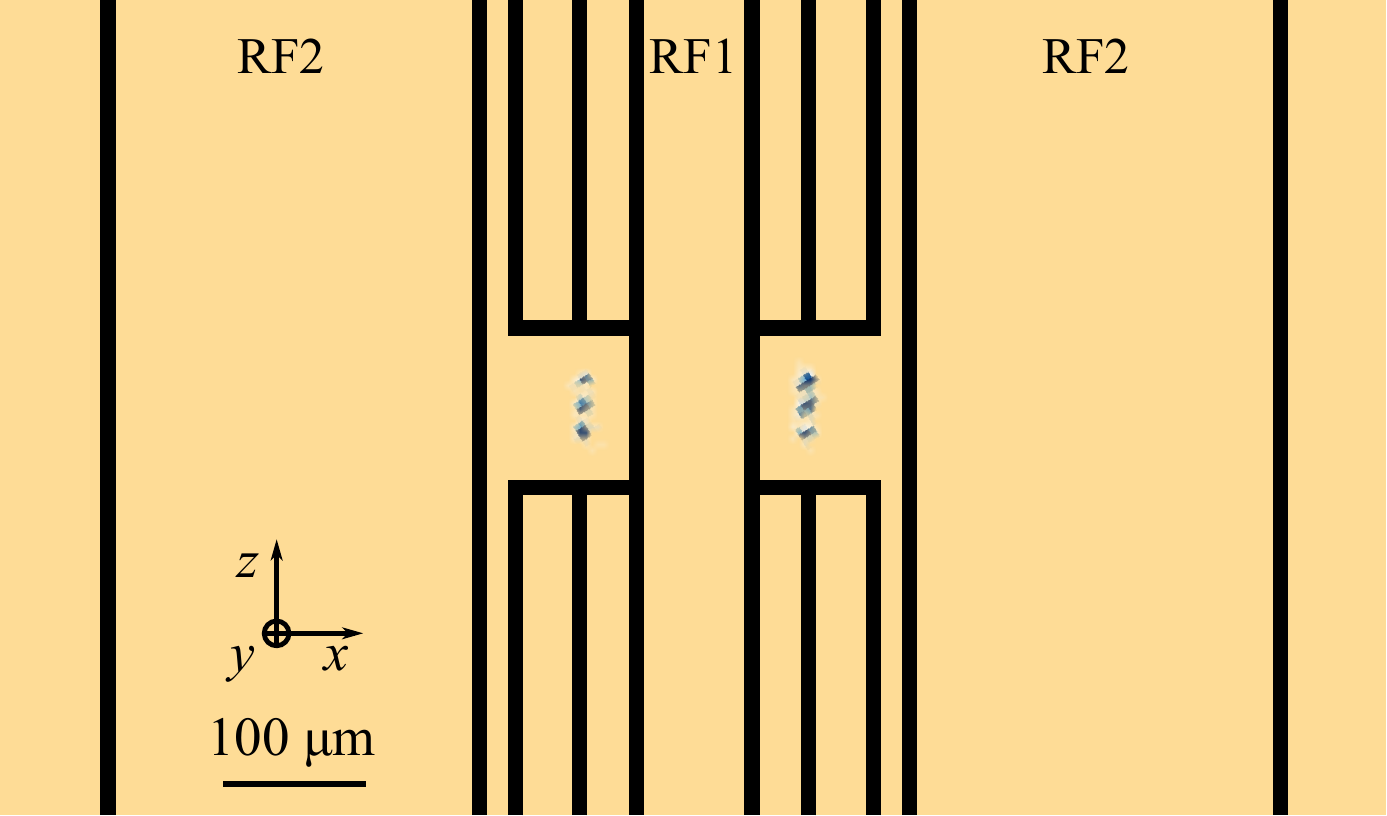}
    \caption{Schematic of the trap used in the RF transport experiments, with $n=3$ ions trapped in each trapping site. Conceptually, the trap design is similar to the one presented in Section~\ref{sec:2S_coupling}. In this case, the inner and outer RF electrodes, RF1 and RF2, have widths of \qty{75}{\micro\meter} and \qty{255}{\micro\meter} and are separated by \qty{115}{\micro\meter}. For RF voltage ratios $\zeta=V_\m{RF1}/V_{\m{RF2}}>0.8$, the RF electrode design generates two RF minima separated in the $x$-axis, about \qty{130}{\micro\meter} above the trap surface.
    The unlabelled electrodes are DC electrodes used for generating axial confinement of ions in the two sites, and for local micromotion compensation.}
    
    \label{fig:4Strap}
\end{figure}

\begin{figure*}
        \centering        \includegraphics[width=1\linewidth]{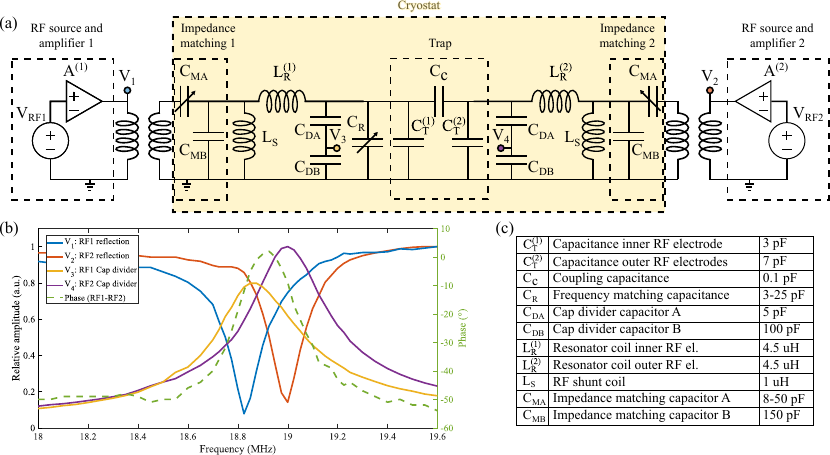}
        \caption{(a) RF circuitry. Two independent RF sources drive two distinct in-cryo resonators. The resonator circuit 1 drives the inner RF electrode C$_\m{T}^{(1)}$, while resonator 2 drives the outer RF electrode C$_\m{T}^{(2)}$.
        A variable capacitor C$_\m{R}$ is soldered in parallel to one of the trap electrodes, allowing us to match the resonance of the two resonators. (b) Resonator characterization. The colored circles in (a) denote the location where the displayed measurements were performed, with respect to ground.  
        The outer RF electrode resonator has a Q factor of 170, and the inner has a Q factor of 120. (c) Table of parameter values for schematic (a). The values of the of the shunt coil L$_\m{S}$ and of the capacitances are taken from the components' data sheets, except for C$_\m{C}$ which is deduced from electrostatic simulations of the trap in Fig.~\ref{fig:4Strap}.
        The values of L$_\m{{1,2}}$ are measured with an LCR meter.}
        \label{fig:resonator}
\end{figure*}

To realize RF shuttling experimentally, we operate the trap shown in Fig.~\ref{fig:4Strap} in a cryostat, equipped with two independent RF resonators which are driven by two phase-locked RF sources, similar to the system presented in Ref.~\cite{Tanaka2021}. One of the two resonators drives the inner RF electrode, labelled RF1 in Fig.~\ref{fig:4Strap}, while the other drives the outer RF electrodes, RF2. The diagram and characterization of the RF circuitry that supplies these electrodes is shown in Fig.~\ref{fig:resonator}. 

An RF voltage signal ($V_{\m{RF1}}$ for the inner RF electrode and $V_{\m{RF2}}$ for the outer RF electrode) is fed into an RF amplifier A and then delivered to the RF resonator via a transformer, to decouple the ground of the amplifier from the ground of the cryostat, to which the rest of the circuit is anchored.
A variable capacitor C$_{\m{MA}}$, and a capacitor C$_{\m{MB}}$, placed respectively in series and in parallel to the input signal, are used to match the impedance of the resonator circuit to the \qty{50}{\ohm} impedance of the cable connecting the amplifier to the cryostat. A coil with inductance L$_\m{S}$ is used to attenuate any DC offset voltage.
Afterward, the RF signal reaches the resonator coil L$_\m{R}$, and the trap electrode C$_\m{T}$, which together form a resonator circuit. Being close to each other on the trap surface, the inner and outer RF electrodes are capacitively coupled, with coupling capacitance C$_{\m{C}}$.
The RF signal on the trap is monitored via a capacitive voltage divider placed in parallel to the trap electrode, consisting of the capacitors C$_{\m{DA}}$ and C$_{\m{DB}}$.
The resonance frequency of the resonator consisting of the inner RF electrode can be tuned with a variable capacitor C$_R$, which is placed in parallel to the trap electrode. This allows us to match the resonance frequency of the two resonators.

After tuning both the resonance frequency and impedance matching of the two resonators, we measure the transmission, reflection, and phase difference of $V_{\m{RF1}}$ and $V_{\m{RF2}}$ while the cryostat is at a temperature of $\approx \qty{10}{\kelvin}$. The measurements are performed using a single RF source connected to the two resonators via an RF splitter, to ensure that a signal with the same phase is fed at the input of both resonators. The reflection of the signal is measured through a bi-directional coupler inserted between the output of the amplifiers and the transformers, while the transmission is monitored via the in-cryo capacitive dividers.

The results are shown in Fig.~\ref{fig:resonator}(b) as a function of the signal frequency. Both the transmission and the reflection curves indicate that the resonance frequency of the two resonators differs by approximately \qty{180}{\kilo\hertz}. Since both gain curves have a full-width-half-maximum of $\approx \qty{300}{\kilo\hertz}$, we can drive the resonators at a frequency corresponding to the average of the two resonance frequencies (\qty{18.95}{\mega\hertz}), while still providing sufficient voltage gain for both.  

\section{MS gate in presence of closely spaced motional modes}
\label{sec:appendix_MS}
In Section \ref{sec:radial_entanglement} we have implemented an M\o lmer-S\o rensen (MS) gate between two ions located in two distinct linear traps. 
In this Appendix, we present the details of performing an MS gate in the presence of two modes of oscillation which are closely spaced in frequency, such that the laser light non-negligibly couples to both modes simultaneously while the gate is driven.

An MS gate is a geometric-phase gate in which ion-ion entanglement is generated by applying a bichromatic laser pulse with a specific frequency detuning from a collective vibrational mode transition, and a target pulse area, dependent on the chosen laser detuning \cite{Sorensen1999}.

While the pulse is applied, both modes of motion of the ions undergo a circular trajectory in phase space during which the spin and motional degrees of freedom of the targeted ions are coupled. As mentioned in Section \ref{sec:radial_entanglement}, to eliminate spin-motion entanglement, the laser pulse length has to be chosen such that the motional modes phase-space trajectories are closed at the end of the gate. The gate time \textcolor{blue}{$t$} \sout{$\tau$} must thus simultaneously satisfy $t = 2 \pi n_m / \delta_m$ for each motional mode m \cite{Sorensen1999}, where $\delta_m$ is the frequency detuning of the laser from the motional sideband transition. Moreover, to obtain a maximally entangled state at the end of the gate, the laser power has to be chosen so that the phase-space area $A$ subtended by the circular trajectories of all the modes participating to the gate is $A = \pi/4$ \cite{Sorensen1999}.

\begin{figure}
    \centering
    \includegraphics[width=1\linewidth]{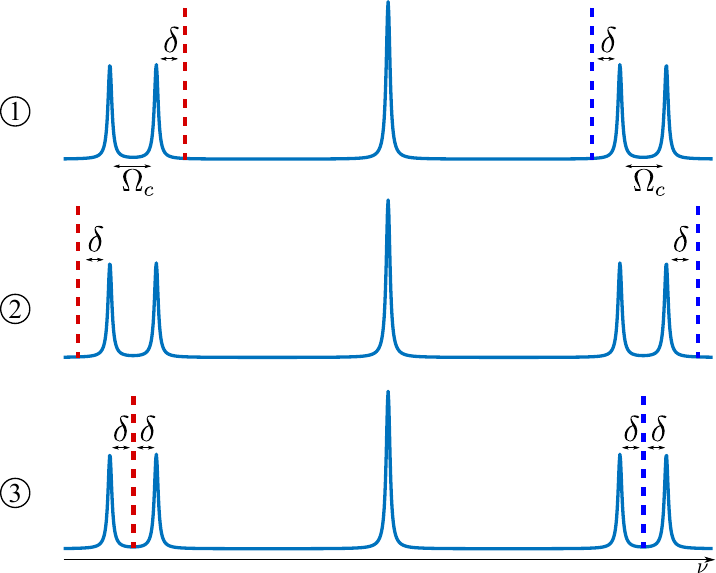}
    \caption{Different approaches to performing a standard MS gate on two closely spaced axial modes. The center peak represents the carrier transition, while the ones symmetric to the center represent the sideband transitions, with a separation of $\Omega_c$. The dashed lines represent the frequencies of the two tones of the bichromatic laser beam. $\delta$ is the laser detuning from the nearest mode.}
    \label{fig:MS_two_modes}
\end{figure}

We consider now a system of two ions with two axial modes of motion whose frequency separation is of the order of the laser detuning $\delta$. As shown in Fig.\ref{fig:MS_two_modes}, with such a motional mode spectrum we can distinguish three different ways of driving an MS gate, depending on the laser detuning with respect to the sideband transitions: red detuned to both modes, blue detuned to both modes, and in between the two. Each of these three scenarios presents a trade-off in terms of gate time and laser power.

Concerning gate time, as mentioned above, the phase-space trajectories of both the stretch and common modes have to close simultaneously at the end of the gate, to avoid unwanted residual spin-motion entanglement \cite{MS_PRA}. For the first two scenarios, this happens just if the laser detuning is set to $\delta = n\Omega_c$ with an integer $n$,
red or blue detuned to both modes respectively, where we recall that $\Omega_c $ is the frequency spacing between the two modes.
For the third case, instead, the condition that satisfies closing both phase-space loops while minimizing the gate time is given by $\delta = \Omega_c / 2$, i.e. if the gate is driven exactly in the middle between the two modes, still restricting the gate time to be twice as large compared to the previous cases.

Concerning laser power, the sum of the areas subtended by the phase-space trajectories for both stretch and common mode have to add up to $A = \pi/4$ to create a maximally entangled state at the end of the gate \cite{Sorensen1999}.
For the first two cases, as the laser detuning has the same sign for both modes, the stretch and common modes will have opposite contributions to the area enclosed by the modes' trajectories in phase-space \cite{MS_Brown_PRA2023}, as the two ions move in-phase for the common mode, and out-of-phase for the stretch mode. For the third case instead, as the laser detuning has opposite sign with respect to the two modes, the two contributions will add up, resulting in less laser power required to create a maximally entangled state at the end of the gate compared to the first two scenarios.

In the rest of the appendix, we quantitatively compare the three cases, focusing on the role that mode heating has on the resulting gate fidelity.
We thus conducted MS gate simulations on a system of two ions, considering heating rate as the only source of decoherence in the system. The laser-ion interaction\cite{MS_PRA} is described by the following interaction Hamiltonian 
\begin{equation}
    \begin{split}
        \mathcal{H}_1 = \frac{\Omega}{2} \sum_{i=1,2}\sum_{m=1,2} &\sigma_+^{(i)} {\rm e}^{{\rm i}\eta(a^\dagger_m + a_m)} ( {\rm e}^{-{\rm i}(\omega_{eg} - \omega_{ax,m} - \delta_m)t} + \\ 
        &+ {\rm e}^{-{\rm i}(\omega_{eg} + \omega_{ax,m} +\delta_m)t}) + {\rm H.c.},
    \end{split}
\end{equation}
where $\sigma_+^{(i)}$ describes the electronic excitation of the $i$-th ion, and $a^\dagger_m$ and $a_m$ are the ladder operators relative to the motional mode $m$. Moreover, the parameter $\omega_{eg}$ is the carrier transition frequency, $\omega_{ax,m}$ the motional mode frequency, $\delta_m$ the laser detuning from the $m$-th motional mode, $\eta$ the Lamb-Dicke factor, which we consider equal for both ions,  and $\Omega$ is the
resonant Rabi frequency of the ith ion in the laser field.
We consider the two ions to have an axial frequency $\omega_{ax} =  2\pi$ \qty{740}{\kilo\hertz}, with a mode splitting $\Omega_c=2\pi$ \qty{5.2}{\kilo\hertz}, and the stretch mode of oscillation to be at a lower frequency compared to the common mode, as would be the case for the axial modes of two radially coupled ions. Moreover, we consider the stretch and common mode heating rates the same as the ones measured in Sec.\ref{sec:radial_heating} for an axial frequency of $\omega_{ax} = 2\pi$ \qty{740}{\kilo\hertz}. The gate infidelity results for the three cases described in Fig.\ref{fig:MS_two_modes}(b) are listed in Table \ref{tab:fidelity_values}.
\begin{table}[]
\centering
\caption{Summary of MS gate infidelity for a system of 1 ion per well, and radially coupled axial modes.}
\label{tab:fidelity_values}
\begin{tabular}{c|c|c|c}
\begin{tabular}{@{}c@{}} Heating \\rates (ph/s) \\(STR,COM)\end{tabular} & \begin{tabular}{@{}c@{}} Gate infidelity\\Case \circled{1}\\($\tau$ = 190 \unit{\micro\second},\\$\Omega_{sb} \approx$ 3.7 \unit{\kilo\hertz})\\ \end{tabular} & \begin{tabular}{@{}c@{}} Gate infidelity\\Case \circled{2}\\($\tau$ = 190 \unit{\micro\second},\\$\Omega_{sb} \approx$ 3.7 \unit{\kilo\hertz})\\ \end{tabular} & \begin{tabular}{@{}c@{}} Gate infidelity\\Case \circled{3}\\($\tau$ = 380 \unit{\micro\second},\\$\Omega_{sb} \approx$ 0.9 \unit{\kilo\hertz})\\ \end{tabular} \\ \hline
(2.6,18)    & $\approx$ 0.13$\%$   & $\approx$ 0.35$\%$     & $\approx$ 0.2$\%$  \\                         
\end{tabular}
\end{table}

For cases \circled{1} and \circled{2} the gate time $t$ is half that of case \circled{3}, as the laser detuning $\delta$ is twice as large for the first two scenarios compared to the third.
This comes at the expense of more laser power for two reasons. First, the required sideband Rabi frequency $\Omega_{sb}$ is linearly proportional to the detuning $\delta$ \cite{MS_PRA}. Second, as pointed out previously, for the case \circled{3} the geometric phases accumulated by the stretch and common modes add up, while they subtract for the cases \circled{1} and \circled{2}. The combination of these two effects results in a total factor of 4 more laser power for the first two scenarios compared to the third. 
Concerning gate fidelities, we can observe how implementing the MS gate as explained in case \circled{1}, and thus setting the laser bichromatic driving farther away from the common mode, which displays a higher heating rate, improves the gate fidelity by a factor of almost 3 compared to case \circled{2} and
by a factor of almost 2 compared to case \circled{3}.

\section{Effective Hamiltonian for transversal gates between radially separated wells}
\label{sec:tv_inter_well_gate}
\label{appendix:transversal_gates}

We can derive an effective Hamiltonian describing the interactions for the considered parameter regime. As discussed in in Sec.~\ref{sec:double_well_physics} and App.~\ref{sec:appendix_MS}, we find two closely spaced axial modes that are separated from higher-frequency modes, meaning that we can drive them by near-resonant lasers without affecting the other higher-frequency modes.
By tuning the difference between the laser frequencies $\Delta \omega_L = \omega_1 - \omega_2$ closely to these axial vibrational frequencies, we can neglect all non-resonant terms and obtain the Hamiltonian $H_d = \frac{\Omega}{2} \sum_j Z_j {\rm e}^{{\rm i} \Delta \vec{k}_L \cdot \vec{r}_j^0} {\rm e}^{{\rm i}(\Delta\vec{k}_L \cdot \Delta \vec{r}_j - \Delta \omega_L t)} + \mathrm{H.c.}. $
Here, $\Omega$ is the gate amplitude of the two-photon carrier transition, $Z$ is the Pauli-Z matrix, $\vec{r}_j^0$ is the equilibrium position of ion $j$, $\Delta \vec{r}_j$ is the displacement of ion $j$ from its equilibrium position, and the effective wavevector \hbox{$\Delta\vec{k}_L = \vec{k}_1 - \vec{k}_2 = k (0, \sin{\theta}, \cos{\theta})$} lies in the $yz$-plane. 
As customary, one can expand this expression for small Lamb-Dicke parameters $\eta_{\alpha} = \Delta k_{L, z} \sqrt{\hbar / 2 m \omega_{\alpha}^{\mathrm{ax}}}$, where $m$ is the ion mass. In the rotating wave approximation we obtain \hbox{$H_d = \frac{\Omega}{2} \sum_{j, m} {\rm i e}^{{\rm i} \vec{\Delta k}_L \cdot \vec{r}_j^0} \eta_{m} {\nu}^{(m)}_{j} Z_j a_{m}^{\dag} {\rm e}^{{\rm i}\delta_{m}t} + \mathrm{H.c.}$}. Here, \hbox{$|\delta_m| = |\omega_m^{\mathrm{ax}} - \Delta \omega_L|$} is the detuning between the laser beatnote $\Delta\omega_L$ and the axial mode $m$, and 
$\nu^{(m)}_{j}$ is the amplitude of the axial oscillations of ion $j$ in the corresponding   axial mode $m$. 

Using a Magnus expansion, one readily finds that upon closure of the phase-space trajectories, the above state-dependent dipole forces lead to
an effective Hamiltonian~\cite{mintert2001ion, bermudez2012quantum} that reads
\begin{align}
    H_{\mathrm{eff}} = \sum_k \sum_{j \neq k} J_{jk}^{\mathrm{eff}} Z_j Z_k
    \label{eq:app_effective_Hamiltonian}
\end{align}
with the effective interaction $J_{jk}^{\mathrm{eff}}$ between ion $j$ and $k$ is
\begin{align}
    J_{jk}^{\mathrm{eff}} &= \chi_{jk} \cos{(\Delta\vec{k}_L \cdot \vec{r}_{jk}^0)}, \label{eq:effective_interaction_J_jk}\\ \chi_{jk} &= -\sum_m \frac{\Omega^2 \eta_{m}^2}{4m  \delta_m} \nu^{(m)}_j\nu^{(m)}_{k}. \label{eq:effective_interaction}
\end{align}

The effective interaction therefore depends on the geometric factor $J_{jk}^{\mathrm{eff}} \propto \cos{(\Delta\vec{k}_L \cdot \vec{r}_{jk}^0)}$, which lies at the core of the idea for the geometric cancellation as discussed in the main text. Instead of shaping the $J_{ij}^{\rm eff}$ matrix by allowing for individual Rabi frequencies $\Omega\to\Omega_j$, or multiple beams to control the detunings $\delta_{\alpha}\to\delta_{\alpha,j}$, the condition for vanishing interaction $J_{jk}^{\mathrm{eff}}=0$,~i.e. geometric cancellation, is \hbox{$\Delta k _{L, z} = (2p + 1) \cdot \frac{\pi}{2 |z_1 - z_2|}$}, which corresponds to an angle \hbox{$\theta = \arccos{( (2p + 1)\pi/ (2 k|z_j - z_k|))}$} for the effective wavevector \hbox{$\Delta\vec{k}_L = k \cdot (0, \sin{\theta}, \cos{\theta})$}.

\clearpage
\bibliographystyle{apsrev4-2-titles}
\bibliography{main.bib}
    
\end{document}